\documentclass[a4paper,fleqn,usenatbib]{mnras}

\usepackage{newtxtext,newtxmath}
\usepackage{threeparttable}

\usepackage[T1]{fontenc}
\usepackage{ae,aecompl}

\usepackage{graphicx}	
\usepackage{amsmath}	

\usepackage{amssymb}	
\usepackage{subcaption}
\usepackage{lscape}
\usepackage{xcolor}
\usepackage{makecell}

\hypersetup{citecolor=magenta}

\newcommand{\HI}{H\textsc{i}}	
\newcommand\todo[1]{\textcolor{red}{\textbf{#1}}}
\newcommand\vtwo[1]{\textcolor{black}{#1}}
\newcommand\vthree[1]{\textcolor{black}{#1}}
\newcommand\refrep[1]{\textcolor{black}{#1}}

\graphicspath{{./}{Figures/}}

\title[The \HI\ in MHONGOOSE dwarf galaxy UGCA~320]{Tracing neutral hydrogen in UGCA~320: A MHONGOOSE perspective on an edge-on dwarf galaxy in a group environment}

\author[Nikki Zabel et al.,]{Nikki Zabel,$^{1}$\thanks{E-mail: nikki.zabel@uct.ac.za}
D.J. Pisano,$^{1}$
Sushma Kurapati,$^{2}$
Omri Scannell,$^{1}$
Notahiana Ranaivoharimina$^{1}$ \newauthor
Julia Healy$^{3, 4}$,
W.J.G. de Blok$^{2,1,5}$, 
Peter Kamphuis,$^{6}$
Adebusola B. Alabi,$^7$
S. Ilani Loubser,$^{7,8}$ \newauthor
and Moses K. Mogotsi$^{9,1}$
\\
$^{1}$Department of Astronomy, University of Cape Town, Private Bag X3, Rondebosch 7701, South Africa \\
$^{2}$Netherlands Institute for Radio Astronomy (ASTRON), Oude Hoogeveensedijk 4, 7991 PD Dwingeloo, the Netherlands \\
$^{3}$Jodrell Bank Centre for Astrophysics, School of Physics and Astronomy, University of Manchester, Oxford Road, Manchester M13 9PL, UK \\
$^4$United Kingdom SKA Regional Centre (UKSRC), UK \\
$^5$Kapteyn Astronomical Institute, University of Groningen, PO Box
800, 9700 AV Groningen, The Netherlands \\
$^6$Ruhr University Bochum, Faculty of Physics and Astronomy, Astronomical Institute
(AIRUB), 44780 Bochum, Germany \\
$^{7}$Centre for Space Research, North-West University, Potchefstroom Campus, Potchefstroom 2520, South Africa \\
$^{8}$National Institute for Theoretical and Computational Sciences (NITheCS), Potchefstroom 2520, South Africa \\
$^9$South African Astronomical Observatory, PO Box 9, Observatory Cape Town 7935, South Africa \\
}

\date{Accepted 2026 March 26. Received 2026 March 05; in original form 2025 December 09}

\pubyear{2026}

\begin{document}
\label{firstpage}
\pagerange{\pageref{firstpage}--\pageref{lastpage}}
\maketitle

\begin{abstract}
We present a detailed analysis of the neutral atomic gas (\HI) in \vthree{the} dwarf galaxy UGCA~320, observed with the MeerKAT telescope as part of the MHONGOOSE (MeerKAT \HI\ Observations of Nearby Galactic Objects: Observing Southern Emitters) programme. \vthree{In a small group consisting of three dwarf galaxies, all of which contain \HI, it is the most massive.} \vthree{Detailed kinematic modelling shows that UGCA~320 contains a substantial amount of (kinematically) anomalous gas ($\gtrsim$20\%), at least $\sim$30\% of which is likely the result of \vthree{a tidal} interaction with its neighbour UGCA~319. It also reveals that UGCA~320 likely harbours a star-formation driven outflow, and that $\sim$10\% of its \HI\ is extra-planar and has a filamentary structure. Although UGCA~320 aligns with established scaling relations from the literature, its neutral hydrogen content is notably complex -- shaped by its immediate environment. This underscores the importance of deep, resolved observations and detailed kinematic analyses to capture the nuances of galaxy evolution.}
\end{abstract}




\section{Introduction}
\label{sec:intro}
It has been known for some time that for present-day galaxies to maintain their star formation external supplies of fresh fuel are required \citep[e.g.][]{Kennicutt1998, Bigiel2008, Leroy2013}. \refrep{The acquisition of such external gas can occur through wet mergers, where a gas-rich lower-mass companion is ``absorbed'' by the main galaxy, or, to a certain extent, through major mergers. Additionally, gas can be accreted from the cosmic web.} It has been estimated that galactic cannibalism alone is not sufficient to sustain the star formation in higher-mass galaxies \citep{Sancisi2008, DiTeodoro2014}, and that mergers only become important in galaxy groups and clusters \citep{Voort2011}. Thus, accretion must play a role in the maintenance of star formation in galaxies. According to simulations, \vthree{a non-negligible amount of this accretion likely falls into the category of ``cold mode accretion'', where cool gas flows into the galaxy along the filaments of the cosmic web \citep[e.g.][]{Brooks2009, Keres2009, Voort2011}. Although the prevalence of cold mode accretion is a topic of debate, particularly at $z=0$, it is more likely to occur in low-mass haloes \citep[and references therein]{Kamphuis2022}. In particular, \citet{Kamphuis2022} showed that the \HI\ observed \refrep{in and around galaxies} at $z=0$ is insufficient to maintain ongoing star formation. This suggests that, if cold-mode accretion does take place, it must do so through ionised hydrogen.} To-date, there has been no definite observational evidence of ongoing cold accretion from the cosmic web, mainly because the column densities of the accreted gas were too low to be observed at the required resolution ($M_\text{\HI} \sim 10^{17} - 10^{18} \text{cm}^{-2}$, \citealt{Voort2019, Ramesh2023}). However, there have been several observations of ``extra-planar gas'' (EPG) around galaxies, neutral gas surrounding the stellar disc, that in some cases can have \vthree{a maximum vertical extent} of up to well over 10 kpc \citep[e.g.][]{Swaters1997, Oosterloo2007, Sancisi2008, Lucero2015, Kurapati2025}. \citet{Marasco2019} even find that EPG is common in nearby spiral galaxies (they detect it in over 85\% of their sample). EPG can account for up to 30\% of the neutral gas in galaxies \citep[e.g.][]{Oosterloo2007, Gentile2013, Vargas2017}, with typical values around 15\% \citep{Marasco2019}. This gas is often lagging in velocity compared to the gas in the disc \cite[e.g.][]{Oosterloo2007, Kurapati2025}. The most common explanation for the presence of EPG is galactic fountains: large-scale outflows of gas and other constituents of the interstellar medium (ISM) into the areas surrounding the galaxies, which may eventually cool and ``rain'' back down onto the galaxy disc \citep[e.g.][]{Shapiro1976, Bregman1980, Swaters1997, Oosterloo2007, Fraternali2008, Vargas2017, Marasco2019}. However, while they likely account for a significant fraction of the observed EPG, galactic fountains may not explain all of it, and other mechanisms, such as accretion, mergers, and tidal interactions, could still play a role in the assembly of EPG reservoirs. In order to study the EPG around galaxies, and its origin, in detail, high-resolution, high-sensitivity observations of the neutral gas surrounding a diverse sample of galaxies are required.

\vthree{The MeerKAT \HI\ Observations of Nearby Galactic Objects: Observing Southern Emitters (MHONGOOSE) survey} is a Meerkat Large Survey Project that acquired ultra-deep resolved \HI\ observations of 30 nearby dwarf and disc galaxies with stellar masses ranging from 10$^7$ M$_\odot$ to just under 10$^{11}$ M$_\odot$ \citep{Blok2024}. The main goal of the survey is to detect and characterise any low-column density gas around these galaxies, constrain where it is flowing from and to, and to connect these findings with other star formation properties. Highlights from previous studies of MHONGOOSE galaxies include the discovery of a region of clumpy anomalous gas associated with NGC 5068, which is possibly an accretion feature \citep{Healy2024}, the discovery of a gas-rich low-surface brightness galaxy in the Dorado Group -- the lowest stellar mass \HI-rich galaxy observed outside the Local Group \citep{Maccagni2024}, the discovery of a significant number of low-mass companions surrounding UGCA250, a substantial, lagging halo of EPG around it, and evidence for tidal interactions, which are responsible for a fraction of the EPG \citep{Kurapati2025}. A stacking experiment focused on the surroundings of six MHONGOOSE galaxies has shown that the column densities of \HI\ directly surrounding galaxy discs may be lower than predicted by simulations, making it challenging to directly observe such features \citep{Veronese2025}. \vthree{\citet{Marasco2025} have shown that Milky Way-like galaxies in the TNG50 and FIRE-2 simulations have stellar masses, SFRs, and \HI\ content that are in agreement with MHONGOOSE observations, though some atomic-to-molecular hydrogen partitioning recipes overestimate their H$_2$ content, and simulated galaxies tend to have more irregularities/complexities and low-density gas compared to observed galaxies.} 

\refrep{In this work, UGCA~320 (J1303-17B) is studied in detail as part of the broader effort of MHONGOOSE to characterise low-density \HI\ and star formation processes in a wide range of galaxies. The papers mentioned above explore various aspects of the MHONGOOSE framework. Within this context, UGCA~320 falls in the mid-mass range of the sample and toward the higher-mass end of the dwarf galaxy regime. It is part of a small group of galaxies, of which one is a close companion. Additionally, similarly to UGCA~250, its disc is viewed edge-on, enabling a clear view of its extra-planar gas. Thus, UGCA~320 offers a unique opportunity to study the nuances of the evolution of low-density \HI\ in higher-mass, star forming dwarf galaxies in a group environment.}

\subsection{UGCA~320}
\label{subsec:ugca_320}

\begin{figure}
  \begin{center}
    \includegraphics[width=0.5\textwidth]{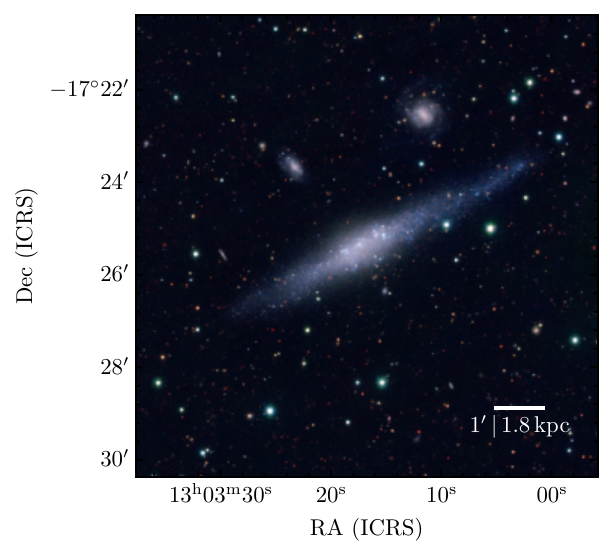}
  \end{center}
  \caption{\vthree{False colour optical image of UGCA~320 created using OmegaCam \textit{g}- and \textit{r}- band, and the DECam \textit{z}-band images. North is up and west is to the right.} A scale bar is shown in the bottom-right corner of the image. UGCA 320's \vthree{stellar} disc is asymmetric: it shows a thickening around the brightest region, which is slightly off-centre, and the west side of the disc is thinner and more elongated compared to the east side. The two objects north of UGCA~320 are background galaxies.}
  \label{fig:optical}
\end{figure}

UGCA~320 is among the lower-mass galaxies in the MHONGOOSE survey, with log$\left(M_\star/\text{M}_\odot\right)~=~7.91$ \citep{Blok2024}. Note, however, that this is a Wide-field Infrared Survey Explorer (\textit{WISE}, \citealt{Wright2010}) derived stellar mass measurement, and the correlation between \textit{WISE} band 1 luminosity and stellar mass is poorly constrained below log$\left(M_\star/\text{M}_\odot\right)~=~8.5$. Moreover, the W1 -- W2 colour of UGCA320 seems unphysical, which implies that the W2 colour correction might not be meaningful. K-band magnitude measurements by \citet{Karachentsev2017} suggest that its mass might be closer to log$\left(M_\star/\text{M}_\odot\right)~=~8.75$. However, an accurate stellar mass measurement is of limited importance for the analysis in this work, and adopting the higher value does not result in a significant difference in position on the relevant scaling relations (see below). \refrep{Thus, in the interest of consistency, we adopt the MHONGOOSE value of log$\left(M_\star/\text{M}_\odot\right)~=~7.91$ here.} UGCA~320's general properties are summarised in Table \ref{tab:gen_props}. It is viewed almost edge-on, as can be seen in the optical image shown in Figure \ref{fig:optical}. This is a \vthree{false colour image created using \textit{g}- and \textit{r}-band images from OmegaCam \citep{Kuijken2002, Kuijken2011}, mounted on the VLT (Very Large Telescope) Survey Telescope (VST, \citealt{Arnaboldi1998}), and a \textit{z}-band image from the Dark Energy Camera (DECam, \citealt{Flaugher2015}).} UGCA~320's stellar disc is asymmetric: it bulges around the brightest region in the south-east, and has a long, thin extension towards the north-west, possibly the remnant of a past tidal interaction \citep{Alabi2025}. The two galaxies north of it are background galaxies. UGCA~320 has a star formation rate between \vthree{SFR $\sim 0.025$ and $0.095\text{ M}_\odot \text{ yr}^{-1}$: the lower value is estimated from the H$\alpha$ flux in the Multi-unit spectroscopic explorer (MUSE) field-of-view (FoV, \citealt{Alabi2025}), while the higher value is \textit{WISE}-derived \citep{Blok2024}. The \textit{WISE}-derived value is close to the FUV flux derived SFR measurement from \citet{Karachentsev2017}, who find a SFR of 0.12 M$_\odot \text{ yr}^{-1}$.} While the MUSE FoV covers the most prominent H\textsc{ii} regions, and H$\alpha$ is a well-established tracer of recent star formation, a significant portion of the stellar disc is not covered by these observations. \refrep{Thus, this number could be an underestimate, and should be interpreted with caution.}

\begin{figure}
  \centering
  \includegraphics[width=0.5\textwidth]{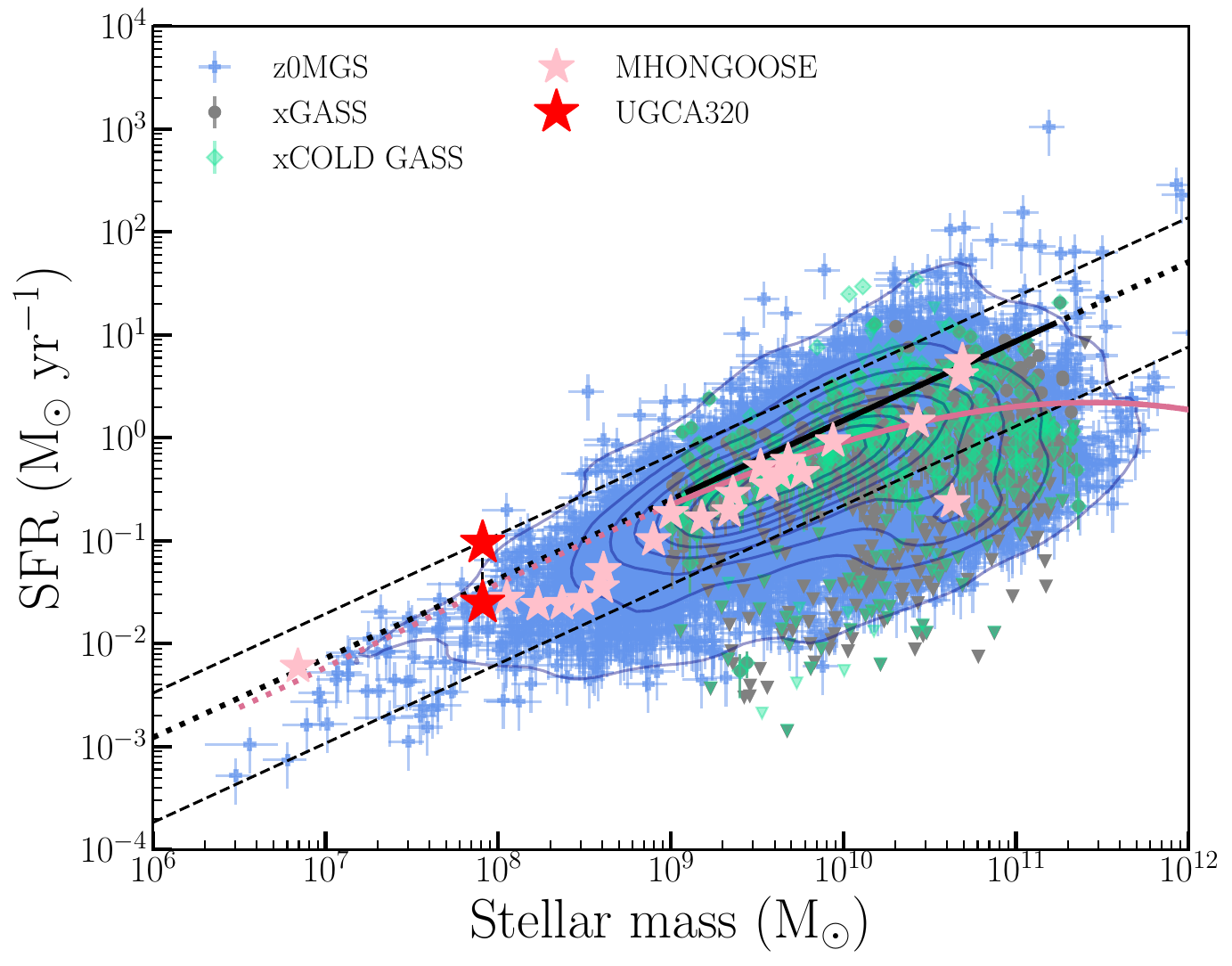}
  \caption{\refrep{The position of UGCA~320 with respect to the star formation main sequence. UGCA~320 is shown as two red stars representing the SFR value from MHONGOOSE \citep{Blok2024} and a lower measurement from H$\alpha$ flux within the MUSE FoV \citep{Alabi2025}. The remaining MHONGOOSE galaxies for which reliable SFR measurements are available are represented by pink stars \citep[Table 1 and Figure 3]{Blok2024}. The SFMS from the extended CO Legacy Database for the GALEX Arecibo SDSS Survey (xCOLD GASS, \citealt{Saintonge2017}) is shown as a pink line, and that from \citet{Elbaz2007} as a black line. The lines are solid where the relation is empirically derived, and dotted where extrapolated. The 1$\sigma$ confidence interval for the relation from \citet{Elbaz2007} is represented by a black line. For reference, galaxies from the z = 0 Multiwavelength Galaxy Synthesis (z0MGS, \citealt{Leroy2019}) are shown as blue plus signs, and darker blue contours are spaced evenly between the maximum kernel density estimate value of this sample and 1\% of the maximum. Galaxies from the extended Galaxy Evolution Explorer (GALEX) Arecibo SDSS (Sloan Digital Sky Survey) survey (xGASS, \citealt{Catinella2018}) are shown as grey dots, and galaxies from xCOLD GASS \citep{Saintonge2017} as green diamonds. For the sake of clarity, values with uncertainties of $>$50\% in either stellar mass or star formation rate were omitted from the comparison samples, and downward pointing triangles indicate upper limits.  UGCA~320 lies just above the (extrapolated) scaling relations from the literature (within 1$\sigma$ from the relation from \citet{Elbaz2007}) and the majority of the z0MGS sample around its stellar mass. Thus, it has normal to high star formation for its stellar mass.}}
  \label{fig:sfms}
\end{figure}

Figure \ref{fig:sfms} shows the position of UGCA~320 on the star formation main sequence (SFMS). 
\refrep{It} lies close to the scaling relations from the literature, and has a regular to slightly elevated SFR for its stellar mass. This also emphasises the need to study gas-rich low mass galaxies. In accordance with its ongoing star formation, UGCA~320 has blue colours (see Figure 1 in \citealt{Alabi2025}). \citet{Alabi2025} perform a full \vthree{stellar kinematics and population analysis} of UGCA~320 using data from the Robert Stobie Spectrograph (RSS) on the South African Large Telescope (SALT) and MUSE. They conclude that UGCA~320 is indeed dominated by young stars with ages $<$1 Gyr, although it also harbours a substantial population of old stars with ages $>$10 Gyr. Both the stars and the ionised gas have sub-solar metallicities (between 15\% and 30\% of the solar value).

\vthree{UGCA~320 is part of a small group of three dwarf galaxies in total, all of which are within the footprint of the MeerKAT observations presented in this work (introduced in \S \ref{sec:observations}). Its companions are dwarf irregular UGCA~319, located 32.7 kpc towards its north-west in projection and with a stellar mass of log$\left(M_\star/\text{M}_\odot\right)~=~8.01$ \citep{Karachentsev2017}, and the even smaller LEDA 886203, $\sim$50 kpc in projection towards its north-east. A rough estimate of its stellar mass using data from WISE band 1 (3.4 $\mu$m) and 2 (4.6 $\mu$m), following the recipe in \citet{Jarrett2023}, gives a value of log$\left(M_\star/\text{M}_\odot\right)~=~6.2$. However, as described above, this method is poorly constrained in the relevant stellar mass range, and only a handful of calibration data points exist below log$\left(M_\star/\text{M}_\odot\right)~=~7.5$. Thus, this value serves as a ballpark estimate for context only, and is not used for further analysis.}

\vthree{The distance to UGCA~320 was estimated to be $6.03^{+0.26}_{-0.21}$ Mpc, using the Tip of the Red Giant Branch (TRGB, \citealt{Karachentsev2017}). Similarly, the distance to UGCA~319 was estimated to be $5.75 \pm 0.18$ Mpc. However, the systemic velocities of the \HI\ line are very similar for UGCA~320 and UGCA~319, with a difference of $\leq$ 20 km s$^{-1}$ (Scannell et al., in prep). Therefore, in this work we adopt the distance to UGCA~320 for both galaxies, which is still in agreement with the numbers from \citet{Karachentsev2017}.}

UGCA~320 may also be part of a larger, loose galaxy group or ``association'' centred around NGC 5068, although this is debated in the literature \citep{Garcia1993, Pisano2011, Kourkchi2017}. \vthree{Using an iterative method based on optical velocities and the \HI\ velocity dispersions of candidate galaxies, \citet{Pisano2011} increased the number of members in the loose group first identified by \citet{Garcia1993}.} \citet{Kourkchi2017} confirm the same galaxy association using their members' expected location on various observed scaling relations, following \citet{Tully2015}. \citet{Karachentsev2017}, on the other hand, conclude that UGCA~320 and UGCA~319 are a tight pair of galaxies that are physically isolated from their \vthree{collective} nearest neighbour by 730 $\pm$ 310 kpc along the line of sight, implying a physical distance of at least $\sim$950 kpc (LEDA 886203 was not part of the discussion). A map of this loose group and optical images of its constituents can be seen in Figure 13 in \citet{Healy2024}, which shows that the projected distance between UGCA~320 and NGC 5068 is slightly under 1 Mpc. \\

\vthree{This paper is organised as follows. In \S \ref{sec:observations} we describe the MeerKAT observations and data reduction. In \S \ref{sec:analysis} we present the MHONGOOSE measurements of UGCA~320, including its \HI\ fraction, channel maps, moment maps, position-velocity diagrams, and a radio continuum image. In \S \ref{subsec:tirific} we describe a detailed analysis of the \HI\ distribution and kinematics using tilted ring modelling. In \S \ref{sec:anomalous_gas} we discuss our findings, in particular anomalies in the \HI\ reservoir and their possible origins, and in \S \ref{sec:discussion} we provide a summary of this work.}

\section{Observations \& Data Reduction}
\label{sec:observations}
UGCA~320 was observed as part of the MHONGOOSE survey. A detailed description of the observations and data reduction is presented in \S 4 and \S 5 in \citet{Blok2024}, and is summarised below. 

To meet its ambitious scientific goals, the MHONGOOSE survey targets \vthree{a column density sensitivity of $N_\text{\HI}~=~5~\times~10^{17}~\text{atoms cm}^{-2}$ ($3 \sigma$ over 16 km s$^{-1}$ at 90$^{\prime\prime}$ resolution)}. Achieving this sensitivity requires 55 hours of observation per galaxy. These 55 hours (resulting in the ``full-depth'' data) are spread over 10 ``single track'' observations (five ``rising'' and five ``setting'') of 5.5 hours each. Each single track observation typically consists of 10 minutes of observing a primary calibrator (J1939--6342 or J0408--6545), five two-minute cycles of observing a secondary/phase calibrator, and $\sim$55 minutes on-source. 

While each MHONGOOSE galaxy was observed in 32k narrow band and 4k wide band mode simultaneously, for this \HI\ study we utilise the narrow band observations observations only. This band consists of 32768 channels of 3.265 kHz each, resulting in a total bandwidth of 107 MHz, and a raw velocity resolution of 0.7 km s$^{-1}$ at the rest frequency of the \HI\ line (1420.405 MHz). From this, 10,000 channels were selected (channels 16384 -- 26383, corresponding to 1390.0 -- 1422.7 MHz) and binned by two channels, resulting in a resolution of \vthree{6.53} kHz (corresponding to 1.4 km s$^{-1}$). 

The data were calibrated using the Containerized Automated Radio Astronomy Calibration (\texttt{CARACal}) pipeline \citep{Jozsa2020}, using standard data reduction steps, which \vthree{include} flagging of the target and calibrators for RFI, cross-calibration, self-calibration, continuum-subtraction, and imaging of the continuum and \HI\ line. The final \HI\ cubes were created and deconvolved using \texttt{wcsclean} \citep{Offringa2014}.

\begin{figure}
  \centering
  \includegraphics[width=0.48\textwidth]{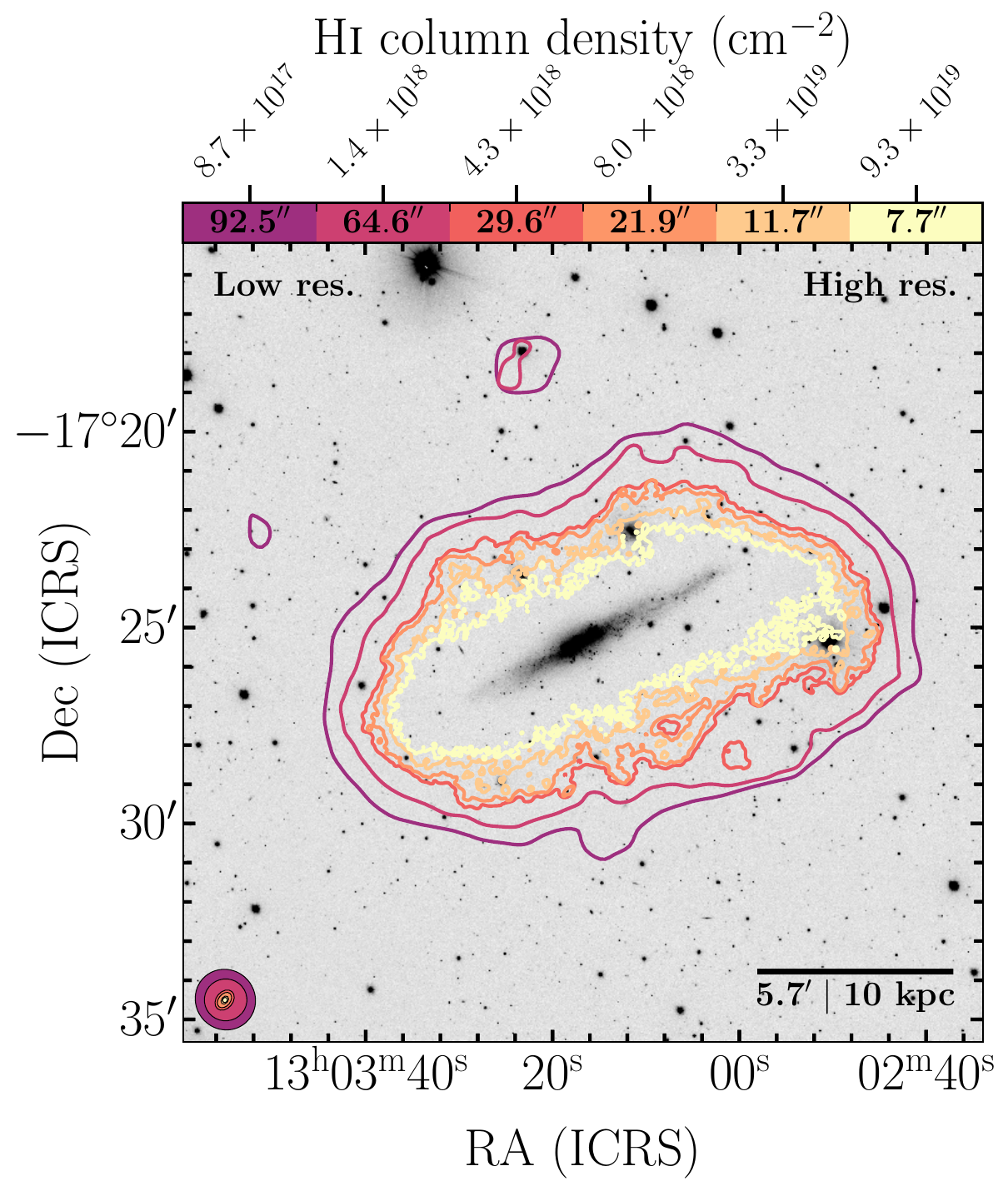}
  \caption{\vtwo{Multi-resolution image showing the resolutions of the six different cubes for UGCA~320, as described in \S \ref{sec:observations} and \citet{Blok2024}, overlaid on an optical \textit{g}-band image from the VST. Each coloured contour represents the 3$\sigma$ threshold in the moment \vthree{zero} map. Their resolutions, \refrep{here defined as the geometric mean of the synthesised beam major and minor axes}, are quoted in the colour bar, and the corresponding beams are shown in the bottom-left corner. The sensitivities at 3$\sigma$ are quoted above the colour bar for each resolution. A scale bar is shown in the bottom-right corner. The different resolution cubes offer a wide range in resolution and sensitivity, providing different views of UGCA~320.}}
  \label{fig:multi_res}
\end{figure}

\refrep{Cubes are imaged at six different resolutions/sensitivities, using different robust weighting parameters and levels of taper, to serve different science cases. Details for each version of the output cube can be found in Table 4 in \citet{Blok2024}. In Figure \ref{fig:multi_res}, 3$\sigma$ contours (at 16 km s$^{-1}$) of the moment zero maps of all six cubes for UGCA~320 are shown, highlighting their difference in resolution and sensitivity. Here, unless mentioned otherwise, we will make use of the \texttt{r15\_t00} cube, created using a robust parameter of 1.5 and no taper. This cube strikes the most suitable balance between resolution and sensitivity for this study, highlighting any detailed structures in the \HI\ disc, while allowing for an in-depth analysis of the EPG. The noise in this cube is $0.154 \pm 0.004$ mJy beam$^{-1}$, corresponding to a column density sensitivity of $N_\text{\HI}^{3 \sigma, 16 \text{ km s}^{-1}}~\text{cm}^2=~2.754 \times 10^{18}$, and the beam size is $34.4^{\prime\prime} \times 25.5^{\prime\prime}$.}


\begin{table}
\centering
\begin{threeparttable}
\caption{General properties of UGCA~320.}
\label{tab:gen_props}
\setlength{\tabcolsep}{2.5mm}
\begin{tabular}{cc}
\hline
\hline
Alt. names & \makecell{J1303-17B, DDO 161, \\ MCG 03-33-030} \\
RA (J2000) & $13^\mathrm{h}03^\mathrm{m}16.7^\mathrm{s}$ \\
Dec (J2000) & $-17^\circ25{}^\prime22.9{}^{\prime\prime}$ \\
log ($M_\star$/M$_\odot$) & \vthree{7.91}$^a$ \\
SFR (M$_\odot$ yr$^{-1}$) & \vthree{0.025 -- 0.095}$^b$ \\
$d$ (Mpc) & $6.03^{+0.29^c}_{-0.21}$ \\ 
\hline
\end{tabular}
\end{threeparttable}

\centering
$^a$\vthree{WISE-derived stellar mass from \citet{Blok2024}.; $^b$The upper measurement is the WISE-derived SFR from \citet{Blok2024}, while the lower measurement is the H$\alpha$-derived SFR within the MUSE FoV from \citep{Alabi2025}.}; $^c$Tip of the Red Giant Branch (TRGB) measurement from \citet{Karachentsev2017}.
\end{table}

\section{Analysis \& Results}
\label{sec:analysis}

\subsection{The \HI\ fraction in UGCA~320}
\begin{figure}
  \centering
  \includegraphics[width=0.5\textwidth]{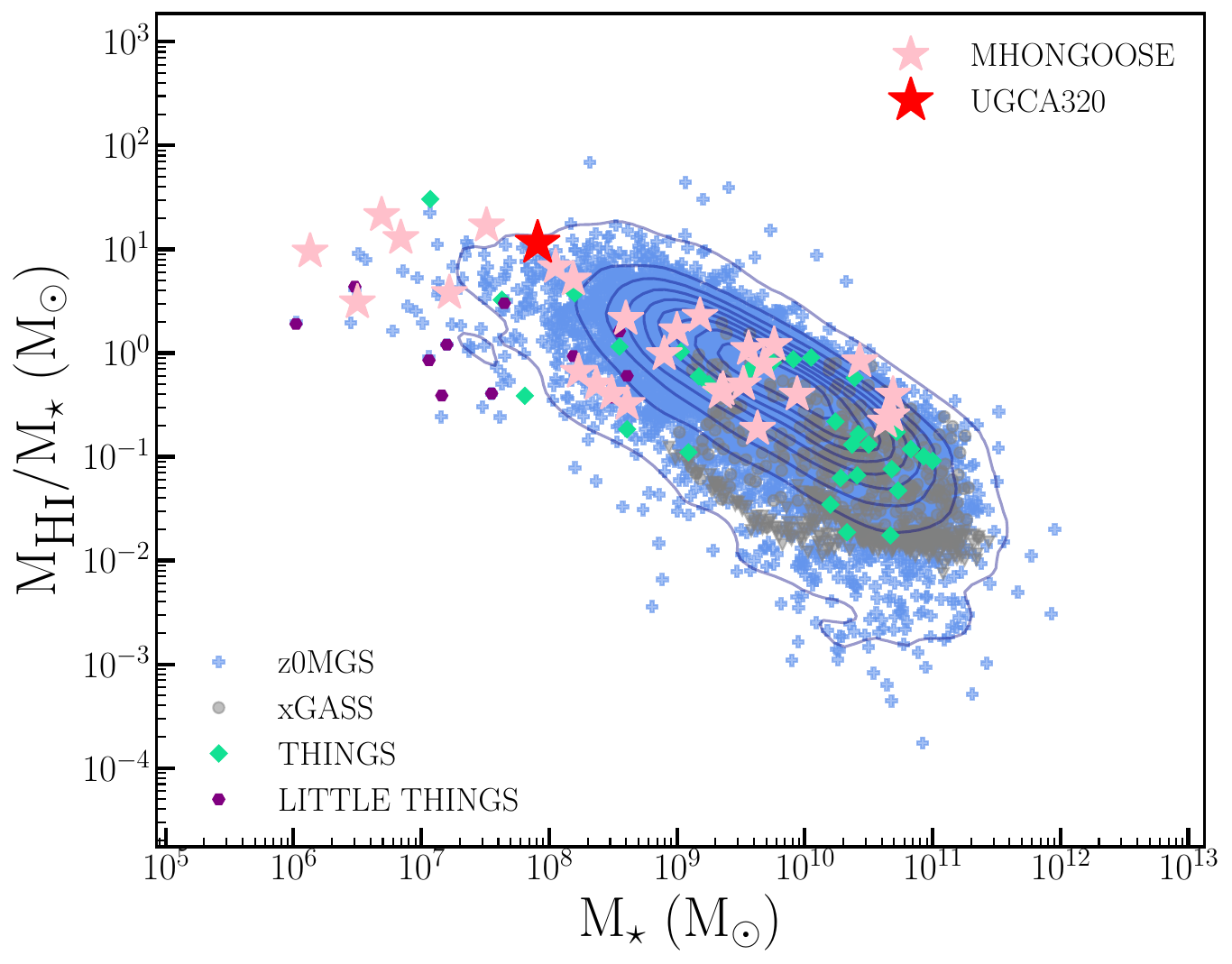}
  \caption{\refrep{The \HI\ fraction of UGCA~320 compared to literature samples. UGCA~320 is shown as a red star, the z0MGS \citep{Leroy2019, Barnes2001} as blue plus signs, with darker contours linearly spaced between the maximum kernel density estimate value and 1\% of it, xGASS \citep{Catinella2018} as grey circles, THINGS \citep{Walter2008} as green diamonds, and LITTLE THINGS \citep{Hunter2012} as purple circles. The remaining galaxies in the MHONGOOSE sample are shown as pink stars. UGCA~320 has a regular to slightly elevated \HI\ fraction compared to galaxies in the comparison samples with similar stellar mass.}}
  \label{fig:hi_frac}
\end{figure}

From our deep MHONGOOSE observations, and following the standard equation (e.g. \citealt{Roberts1978A}), the \HI\ mass of UGCA~320 was \vthree{calculated} to be log$\left( M_\text{HI} / \text{M}_\odot \right)$ = 8.987 (\vthree{quoted from} Table 5 in \citealt{Blok2024}). This implies that the galaxy has a \HI-to-stellar mass fraction of \refrep{$\sim$12}. In Figure \ref{fig:hi_frac} we compare this value to those of other galaxies from legacy surveys. \refrep{This figure shows that UGCA~320 has a regular to slightly elevated \HI\ fraction compared to galaxies from these comparison samples with similar stellar masses.}

\subsection{Channel maps}
\label{sub:channel_maps}
\begin{figure*}
  \centering
  \includegraphics[width=0.99\textwidth]{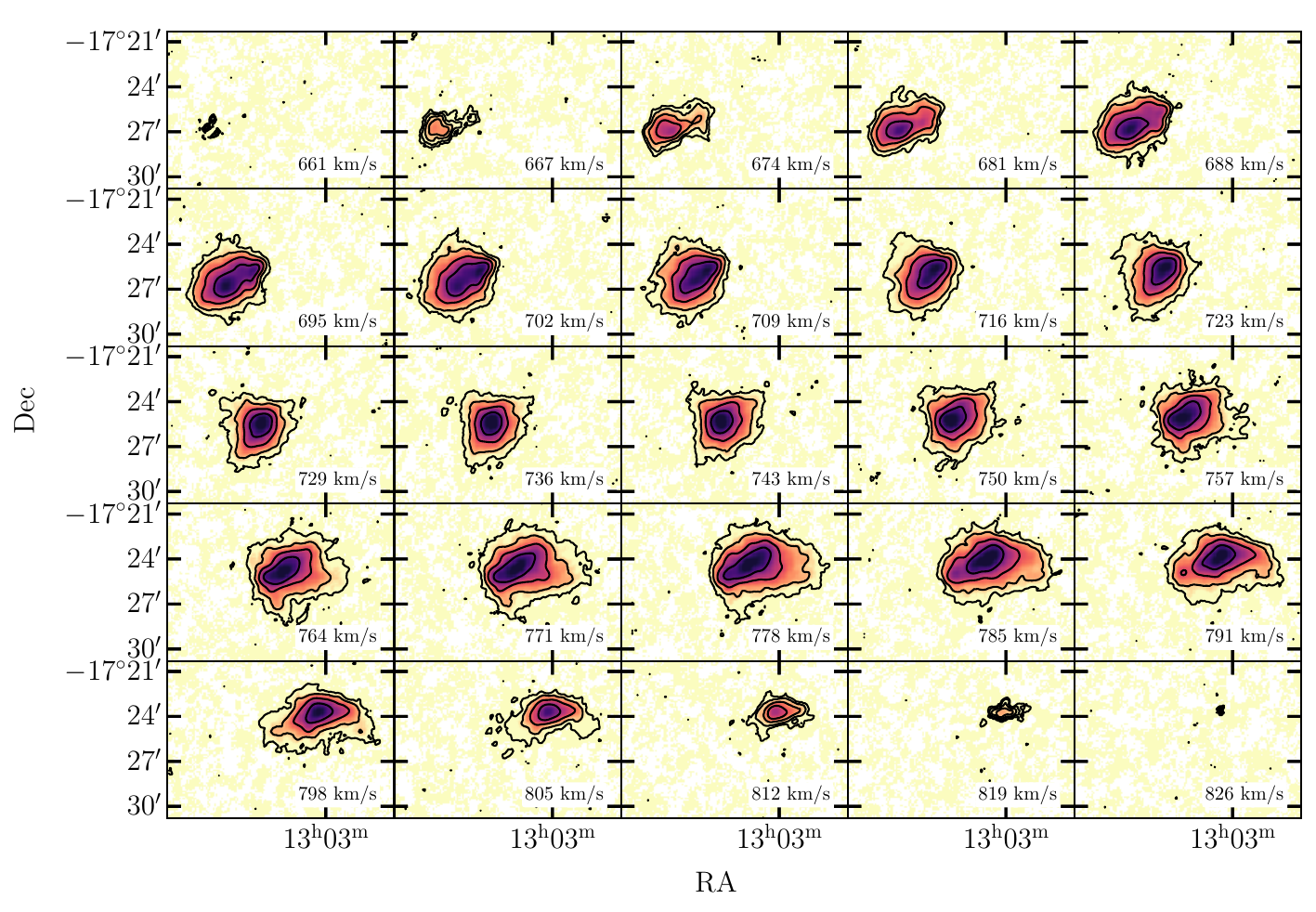}
  \caption{\refrep{Channel maps of UGCA~320. Panels are spaced at 6~--~7~km~s$^{-1}$ intervals (due to rounding), and the velocity of each channel shown is indicated in the bottom right corner of each panel.} The emission in each panel is shown in log scale and at \texttt{r15\_t00} ($25.5^{\prime\prime} \times 34.4^{\prime\prime}$) resolution. Five contours between 3$\sigma$ and the maximum surface brightness are shown in \vthree{black}. The channel maps highlight the relatively large amount of low-surface-brightness gas, and asymmetric features, especially on the receding side of the galaxy.}
  \label{fig:channel_maps}
\end{figure*}

Channel maps for UGCA~320 are shown in Figure \ref{fig:channel_maps}. \refrep{These} highlight the significant amount of low column density extraplanar gas and the asymmetric \HI\ features in this galaxy, particularly on its receding side: an extension of \HI\ in the western direction is clearly visible between 771 -- 791 km s$^{-1}$ in the outer two contours.

\subsection{Moment Maps}
\label{subsec:moment_maps}

\begin{figure*}
    \begin{subfigure}{1\textwidth}
    \centering \hspace{10mm} 
	\includegraphics[width=\textwidth]{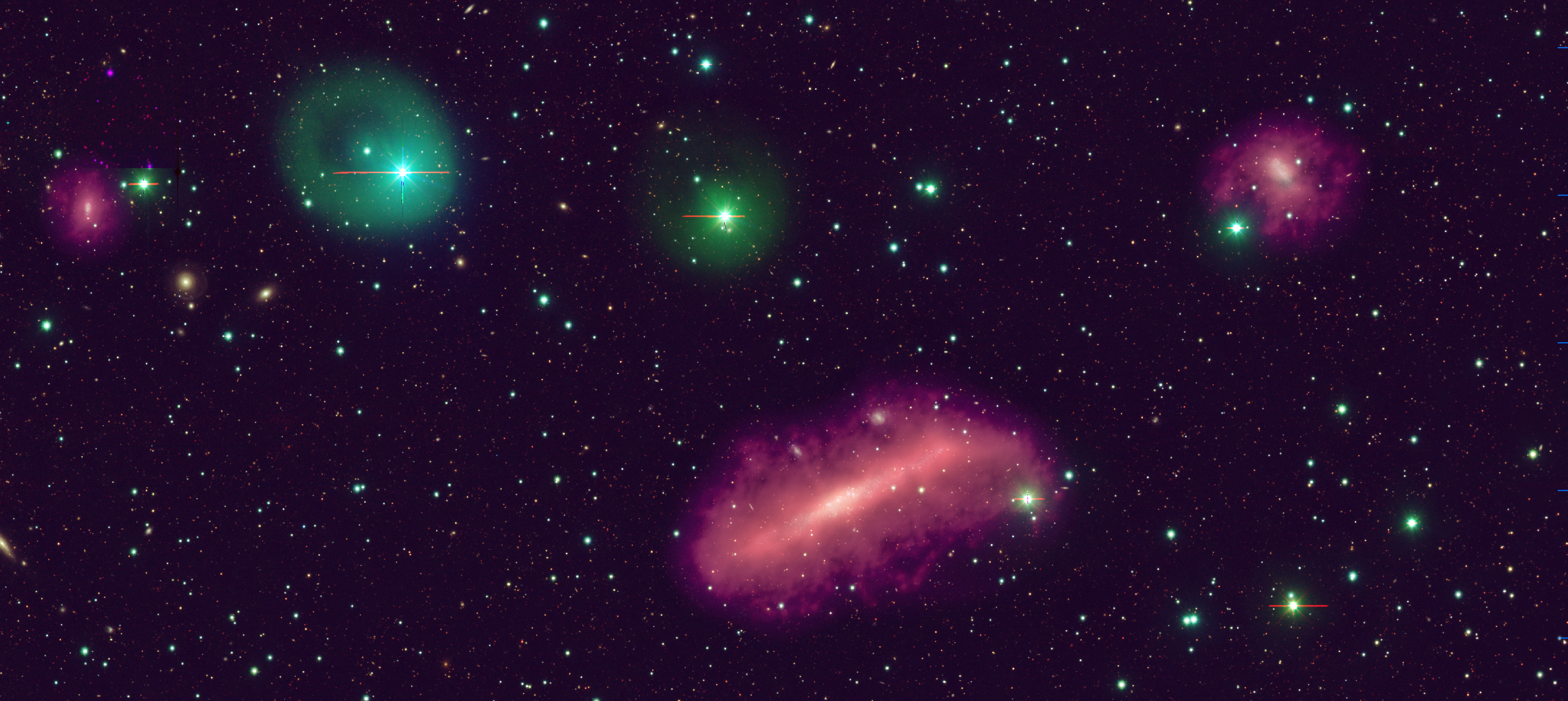}
	\label{subfig:mom0_overlay}
	\end{subfigure} \vspace{7mm}
	
    \begin{subfigure}{1\textwidth}
    \centering
	\includegraphics[width=\textwidth]{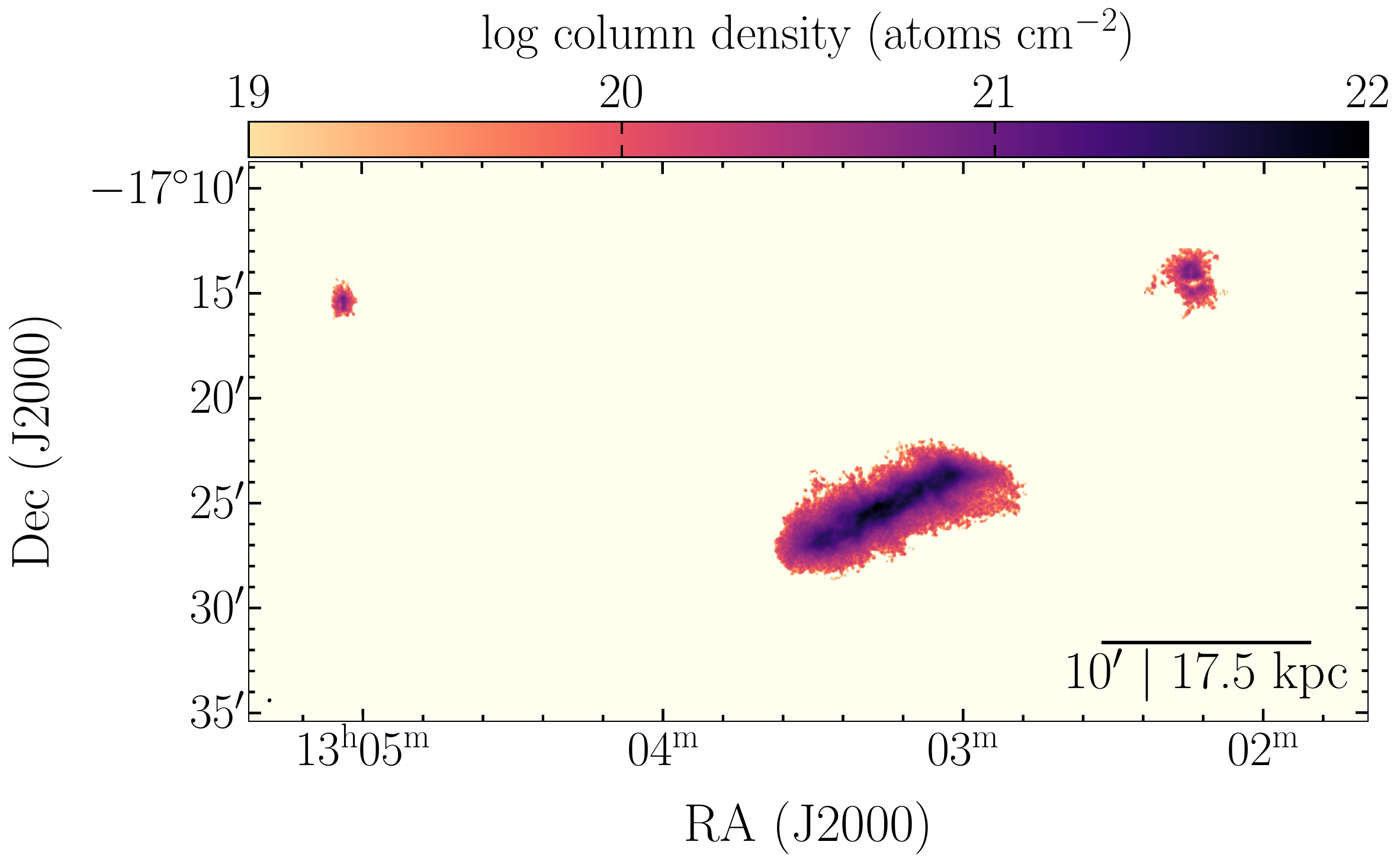}
	\label{subfig:mom0_standalone}
	\end{subfigure}	
	
	\caption{\refrep{Moment zero maps of the \HI\ in UGCA~320 and its two companions, overlaid in red colours on an RGB image (top panel) and standalone (bottom panel). UGCA~319 is located towards the north-west of UGCA~320, and LEDA 886203 to its north-east. The \HI\ image in the top-panel was created using the \texttt{r00\_t00}, ($7.2^{\prime\prime} \times 8.2^{\prime\prime}$), \texttt{r10\_t00} ($18.1^{\prime\prime} \times 26.4^{\prime\prime}$), and \texttt{r05\_t60} resolution ($64.1^{\prime\prime} \times 65.2^{\prime\prime}$) MHONGOOSE cubes. The RGB image was created from DECam \textit{z}-band, and VST/OmegaCam \textit{g}- and \textit{r}-band data (see Figure \ref{fig:optical} for details), following the technique from \citet{English2017}. The green halos around foreground stars are artefacts from the VST imaging and can be disregarded. The \HI\ map in the bottom panel was created using the highest resolution data cube (\texttt{r00\_t00}, $7.2^{\prime\prime} \times 8.2^{\prime\prime}$). The beam is shown in the bottom-left corner, and a scale bar is shown in the bottom-right corner. Note that the beam is very small -- its size is comparable to the dot in ``17.5'' in the scale bar. The high resolution highlights the smaller scale features in the \HI\ reservoirs of the galaxies. UGCA~320 has a patchy column density distribution that looks filamentary at larger scale heights/lower column density. UGCA~319 has a very unusual \HI\ reservoir, whereas the smallest dwarf galaxy in the group, LEDA 886203, has a relatively regular \HI\ distribution.}}
	\label{fig:mom0_group}
\end{figure*}

\texttt{SoFiA-2} \citep{Westmeier2021} was used on the non-primary-beam-corrected cubes to identify the signal in them and define masks for creating the moment maps accordingly. For smoothing, spatial kernels of 0 and 4 pixels were used, and velocity kernels of 0, 9, and 25 channels. The smooth-and-clip (S+C) threshold was set to 4$\sigma$. Chosen values for the remaining \texttt{SoFiA-2} parameters and additional details on the moment map creation process can be found in \S 5.8 of \citet{Blok2024}.

\begin{figure*}
    \begin{subfigure}{0.35\textwidth}
    \centering
	\includegraphics[width=\textwidth]{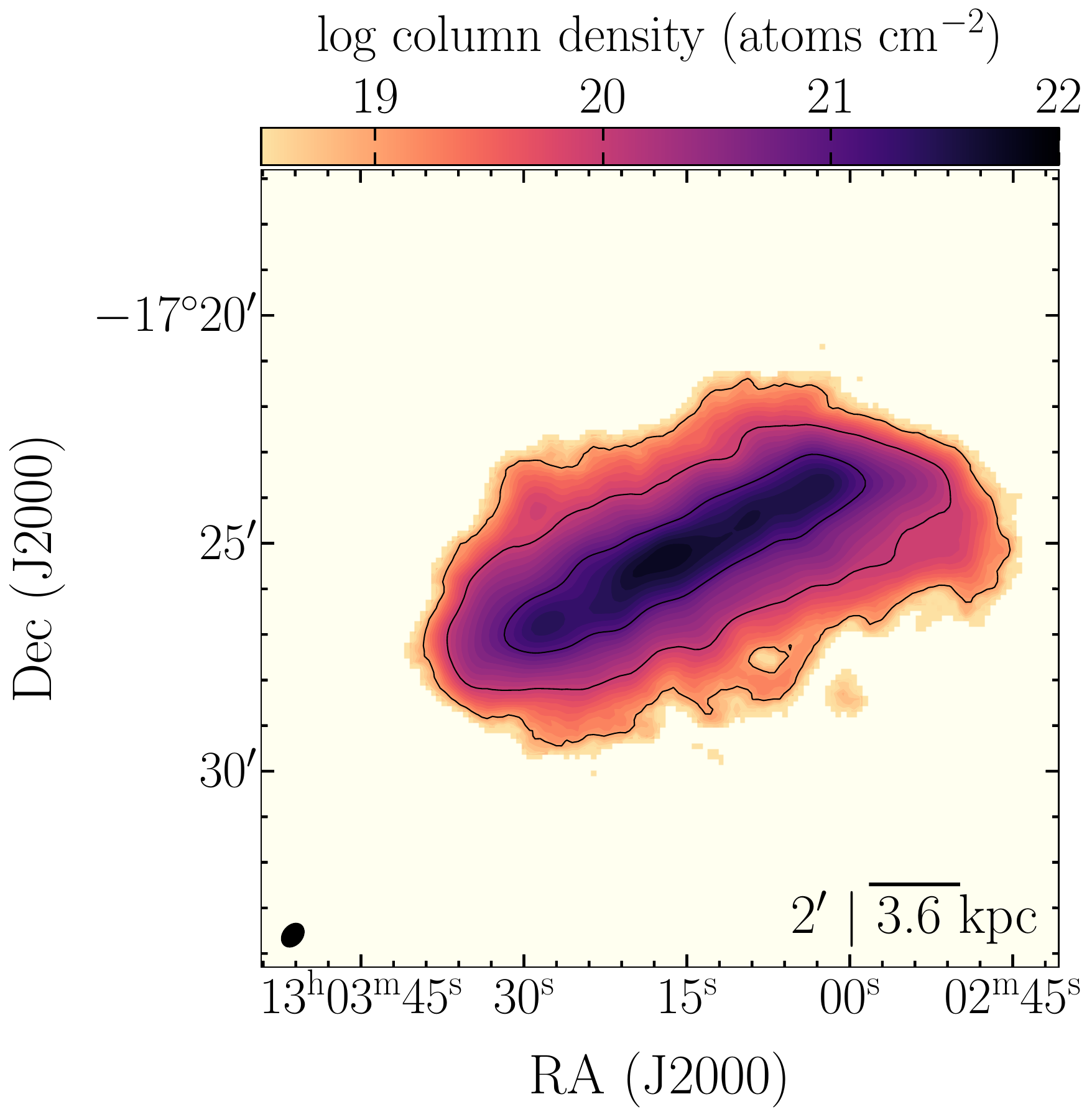}
	\label{subfig:mom0}
	\end{subfigure}
    \begin{subfigure}{0.27\textwidth}
    \centering
	\includegraphics[width=\textwidth]{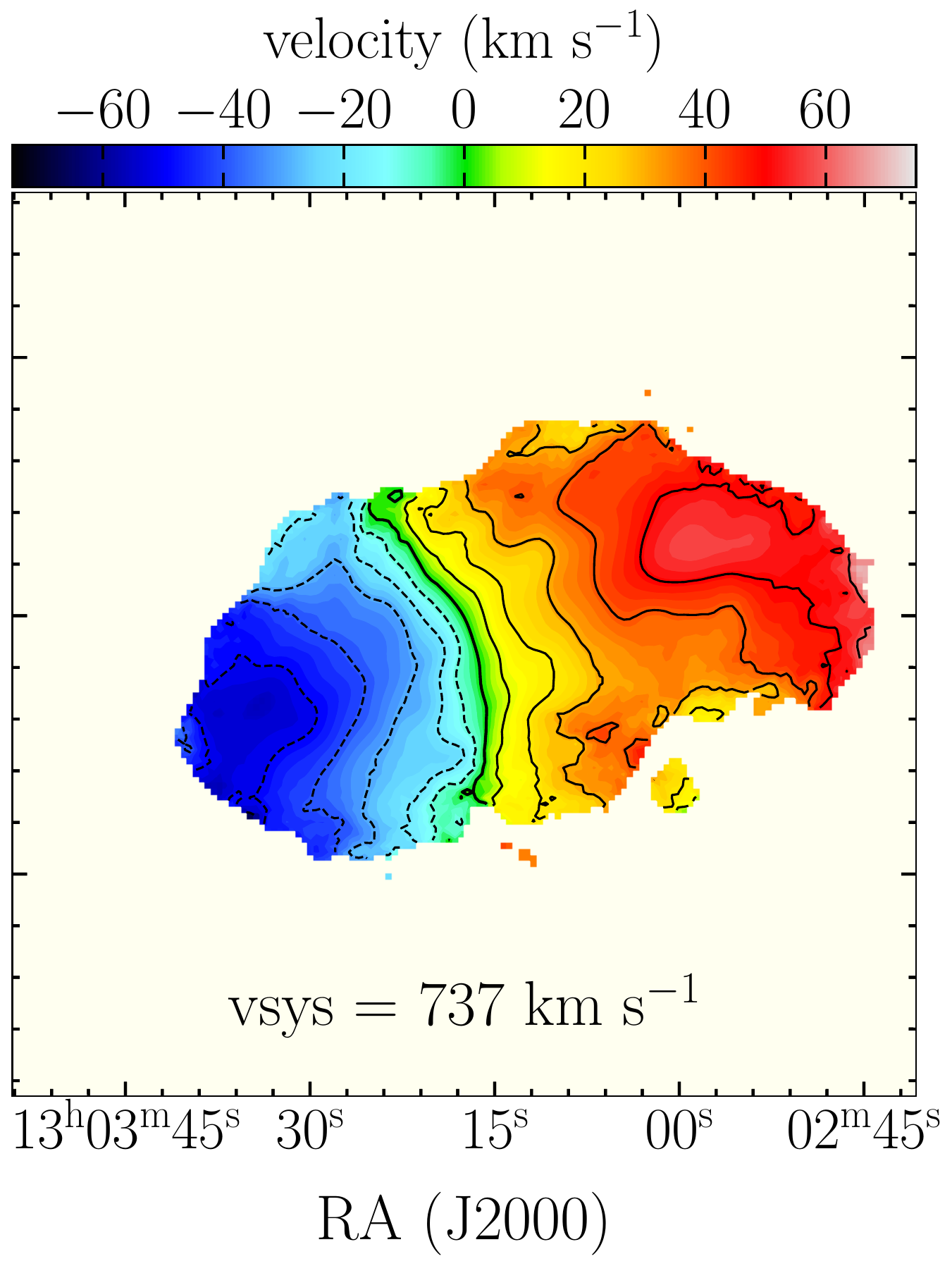}
	\label{subfig:mom1}
	\end{subfigure}	
	\begin{subfigure}{0.32\textwidth}
    \centering
	\includegraphics[width=\textwidth]{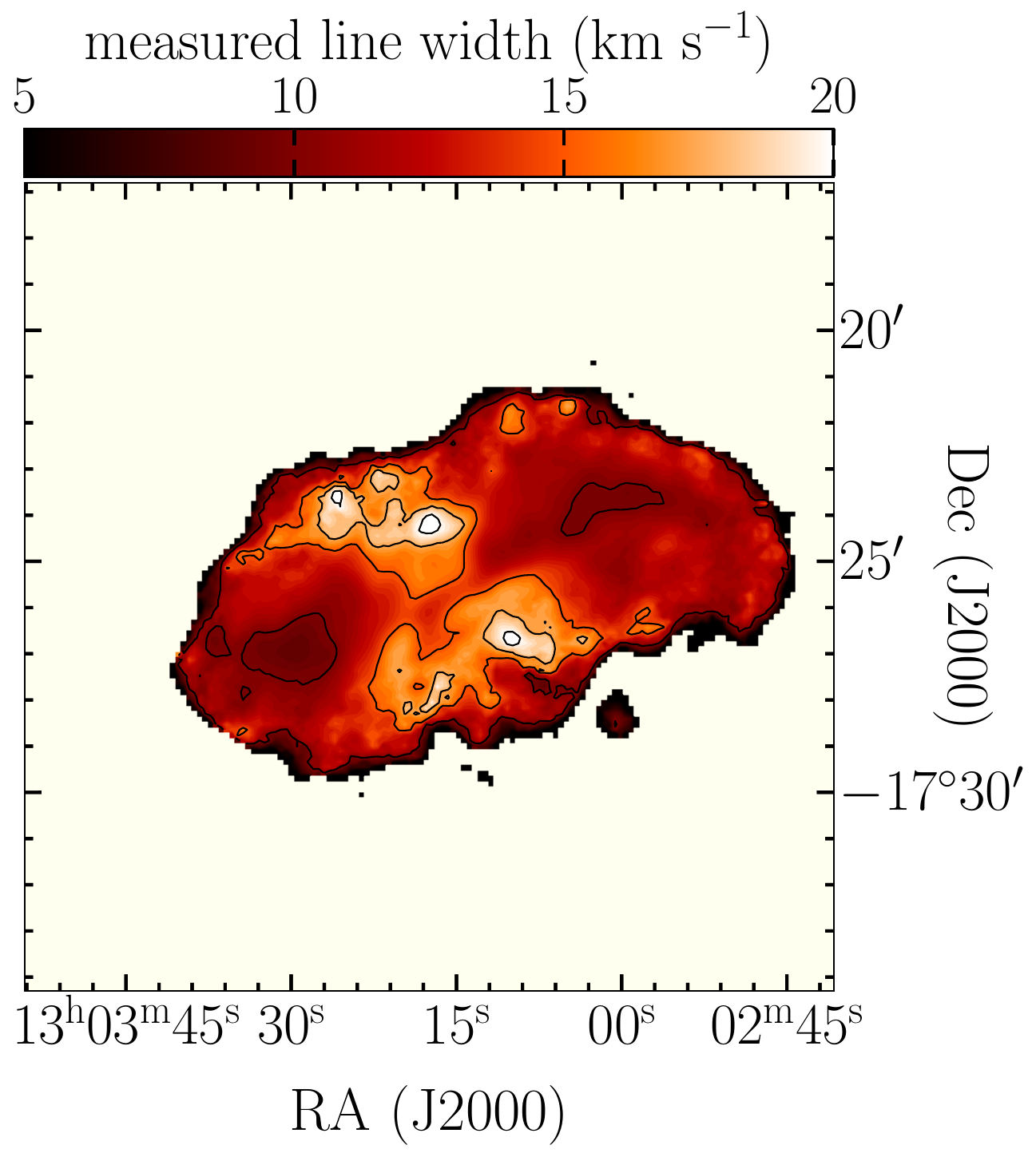}
	\label{subfig:mom2}
	\end{subfigure}
	
	\caption{Moment maps for UGCA~320, centred on the galaxy's optical centre. \refrep{From left to right: moment zero (\HI\ intensity), moment one (velocity), and moment two (observed line width).} The beam of the \texttt{r15\_t00} cube \vthree{($25.5^{\prime\prime} \times 34.4^{\prime\prime}$)} is shown in the bottom-left corner of the moment zero map. A 2$^\prime$ scale bar (corresponding to $\sim$3.6 kpc at the distance of UGCA~320) is shown in the bottom-right corner. \vthree{In the moment zero map (left-hand panel) black contours indicate column densities of 10$^{19}$, 10$^{20}$, and 10$^{21}$ atoms cm$^{-2}$.} The moment one map is shown relative to UGCA~320's systemic velocity of 737 km s$^{-1}$, with isovelocity contours overlaid in black at 10 km s$^{-1}$ intervals from systemic, which is indicated with a thicker contour. Dashed lines indicate velocities negative from systemic. \vthree{In the moment two map (right-hand panel) black contours indicate 10, 15, 17.5, and 20 km s$^{-1}$.} The moment zero map (left-hand panel) reveals a patchy, asymmetric \HI\ distribution with a significant area of relatively low-surface-brightness gas. The moment one map reveals an asymmetric, U-shaped velocity map. The moment two map reveals areas with increased observed line widths on either side of the disc at larger scale heights.}
	\label{fig:moment_maps}
\end{figure*}

In Figure \ref{fig:mom0_group} we show moment zero maps of UGCA~320 and its two nearest neighbours, comprising the small group of galaxies introduced above. UGCA~320 is located in the centre/south, UGCA~319 towards its north-west, and LEDA 886203 in the north-east. In \refrep{the top panel} the \HI\ moment zero map is overlaid on an optical false-colour image. 
This overlay highlights the extra-planer \HI\ gas in UGCA~320, and its filamentary structure. In \refrep{the bottom panel} we show the standalone \HI\ moment zero map at the highest MHONGOOSE resolution (\texttt{r00\_t00}, \vthree{$7.2^{\prime\prime} \times 8.2^{\prime\prime}$}) to emphasise the detailed \vthree{morphologies} of UGCA~320 and its low-mass companions. 
UGCA~320 has an asymmetric morphology both in the optical and in \HI, with the brightest region offset to the south-east side of the disc and an extension towards the opposite side \vthree{in both wavelengths}. In \HI\ this extension is appended with a feature extending towards the south of the disc. Looking carefully at this feature, it appears to consist of two separate structures: one extending from the very end of the \HI\ disc, and one from slightly closer to the galaxy centre. The distribution of the \HI\ in the disc is somewhat clumpy, with multiple local column density minima and maxima throughout the disc. UGCA~319 has an unusual \HI\ morphology, consisting of several distinct clumps, and a clear separation between the north and south of the \HI\ reservoir. UGCA~319 will be the subject of a separate future paper. While the exact distance to and stellar mass of LEDA 886203 are unknown, the frequency of the \HI\ line reveals that its \vthree{radial velocity} is very similar to that of the other two group members \vthree{(Scannell et al., in prep.)}. Based on this, and its optical size (Figure \ref{fig:mom0_group}), we estimate that it is likely significantly less massive than UGCA~319. Despite this, its well-resolved \HI\ distribution looks relatively regular, especially compared to both other galaxies. This could have important implications for the group dynamics, which we will discuss in more detail in \S \ref{sec:anomalous_gas}.

In Figure \ref{fig:moment_maps} we show the moment zero (column density), moment one (intensity-weighted velocity), and moment two (observed line width) maps for UGCA~320 at \texttt{r15\_t00} resolution \vthree{($25.5^{\prime\prime} \times 34.4^{\prime\prime}$)}. 
The velocity map follows the asymmetry of the moment zero map, with the systemic velocity coinciding with the highest column density \HI\ in the centre of the galaxy, and the receding side of the disc being more extended than the approaching side. The velocity field has a U-shape, where the velocities at larger scale \vthree{heights} ``bend'' towards the approaching side of the disc, and are lower than the absolute velocities in the disc at the same radius. Furthermore, \vthree{a warp-like feature is clearly visible, indicating an inclination warp and/or lagging gas at larger scale heights.} The moment two map reveals two pronounced regions with increased measured line widths on either side of the galaxy centre (i.e. above and below the disc). Measured line widths in these areas increase to $>20$~km~s$^{-1}$, whereas values in the majority of the plane remain between 5 and 10 km~s$^{-1}$. This is peculiar, as the gas associated with the plane usually has a higher velocity dispersion compared to gas at larger scale heights \vthree{(e.g. \citealt{Tamburro2009})}. We will explore possible origins of these increased measured line widths in \S \ref{sec:anomalous_gas}. \vthree{Versions of Figure \ref{fig:moment_maps} at all six resolutions are presented in Appendix \ref{app:moment_maps}, Figure \ref{fig:moment_maps_multi_res}.}  

\begin{figure*}

    \begin{subfigure}{0.95\textwidth}
    \centering
	\includegraphics[width=1\textwidth]{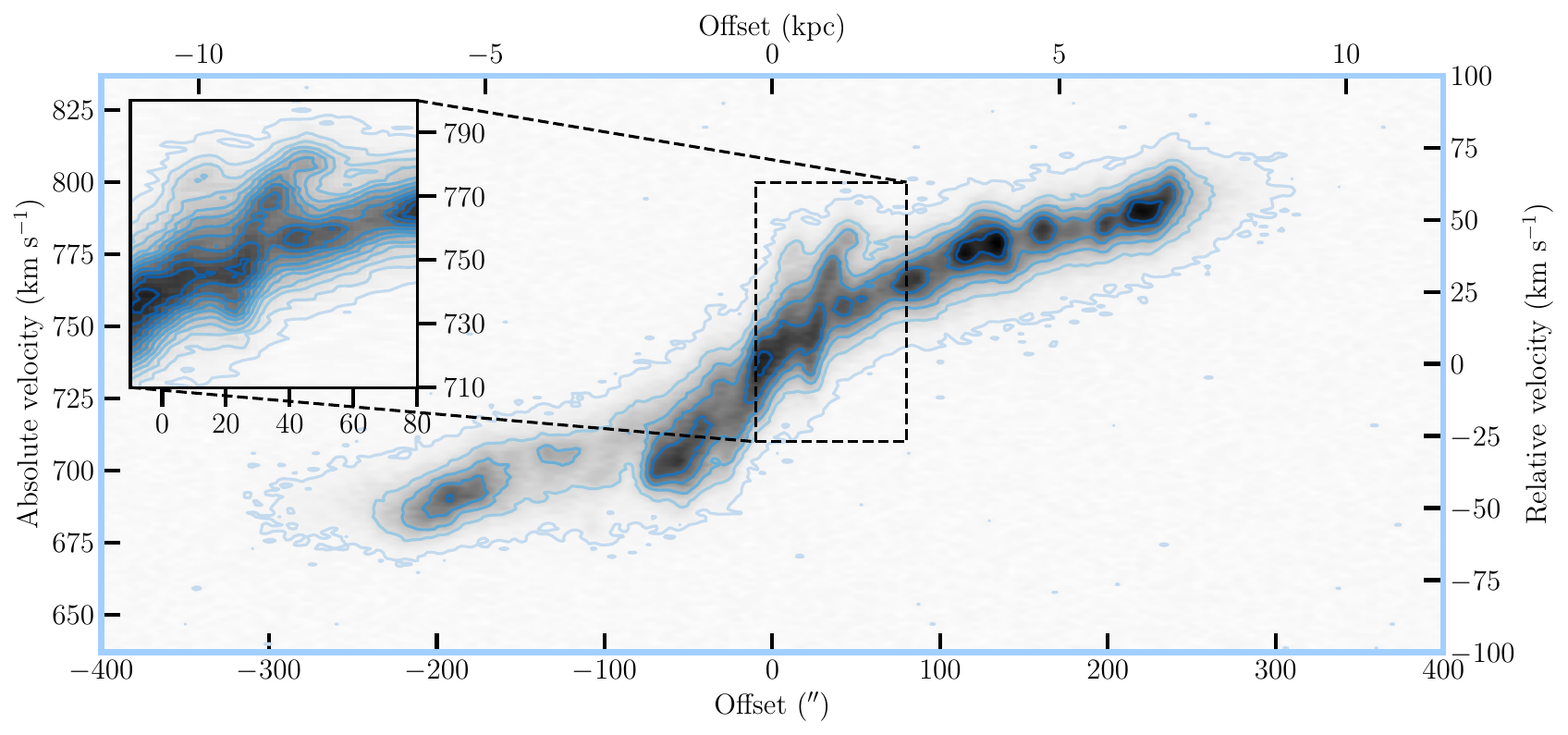}
	\label{subfig:pvd_major}
	\end{subfigure} 
	
    \begin{subfigure}{0.56\textwidth}
    \centering
	\includegraphics[width=1\textwidth]{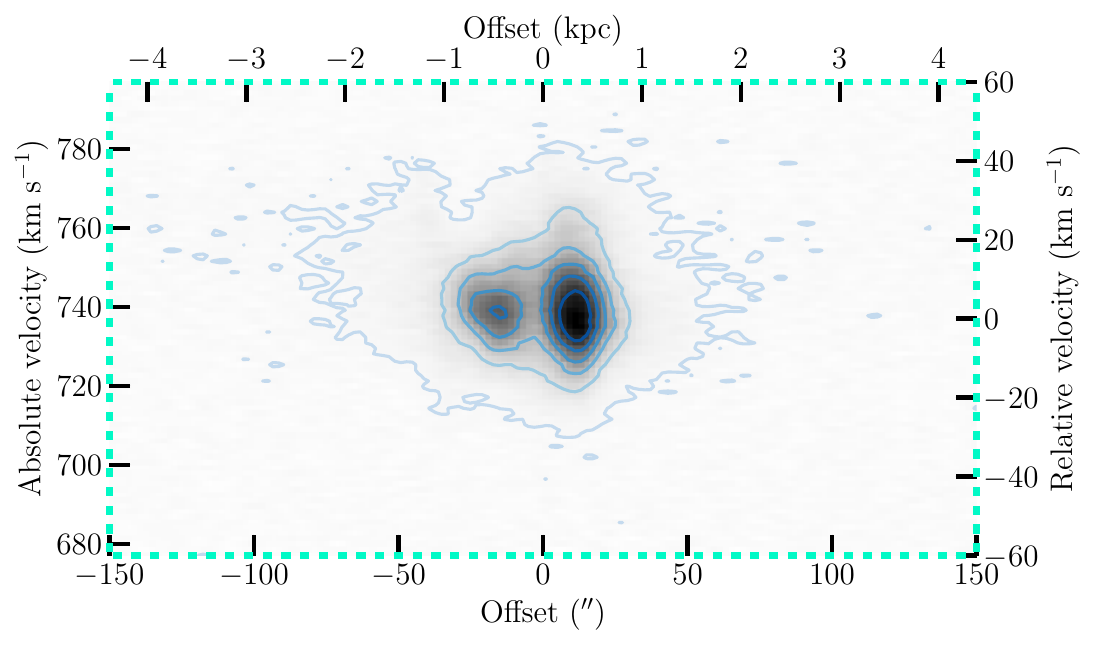}
	\label{subfig:pvd_minor}
	\end{subfigure}	
	\begin{subfigure}{0.42\textwidth}
    \centering
	\includegraphics[width=1\textwidth]{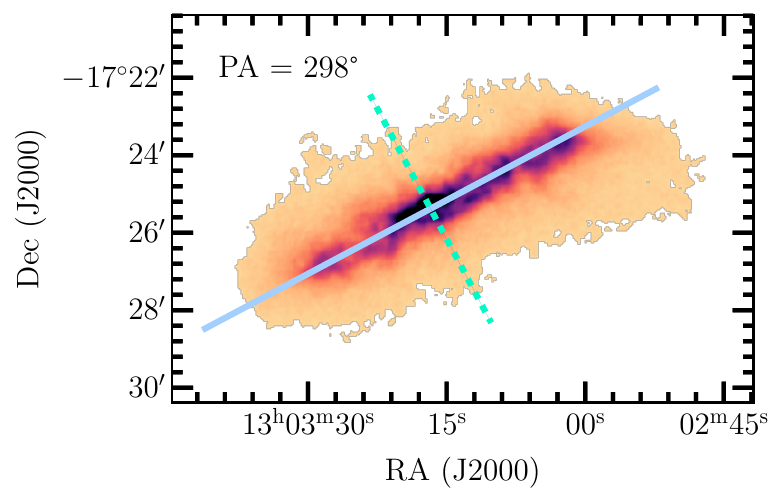}
	\label{subfig:pvd_paths}
	\end{subfigure}	
	
	\caption{\refrep{Position-velocity diagrams of UGCA~320. The \texttt{r00\_t00} cube ($7.2^{\prime\prime} \times 8.2^{\prime\prime}$) was used to highlight the more subtle clumps and features in the disc. The top panel shows the PVD along the major axis, and the bottom-left panel that along the minor axis. The corresponding paths are shown in the bottom-right panel, overlaid on the moment zero map derived from the \texttt{r00\_t00} cube ($7.2^{\prime\prime} \times 8.2^{\prime\prime}$). The line colours and styles of the PVD paths match those of the corresponding PVD panels (e.g. solid blue for the major axis and green dashed for the minor axis). The position angle of 298\textdegree\ is indicated in the top-left corner of this panel. The PVDs highlight the irregular and asymmetric nature of the \HI\ in UGCA~320.}}
	\label{fig:pvds}
\end{figure*}

\subsection{Position-Velocity Diagrams}
\label{subsec:pvds}
In Figure \ref{fig:pvds} we show position-velocity diagrams (PVDs) along the major \refrep{(top panel)} and minor \refrep{(bottom-left panel)} axes of UGCA~320. In order to highlight details in the structure of the galaxy, we use the \texttt{r00\_t00} cube \vthree{($7.2^{\prime\prime} \times 8.2^{\prime\prime}$)}. \refrep{The paths along which the PVDs were extracted are shown in the bottom-right panel.} The PVD is one beam major axis across, which \vthree{corresponds to} 240 pc at the distance of UGCA~320. The PVD along the major axis highlights the irregular and asymmetric \HI\ distribution. The receding side of the disc contains the higher column density \HI, and \vthree{both} sides have a patchy distribution. An irregular feature is visible in the central parts of the disc, which is highlighted in a zoom-in panel in the top left corner \refrep{of the panel}. A low-velocity blob at $\sim-75^{\prime\prime}$ and $v \sim 700$ \vthree{km s$^{-1}$} is possibly part of the same structure. \vthree{These features could indicate a bar-like structure, or possibly spiral arms \citep[Figure 9]{Kamphuis2013}.} \refrep{The PVD along the minor axis highlights the asymmetry of the \HI\ reservoir also in this direction}, with the majority of the \HI\ concentrated on the north side of the disc (the right-hand side \refrep{of the PVD}), and a discontinuity between the north and south sides of the disc. The extended low column density outskirts of the \HI\ disc are clearly visible in both \refrep{PVDs}. 

\subsection{Continuum}
\label{subsec:continuum}

\begin{figure}
  \begin{center}
    \includegraphics[width=0.48\textwidth]{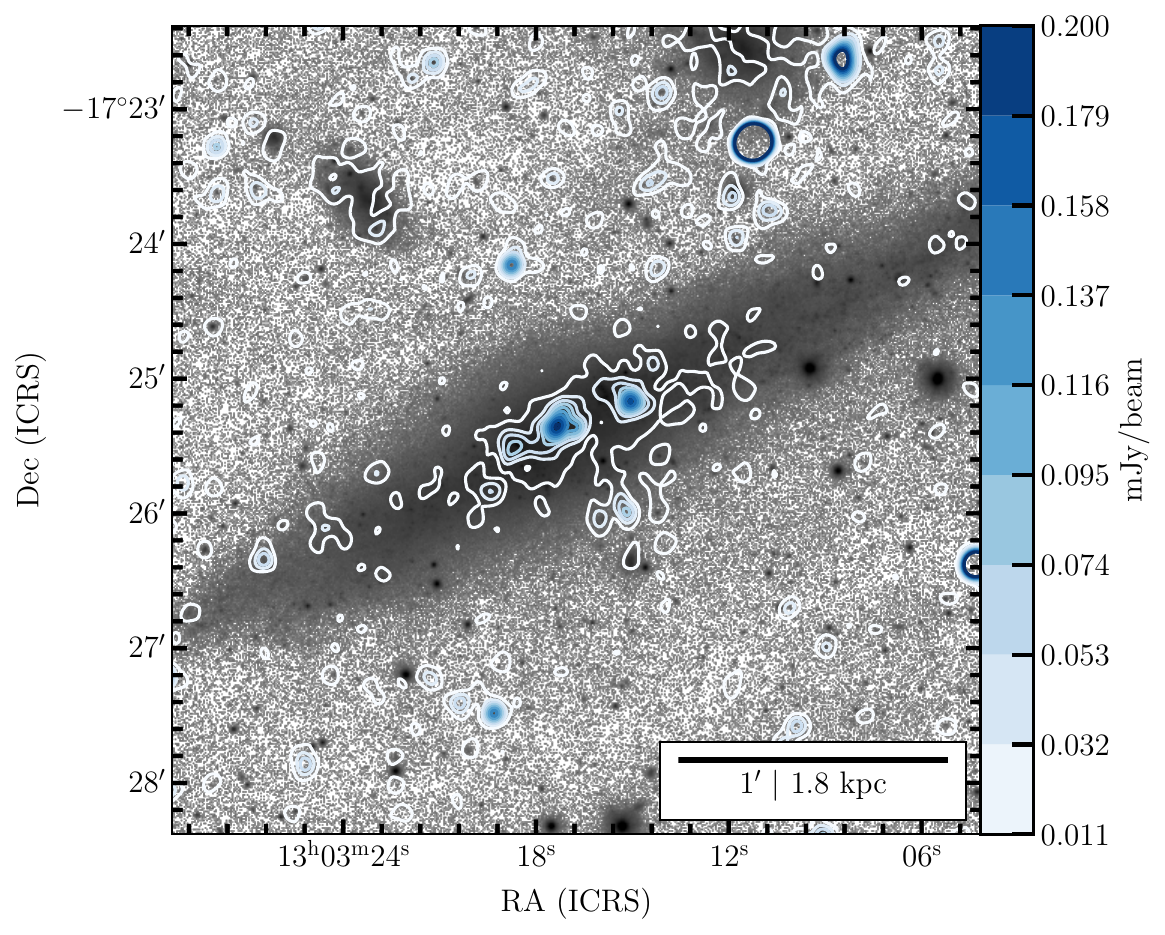}
  \end{center}
  \caption{\refrep{Continuum image of UGCA~320 overplotted on a \textit{g}-band image from OmegaCam (similar to Figure \ref{fig:optical}, here shown in greyscale). A 1$^\prime$ scale bar (corresponding to 1.8 kpc at the distance of UGCA~320) is shown in the bottom right corner.} The continuum in and around UGCA~320 is shown as blue contours. An extended continuum source is detected in the centre of the galaxy.}
  \label{fig:continuum}
\end{figure}



\refrep{Continuum images and measurement sets of the MHONGOOSE galaxies were created as part of the self-calibration procedure which is part of the \HI\ reduction. These continuum images used the frequency range between 1390 and 1422 MHz, covering a total bandwidth of 32 MHz. Details of these steps are provided in \citet{Blok2024}. The resulting continuum measurement sets were combined and additional direction-dependent calibration was applied using \texttt{oxkat} version 0.41 \citep{Heywood2020}. After this, images were created within \texttt{oxkat}. The dirty images were first deconvolved using a clean mask containing all emission above 3$\sigma$, after which a second round of cleaning was performed using a refined clean mask derived from the initial image. The final image was produced with a robust parameter of $r = 0$, to match the highest-resolution \HI\ cube. The resulting images have dimensions of 3 $\times$ 3 degrees, with a pixel size of 1.1$^{\prime\prime}$.}

An extended continuum source is detected in the centre of UGCA~320, associated with the highest \HI\ column densities and brightest \vthree{optical/star forming regions in the disc (Figure \ref{fig:optical}, \citealt{Alabi2025})}. \vtwo{The extent of the source is $\sim 1.5 ^\prime$, which corresponds to $\sim$2.6 kpc at the distance of UGCA~320.} The total \vtwo{flux} density of the continuum associated with UGCA~320 is 0.14 Jy beam$^{-1}$, with a peak SNR of 3.2.

\section{Tilted Ring Modelling}
\label{subsec:tirific}
In the previous section we have seen that UGCA~320 has some notable \HI\ elements, including a warp-like feature, an extended asymmetry, and smaller scale features in the centre. In order to better understand the 3D morphology and kinematics of the \HI\ reservoir, and to what extent it follows or deviates from typical galactic shapes and motions, we further analyse it it using the Tilted Ring Fitting Code (TiRiFiC, \citealt{Jozsa2007}). This code simulates spectroscopic data cubes by representing the 3D geometry of the observed line (usually \HI) as a set of concentric rings, each of which can be assigned its own set of morphological and kinematic parameters, collectively forming the rotating gas disc. By constructing the tilted ring model that matches the \HI\ disc of UGCA~320  most closely, we aim to describe any distinctive kinematic features (previously identified or newly detected), and identify the presence of any anomalous gas. Here we will primarily make use of the \texttt{r15\_t00} cube, (created using a robust parameter of 1.5 and no taper), as it strikes the most suitable balance between resolution and sensitivity for this study, highlighting any detailed structures in the \HI\ disc, while allowing for an in-depth analysis of the EPG.  and investigate its possible origins.

\begin{figure*}

    \centering
	\includegraphics[width=\textwidth]{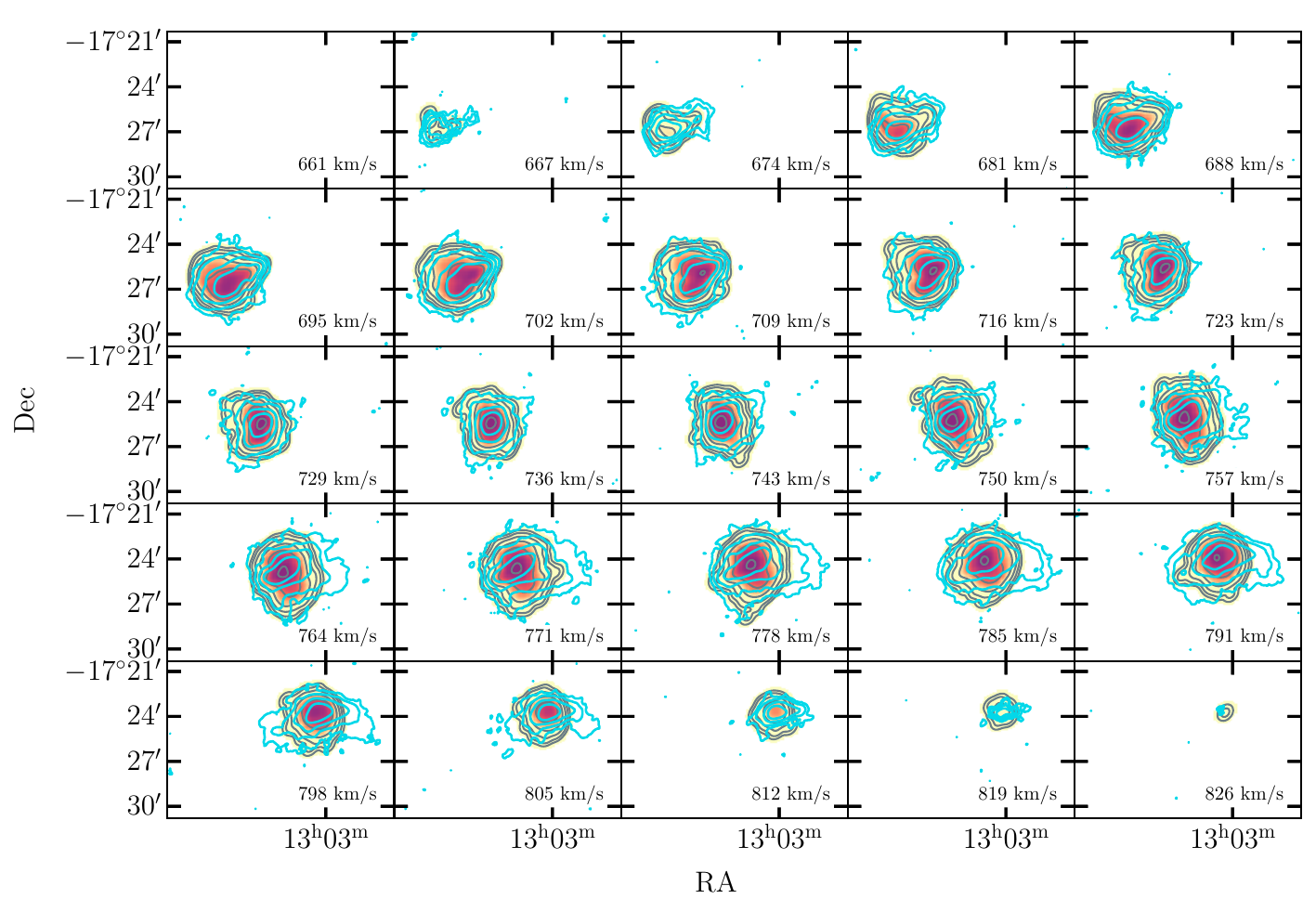}
	
	
	\caption{Channel maps of the best fit \texttt{TiRiFiC} model, shown in log scale. The velocities of each \refrep{represented} channel are indicated in the bottom-right corners of each panel. \refrep{Cyan contours represent the data in these channels. While the model channel maps trace the progression of the bulk of the atomic gas with velocity reasonably well, there are some clear discrepancies, most notably on the receding side of the disc.}}
	\label{fig:channel_maps_model}
\end{figure*}

The essential modelling parameters for \texttt{TiRiFiC} consist of the rotation curve, the surface brightness profile, the inclination, the position angle, the central coordinates, and the systemic velocity. To obtain an initial set of these parameters, from which we will build our final model, we first run Fully Automated \texttt{TiRiFiC} (\texttt{FAT}, \citealt{Kamphuis2015}), a fully automated procedure based on \texttt{TiRiFiC} and tested on \HI\ data cubes. \vthree{The resulting values are input into \texttt{TiRiFiC}, and each parameter is refitted separately before adding complexity to produce the final model}.

We then start the tilted ring modelling process by first identifying the best-fit model for a simple single disc, from which we later build more sophisticated models (e.g. separating the approaching and receding side of the disc to accommodate the asymmetries, more complex morphologies such as a flare or a \vthree{thick disc}, more complex kinematics such as radial motions, etc.). Throughout this process, we use a combination of fitting and manual tweaking of the resulting models. 

Optimisation of the model parameters by \texttt{TiRiFiC} is performed using a $\chi ^2$ minimisation approach. \vthree{Because of the irregular and patchy nature of the \HI\ in UGCA~320 (e.g. Figures \refrep{\ref{fig:mom0_group}} and \ref{fig:pvds}), this approach sometimes yields model parameters that appear erratic (e.g. discontinuities or highly variable/unphysical surface brightness profiles or velocity curves).} Omitting problematic rings from the fit and interpolating over them, or fitting multiple rings at once, can mitigate this issue to a degree. However, we found that it is often necessary to implement \vthree{some} additional manual tweaks to improve the resulting models.
\begin{figure}
  \centering
  \includegraphics[width=0.47\textwidth]{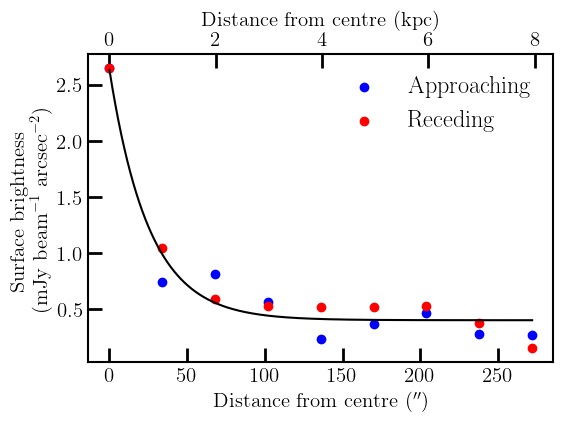}
  \caption{Surface brightness profile of the best-fit \texttt{TiRiFiC} model of UGCA~320. Each location along the x-axis represents a ring, of which the approaching side is indicated by a blue marker, and the receding side by a red one. An exponentially decaying fit to the average surface brightness profile is shown in black for comparison. Both sides have different profiles, which are not strictly decreasing, but follow a more ``wavy'' pattern \vthree{(more pronounced on the approaching side than the receding side)}, reflecting the patchy nature of the \HI\ emission which is also clearly seen in the moment zero map and corresponding model (\refrep{top panel in Figure \ref{fig:residuals}}).}
  \label{fig:sbr_profile}
\end{figure}
\begin{figure}
  \centering
  \includegraphics[width=0.47\textwidth]{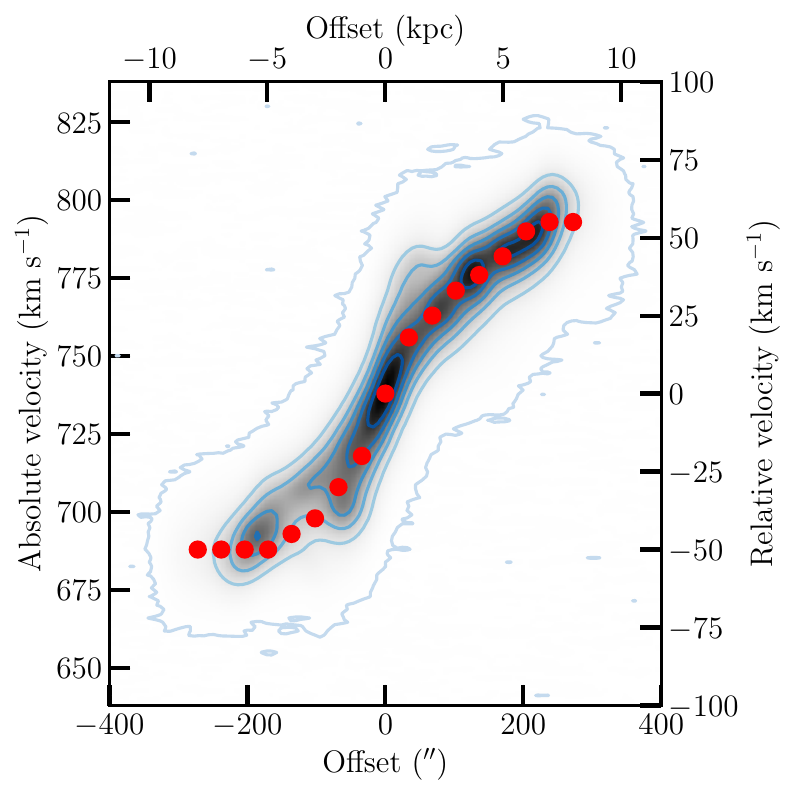}
  \caption{Velocity curve of the final \texttt{TiRiFiC} model overplotted on the PVD (\texttt{r15\_t00}, \vthree{$25.5^{\prime\prime} \times 34.4^{\prime\prime}$}) along the minor axis (similar to Figure \ref{fig:pvds}). The rotational velocity associated with each ring is represented by a red marker. The rotation curve largely follows the PVD, though it flattens in the outer rings, whereas the PVD continues to increase (or decrease).}
  \label{fig:velocity_curve}
\end{figure}
We evaluated the \texttt{TiRiFiC} models by inspecting their moment maps, channel maps, and PVDs -- in particular a series of PVDs parallel to the minor axis. Once no more significant improvements can be made to these metrics using a reasonable set of parameters, we decide on the final model.

\vthree{The final \texttt{TiRiFiC} model consists of a thin disc with a scale height of $\sim$3 kpc, and either an additional thick disc \emph{or} a flare. A thick disc is a vertically extended, flattened layer of gas that rotates coherently with the stellar disc but has a larger scale height than the thin disc, whereas a flare describes a continuous increase in scale height of the thin disc towards its edges. Hence, these scenarios are almost identical in the outer rings, and can look similar depending on projection effects. The two models are sufficiently similar that we do not prefer one over the other; figures in this section are based on the flare model purely for convenience. The central inclination of the modelled gas disc(s) is 83\textdegree, which decreases by up to 6\textdegree\ in the outer rings. The position angle is 298\textdegree, and decreases by $\sim$20\textdegree\ in the very outer ring on the approaching side only. The systemic velocity of the data is adopted, 737 km s$^{-1}$.}

Channel maps for the model are shown in Figure \ref{fig:channel_maps_model}. 
\refrep{Overall, the model channel maps trace the progression of the bulk of the \HI\ emission with velocity in UGCA~320 reasonably well. However, in a significant number of channels the more detailed morphology of the emission is not accurately reproduced. This is particularly noticeable on the receding side of the disc, where the emission of UGCA~320 extends well beyond the boundaries of the model. While the approaching side of the disc is more accurately described by the model, some differences remain. For example, while the model follows a channel progression typical for a rotating \HI\ disc, the data is more asymmetric, with the emission unevenly distributed towards the south side of the channels. We will explore these discrepancies in detail in the remainder of this section.} 

The surface brightness profile of the best fit \texttt{TiRiFiC} model is presented in Figure \ref{fig:sbr_profile}. 
The surface brightness profiles on both sides are roughly exponentially declining. However, neither of them are strictly declining, but they show a more ``wavy'' pattern, that is somewhat different on either side of the disc. This reflects the patchy \HI\ distribution in UGCA~320, which is evident from the moment maps and PVDs shown in Figures \ref{fig:moment_maps} and \ref{fig:pvds}. 

\begin{figure*}

    \begin{subfigure}{0.99\textwidth}
    \centering
	\includegraphics[width=\textwidth]{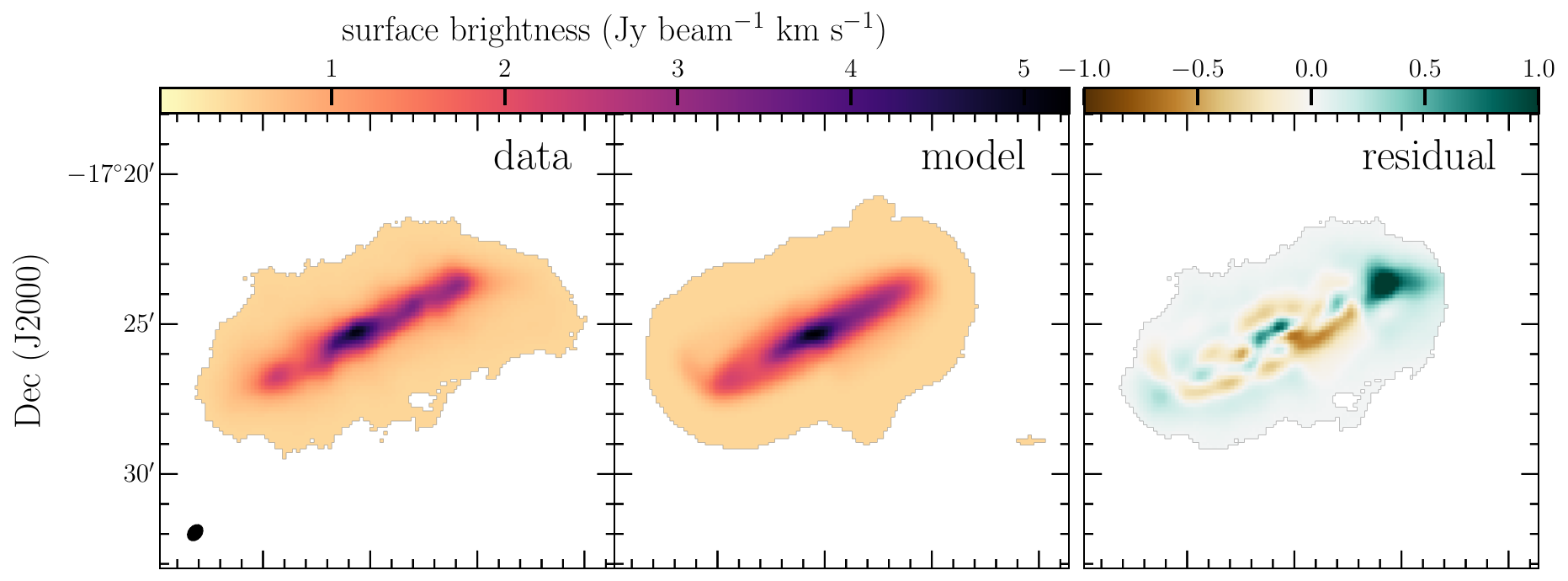}
	\label{subfig:residuals_mom0}
	\end{subfigure}
	
    \begin{subfigure}{0.99\textwidth}
    \centering
	\includegraphics[width=\textwidth]{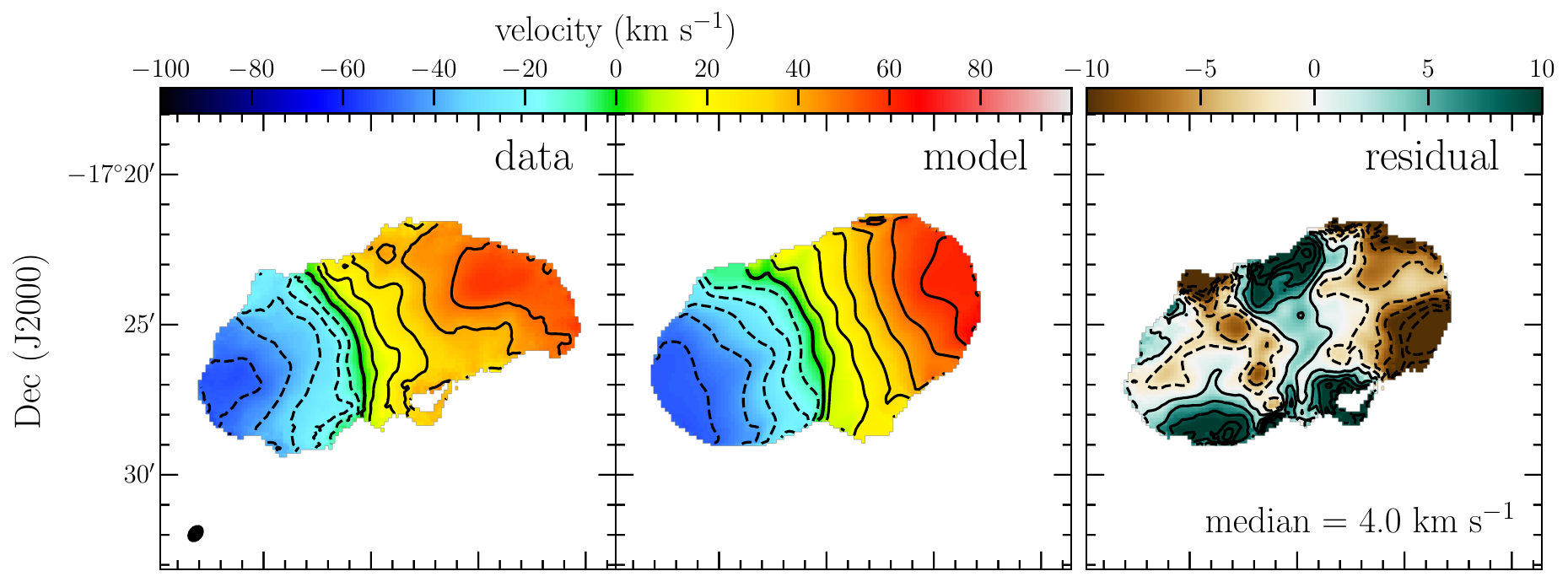}
	\label{subfig:residuals_mom1}
	\end{subfigure}	
	
	\begin{subfigure}{0.99\textwidth}
    \centering
	\includegraphics[width=\textwidth]{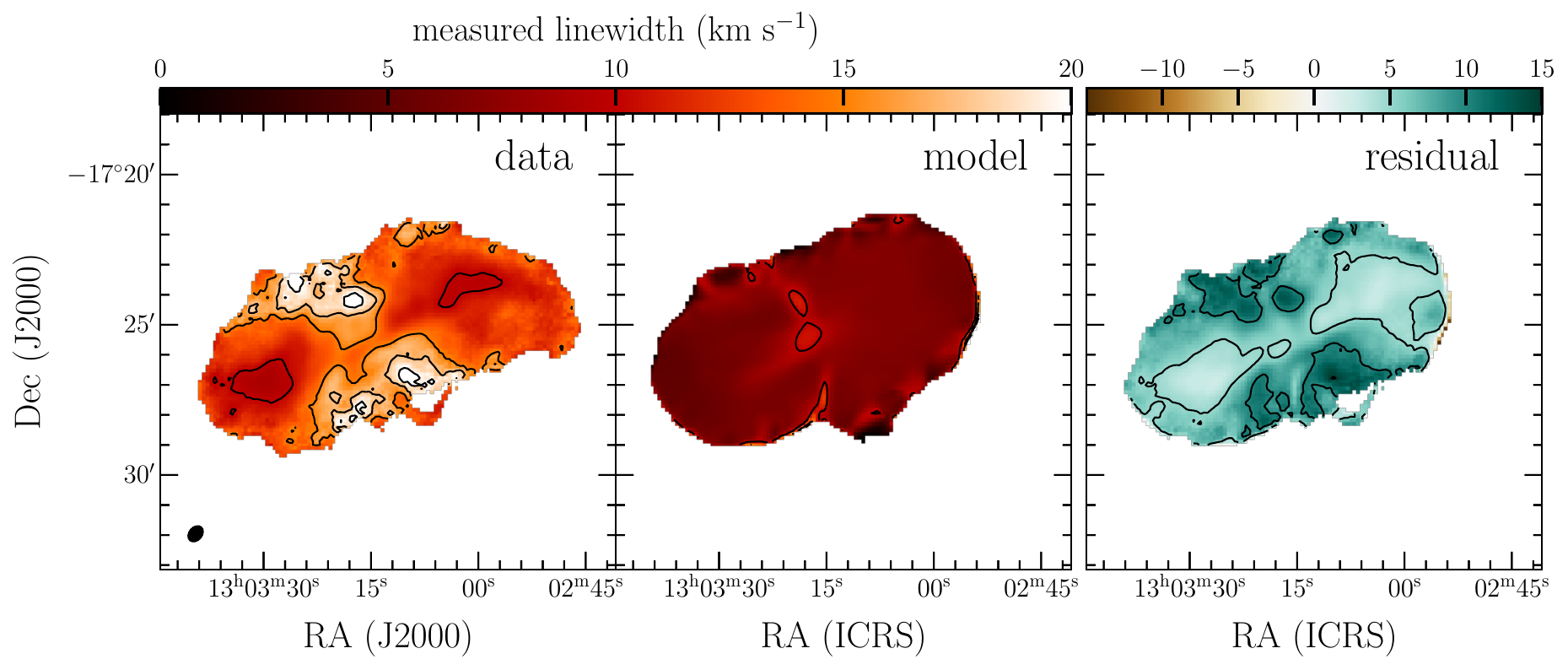}
	\label{subfig:residuals_mom2}
	\end{subfigure}	
	
	\caption{\refrep{Comparison of the \HI\ moment maps of the data (left-hand panels) and models (centre panels), and the corresponding residual maps (right-hand panels), representing the residuals after the model moment map is subtracted from the corresponding moment map from the data. From top to bottom the moment zero, one, and two maps are shown, respectively. In the moment one map (middle row) the iso-velocity contours in the left and centre panels are identical to those in the centre panel of Figure \ref{fig:moment_maps}. Ten contours in the right-hand panel are equally spaced between -15 and 15 km~s$^{-1}$, where dashed contours indicate negative residuals, i.e. the model velocities are higher than the velocities in the data. The beam of the \texttt{r15\_t00} observations ($25.5^{\prime\prime} \times 34.4^{\prime\prime}$) used for the fitting is shown in the bottom left corner of the left-hand panels. Generally the model fits the data well, with residual maps that are largely random with small values. Some features are not reproduced by the model, the most notable of which are the asymmetry on the far end of the receding side of the disc, and anomalous velocities around the galaxy centre.}}
	\label{fig:residuals}
\end{figure*}

\begin{figure*}

    \begin{subfigure}{0.46\textwidth}
    \centering
	\includegraphics[width=\textwidth]{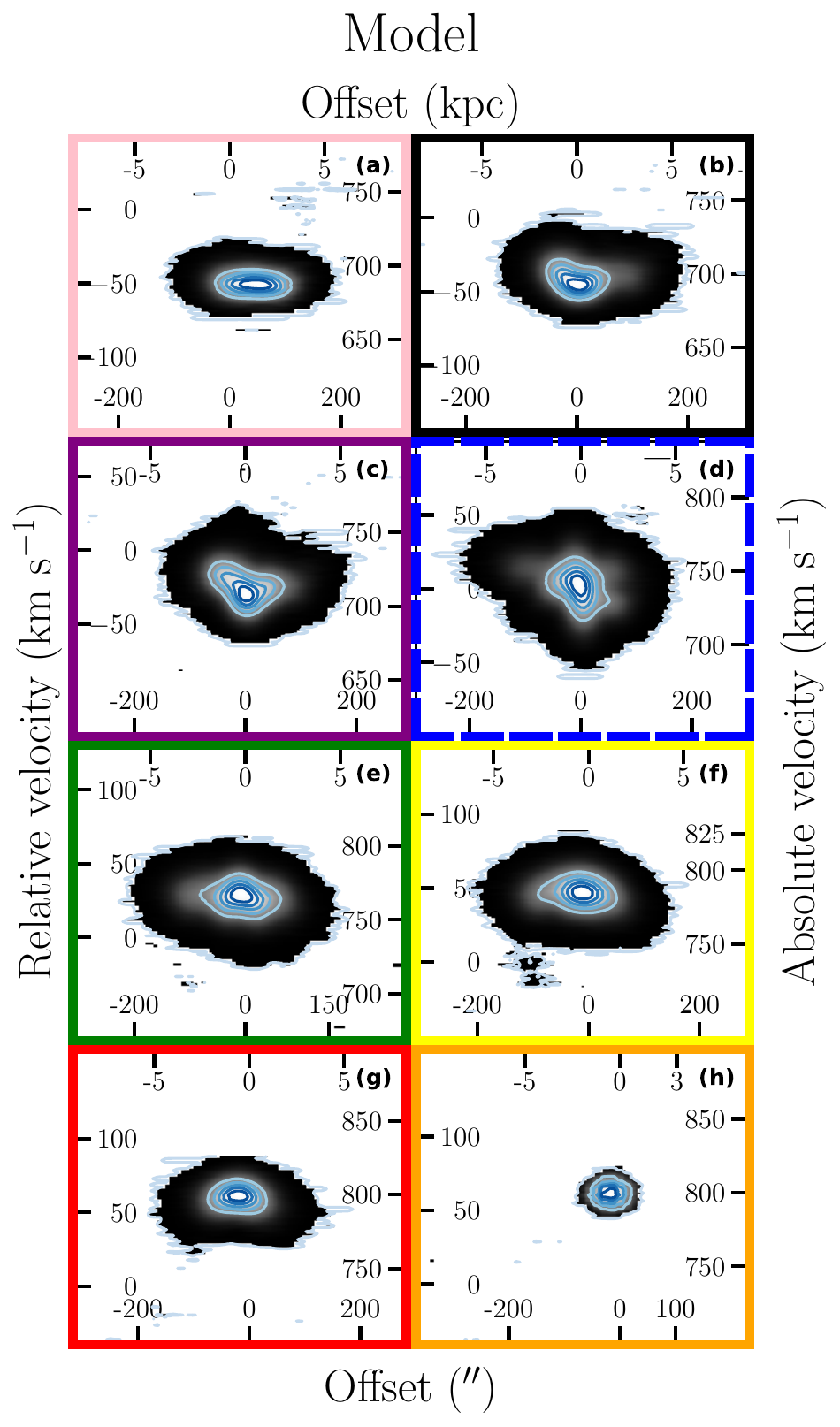}
	\label{subfig:pvd_slices_model}
	\end{subfigure} 
    \begin{subfigure}{0.46\textwidth}
    \centering
	\includegraphics[width=\textwidth]{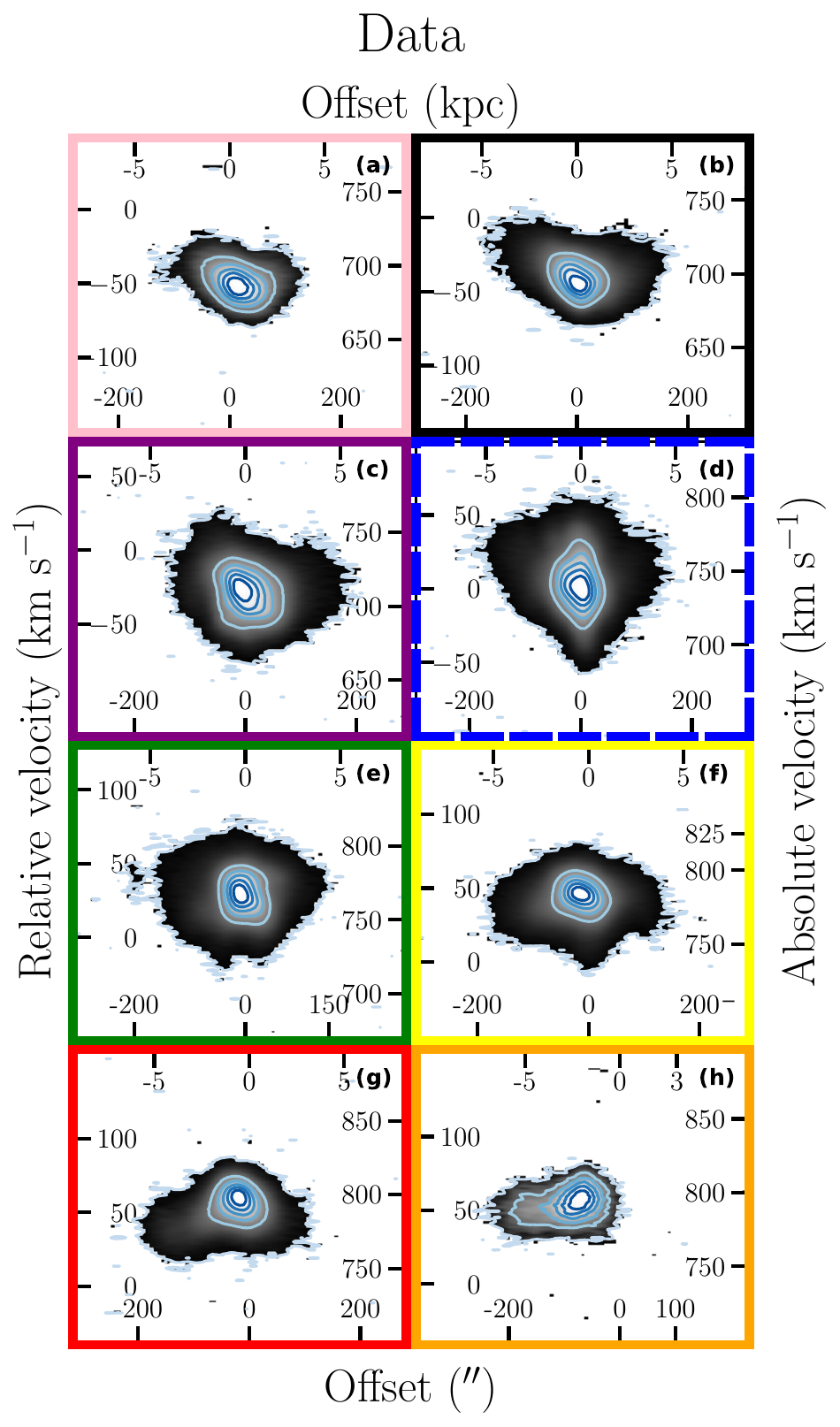}
	\label{subfig:pvd_slices_data}
	\end{subfigure}	
	
	\begin{subfigure}{0.5\textwidth}
    \centering
	\includegraphics[width=\textwidth]{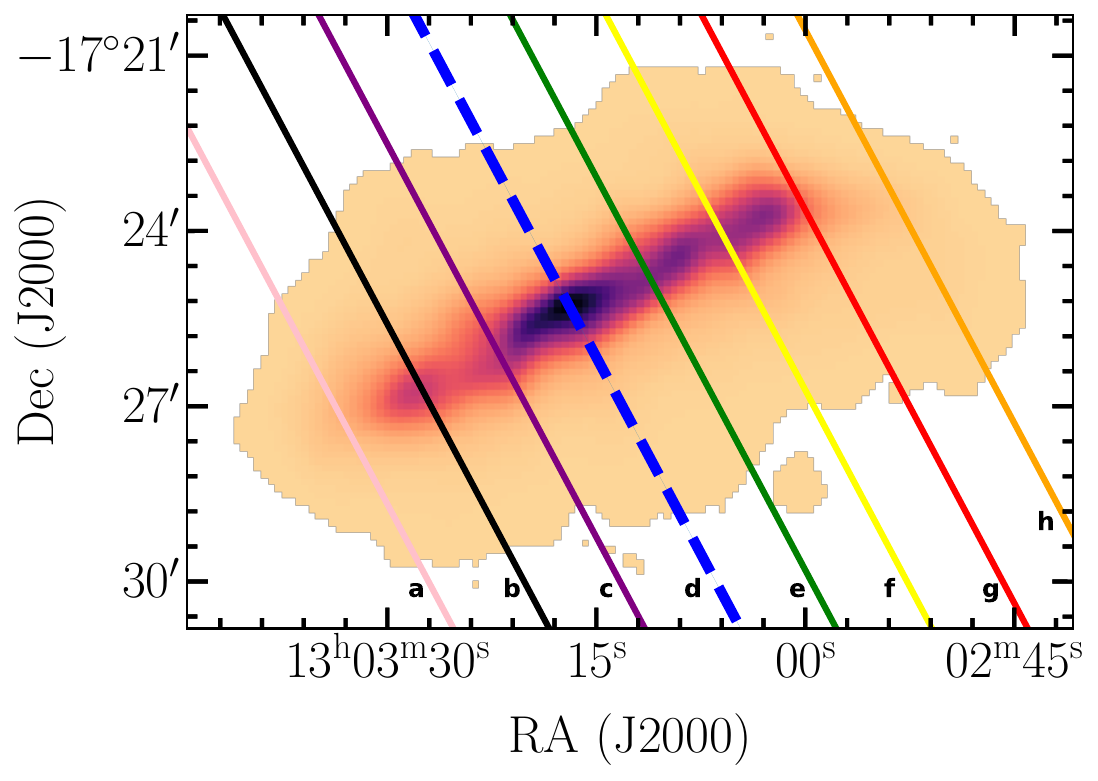}
	\label{fig:pvd_minor_paths}
	\end{subfigure}	
	
	\caption{\refrep{PVD slices along the minor axis for the best-fit \texttt{TiRiFiC} model (top left-hand panel) compared to the data (top right-hand panel). PVDs are one MeerKAT synthesised beam wide, and were taken according to the paths shown in the bottom panel. The sub-figures, read from left to right and from top to bottom, correspond to the paths read from left to right (east to west) in the bottom panel, as indicated by letters in the top right corners of each panel, and to the left of each path in the bottom panel. The bottom panel shows the paths parallel to the minor axis along which the PVDs were created that were used to evaluate the tilted ring models, overlaid on the moment zero map from the \texttt{r15\_t00} cube ($25.5^{\prime\prime} \times 34.4^{\prime\prime}$), which we are modelling here. The thick, dashed, blue line corresponds to the minor axis of the galaxy, and is identical to the green, dashed line in the bottom-right panel of Figure \ref{fig:pvds}. To guide the eye, the sub-figures in the top panels are framed with colours corresponding to those in which the paths are shown in the bottom panel. While the PVDs in the central parts of the galaxy are reproduced reasonably well by the model, they diverge from the data increasingly towards the outskirts. This highlights the complexity of the \HI\ reservoir, which deviates from a tilted ring structure.}}
	\label{fig:pvd_slices_final_model}
\end{figure*}

The velocity curve of the best-fit \texttt{TiRiFiC} model is shown in Figure \ref{fig:velocity_curve}, overlaid on the PVD along the major axis. The PVD is similar to the one shown in \refrep{the top panel of Figure \ref{fig:pvds}} (following the blue solid line in \refrep{the bottom-right panel}), but was extracted from the \texttt{r15\_t00} cube \vthree{($25.5^{\prime\prime} \times 34.4^{\prime\prime}$),} which is used for the fits. 
On the receding side, the model follows the PVD all the way out until the outermost ring. Here, the velocity curve of the model flattens, whereas that of the PVD continues to increase. Similarly, the rotation curve of the model on the approaching side of the disc flattens in the outer three rings, while the PVD continues to increase. On the receding side the velocity curve traces the higher density central areas of the PVD, whereas on the approaching side it traces the extremes of the PVD, \vthree{as is typical for an edge-on disk.}

Moment maps of the final \texttt{TiRiFiC} model, along with their residuals, are shown in Figure \ref{fig:residuals}. 
The residuals of the moment zero map are reasonably small and lack prominent patterns or structures, except for the asymmetry on the receding side, earlier described in \S \ref{subsec:moment_maps}, which here shows up as a strong positive area. From the residuals we estimate that $\sim$80\% of the neutral gas in UGCA~320 is captured by our model (i.e. the absolute values of the residual map correspond to $\sim$20\% of the \HI\ mass of UGCA~320). Of this $\sim$20\%, $\sim$35\% is associated with the asymmetry mentioned above. This corresponds to a total \HI\ mass of log $M_\text{\HI} \sim 7.9$ M$_\odot$, or $\sim$7\% of the total neutral gas mass of UGCA~320. Attempts to model this asymmetry without negatively affecting the remainder of the model proved to be unsuccessful. Therefore, we conclude that this \HI\ feature cannot be reproduced within the limitations of a tilted ring geometry and is anomalous compared to the general rotation. This will be discussed further in \S \ref{sec:anomalous_gas}.

The residuals of the moment one map (\refrep{middle row in Figure \ref{fig:residuals}}) are also small: they are typically well below 10 km s$^{-1}$, and below 5 km s$^{-1}$ in most regions. However, they do exhibit some structure. The approaching side of the disc, close to the systemic velocity, shows three connected areas \vthree{where the} velocities are 5 -- 10 km s$^{-1}$ higher in the model compared to the data. Furthermore, the velocities on the far end on the receding side of the disc are also higher in the model, by up to $\sim$20 km s$^{-1}$. This area is associated with the asymmetry discussed above. \vthree{In the moment one map of the data (\refrep{left panel in the middle row}), the gas appears to lag relative to the gas in the plane (noting possible projection effects), which is not fully captured by the model.} These residuals will be analysed further in \S \ref{sec:anomalous_gas}.

The measured line widths of the model, captured by the moment two map, are close to those of the data in the galaxy disc, but residuals increase with scale height, especially around the galaxy centre \refrep{(bottom row in Figure \ref{fig:residuals})}. Increasing the model velocity dispersions in the outer rings to match the measured line widths in these areas in the data results in an inaccurate 3D structure, which becomes particularly evident when inspecting the PVDs. We show this in \vthree{Figure \ref{subfig:pvd_mom2}}, which shows the PVDs of the model with velocity dispersion values equal to the measured line width from the data. This discrepancy will be explored further in \S \ref{subsec:spectra}, along with any associated non-Gaussian \HI\ line profiles in these regions of UGCA~320.

Figure \ref{fig:pvd_slices_final_model} shows the PVD slices along the minor axis that were used to evaluate the model. The positions of the slices, each the width of the MeerKAT synthesised beam major axis, are indicated in \refrep{the bottom-right panel in Figure \ref{fig:pvds}}. 
The shapes of the PVD slices in the central parts of the disc (panels c (purple), d (blue, dashed), and e (green)) are reproduced well. While evaluating these central PVD slices, it became evident that both a change in position angle and radial motions play important roles in shaping the lower column density gas in these regions. The central PVD slices of the model are somewhat broader than those of the data. This is because the model extends marginally further out compared to the data, which can also be seen in the maps in Figure \ref{fig:residuals}. We do not expect this to affect our analysis of the kinematic structure of the \HI\ reservoir. Additionally, the more central, higher column density gas in these central areas is more ``elongated'' towards more extreme velocities in the data compared to the model. Our attempts to model this, e.g. by increasing the velocity dispersion, \vthree{changing the PA, and/or adding radial motions} in these regions, lead to unrealistic velocities in the outer rings. We will discuss this further in \S \ref{sec:anomalous_gas}. Towards the outer regions of the disc (i.e. the outer 2 -- 3 PVD slices) the difference between the model and the data increases. This is especially evident on the receding side of the disc, where the asymmetric extension (discussed above) becomes clearly visible in panels \textbf{f} - \textbf{h}. On the approaching side, the strong change in position angle is not captured by the model (panel \textbf{a}). Including a change that is sufficient to accurately reproduce the PVD in this area results in the PVDs becoming inaccurate closer to the galaxy centre. It is possible that this position angle change has the same external origin as the anomalous gas on the receding side. We will discuss this further in \S \ref{sec:discussion}.

Overall, the model reproduces the global distribution of \HI\ in UGCA~320 reasonably well, capturing $\sim$80\% of the \HI\ in the system. There are some prominent anomalous gas features that are not reproduced by the model, which we will analyse in more detail in the remainder of this work. In addition to the identification of these structures, the model contains the following distinctive components, informing us about the \HI\ structure in UGCA~320:

\begin{enumerate}
\item Either \textit{a flare or a \vthree{thick disc}}. Although we show the flare model in the main body of the paper, the difference between this model and one that includes a thick disc in addition to a thin disc is marginal enough that we do not favour one over the other. If the galaxy has a flaring \HI\ disc, the scale height increases to $\sim$170\% of the central scale height in the outskirts. If the morphology is closer to a thin + \vthree{thick} disc scenario, the thin disc contains $\sim$90\% of the \HI\, and \vthree{the thick disc}, with a scale height of $\sim$700 pc, the remaining $\sim$10\%. It is also possible that the galaxy has a flaring \HI\ \vthree{thick disc}, combining both scenarios. \vthree{Preliminary results from morphological and tilted-ring analysis of the scale height at the highest resolution ($7.2^{\prime\prime} \times 8.2^{\prime\prime}$, Ranaivoharimina et al., in prep.) suggest the presence of a thin disc and a thick disc that both have flaring features. This analysis reveals a central and peak scale height at the outskirts of $\sim 320$ pc and $\sim 530$ pc, respectively, for the thin disc. For the thick disc, their model shows a scale height of $\sim 1$ kpc at the galactic centre and flares up to $\sim 2$ kpc. Additionally, photometric fits of edge-on galaxies may help distinguish between flaring and thin disc/thick disc scenarios. In fact, this technique provides a geometrical estimate of a scale height and can directly reveal flare signatures. However, projection effects as well as beam smearing may attenuate or obscure the intrinsic flaring feature and inclination warps may mimic the presence of a halo. Thus, consistency in morphological and tilted-ring scale heights is required to confirm the validity of the flare, halo, or both.}

\item An \textit{inclination warp}. The inclination changes by $\sim$6 degrees on the approaching side, and $\sim$13 degrees on the receding side. An inclination warp is needed for an accurate model velocity field: it contributes to reproducing the correct shape of the moment one map, including the ``bend'' in the systemic velocity. Additionally, it is necessary for reproducing the correct observed (projected) scale height of the gas reservoir. The PVDs and velocity field of final model without this inclination warp are shown in Figures \ref{subfig:pvd_no_inc} and \ref{subfig:mom1_no_inc}, respectively. 
\item A \textit{change in position angle} of $\sim$20 degrees in the outer rings of the approaching side of the disc. This is needed to accurately reproduce the shapes of the PVDs along the minor axis in the central regions of the galaxy, especially for the lower density gas around the outskirts, and an accurate velocity field on the approaching side of the disc. A version of the final model without this change in position angle is shown in Figures \ref{subfig:pvd_no_pa} and \ref{subfig:mom1_no_pa}.
\item \textit{Radial motions} that increase with scale height in the outer rings. These are needed to accurately reproduce the kinematics of the gas, as reflected by the moment one map of the model and the shapes of the model PVDs. This is particularly important in the central parts of the disc and on the approaching side: \refrep{without the inclusion of radial motions the residuals are much more asymmetric here}. A version of the final model without radial motions is shown in Figures \ref{subfig:mom1_no_vrad} and \ref{subfig:pvd_no_vrad}.
\item A \textit{decrease in systemic velocity} of 6 -- 7 km s$^{-1}$ in the outer rings on the approaching side of the disc and an \textit{increase} in systemic velocity on the receding side. Without this change in systemic velocity it is impossible to reproduce the ``bend'' in the velocity field that is seen in the moment one map. A version of the final model without this change in systemic velocity is shown in Figures \ref{subfig:pvd_no_vsys} and \ref{subfig:mom1_no_vsys}. \vthree{We will discuss this further in \S \ref{subsec:vsys_change}.}
\item A \textit{shift in central position} along the minor axis (equivalent to a change in the central coordinates of the disc). This is particularly relevant in the outskirts of the receding side of the disc.
\end{enumerate}

\section{Irregular features \& anomalous gas in UGCA~320}
\label{sec:anomalous_gas}

\begin{figure*}
  \centering
  \includegraphics[width=0.8\textwidth]{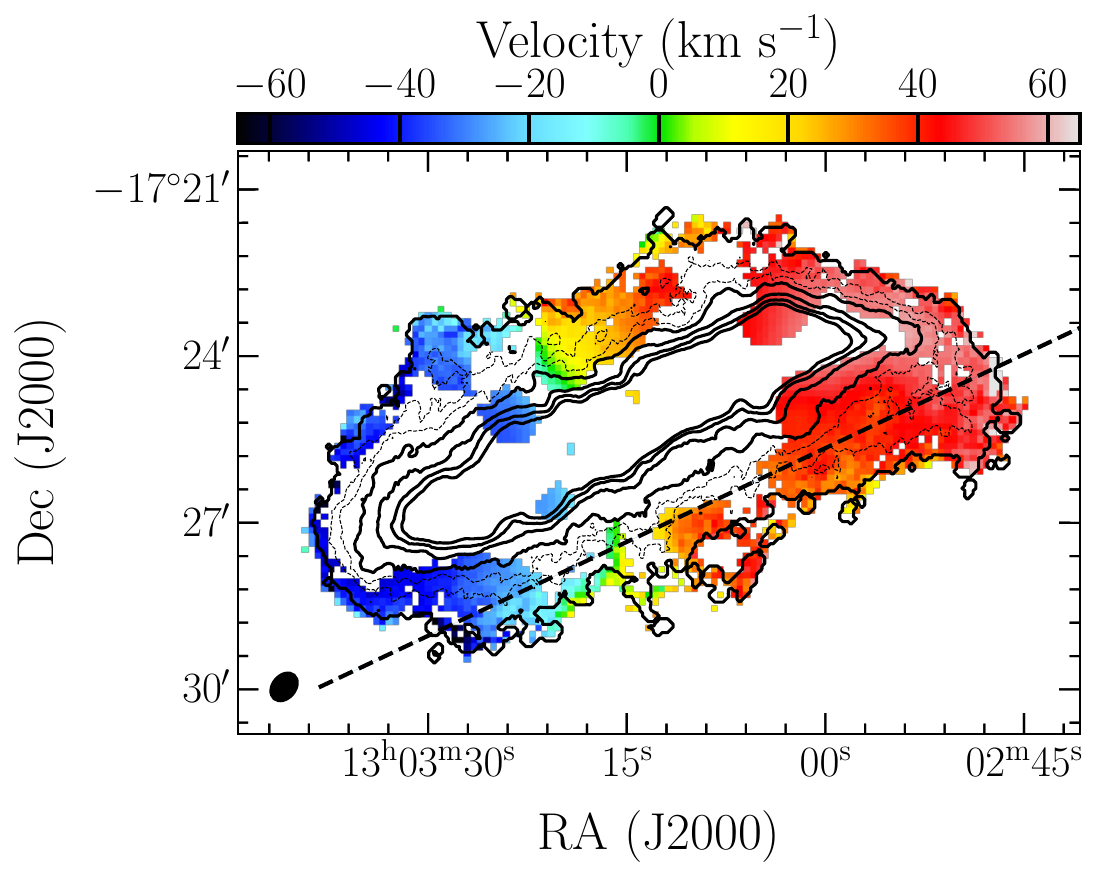}
  \caption{Velocity field of the gas that was not captured by the tilted ring model (i.e. residuals with absolute values $>5$ km s$^{-1}$). The five thick black contours are linearly spaced between \refrep{the minimum} and 10\% of the maximum surface brightness in the intensity map \refrep{presented in Figure \ref{fig:moment_maps}}, and an additional thin, dashed set of three contours are linearly spaced between \refrep{the minimum} and 1.5\% of the maximum surface brightness, to show the structure of the faintest gas in more detail. The synthesised beam of the observations is shown in the lower left corner. The dashed line indicates the path of the PVD presented in \S \ref{subsec:interaction}. While most of the anomalous gas is distributed around the very outskirts of the \HI\ reservoir, there are some prominent irregular features. There are two blobs of gas closer to the disc, on the \vthree{approaching} side, and a large asymmetric feature towards the south of the \vthree{receding} side of the \HI\ reservoir.}
  \label{fig:anomalous_gas}
\end{figure*}

In Figure \ref{fig:anomalous_gas} we show the velocity field of the neutral gas that was not captured by our final model (described in \S \ref{subsec:tirific}), i.e. the gas that has residuals with absolute values $> 5$ km s$^{-1}$ in velocity (see also \refrep{the middle row in Figure \ref{fig:residuals}}). 
Some gas structures clearly deviate from a regularly rotating disc, even with the addition of more complex model parameters such as radial motions and warps. Most of this anomalous gas has low surface densities and is \vthree{coincides} with the faint outskirts of the observed \HI\ reservoir. The most prominent anomalous gas feature is the extension towards the south-west on the outskirts of the \vthree{receding} side of the disc, which starts to be visible at $\sim 10\%$ of the maximum \HI\ surface brightness. Opposite this extension, on the north side of the galaxy, there is another patch of slightly higher surface density gas that is not \vthree{reproduced} by the tilted ring model, which possibly has the same origin. Other aspects that are not reproduced by the model are two relatively high column density blobs of gas on the \vthree{approaching} side of the disc, and a few larger areas of fainter gas at larger scale heights. Lastly, several filaments of \HI, most prominently visible on the south side of the \HI\ maps (see also \refrep{the top panel of Figure \ref{fig:mom0_group}}), are not captured by the model. We will discuss the possible sources of these anomalous gas features in the remainder of this section.

\subsection{\vthree{Change in systemic velocity}}
\label{subsec:vsys_change}
\vthree{In \S \ref{subsec:tirific} we find that a change in systemic velocity in the outer rings is required to adequately model the kinematics of the \HI\ disc. \citet{Serra2024} similarly found that a radially changing systemic velocity is required to reproduce the U-shaped \HI\ reservoir in the outer regions of dwarf galaxy NGC 1427A. They interpret this as an additional, radius dependent velocity component, which they attribute to external sources (ram pressure stripping). A similar difference in systemic velocity of $\sim$8 km s$^{-1}$, between the inner and outer regions of the \HI\ disc, was also found in the massive spiral galaxy NGC 5055 \citep{Battaglia2006, Blok2008, Jovanovic2017}. This is interpreted as the outer parts of the disc being dynamically decoupled from the inner parts, and \citet{Battaglia2006} suggest this may be related to the transition from the stellar disc dominated region to the dark matter halo dominated region. They also find a change in inclination and position angle in the outer regions, similar to what we observe in UGCA~320. Since we find a similar change in the systemic velocity in the outer two rings of the \HI\ disc in UGCA~320, this could be related to the same phenomenon. Alternatively, it could be an additional radius dependent velocity component, similar to NGC 1427A, which in this case could be a remnant of an interaction with UGCA~319. This will be discussed further in \S \ref{subsec:interaction}. }

\subsection{The area around the galaxy centre -- star formation driven outflow?}
\label{subsec:spectra}

\begin{figure*}

    \begin{subfigure}{0.33\textwidth}
    \centering
	\includegraphics[width=\textwidth]{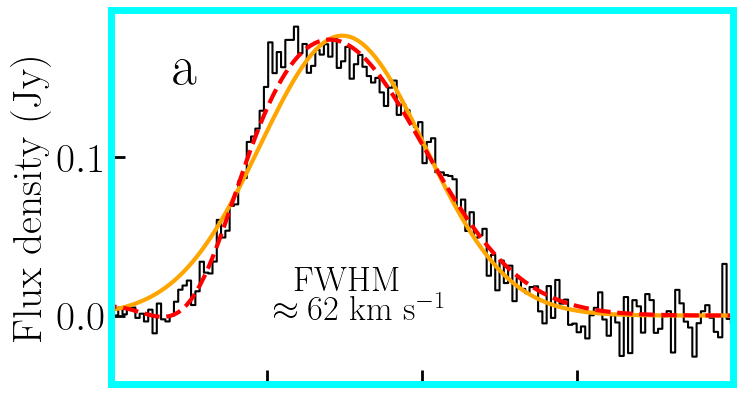}
	\label{subfig:spectrum_a}
	\end{subfigure} \hspace{2mm}
	\begin{subfigure}{0.32\textwidth}
    \centering
	\includegraphics[width=\textwidth]{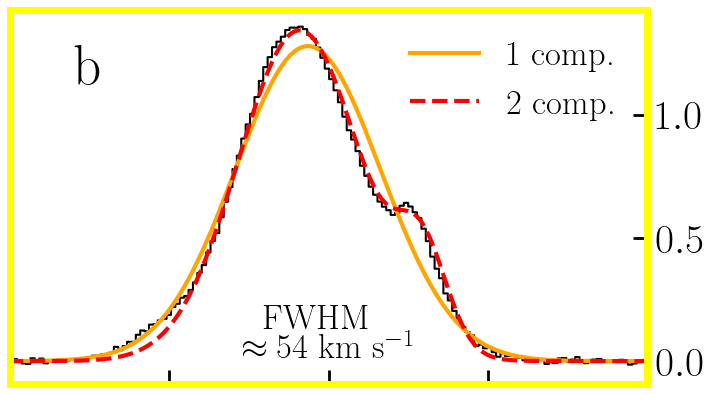}
	\label{subfig:spectrum_b}
	\end{subfigure}	
	
	\begin{subfigure}{0.31\textwidth}
    \centering
	\includegraphics[width=\textwidth]{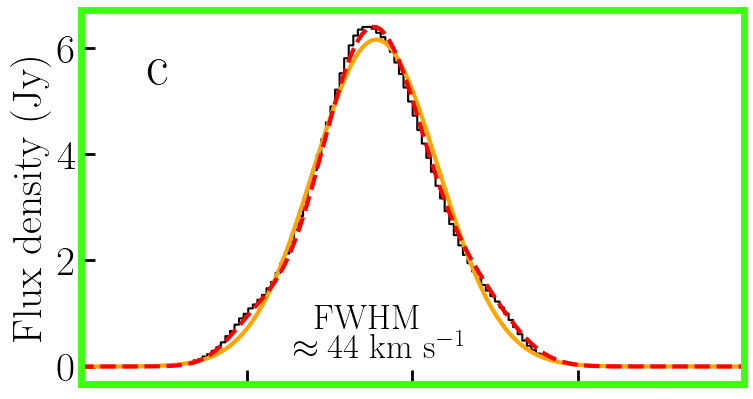}
	\label{subfig:spectrum_c}
	\end{subfigure}  \hspace{2mm}
	\begin{subfigure}{0.29\textwidth}
    \centering
	\includegraphics[width=\textwidth]{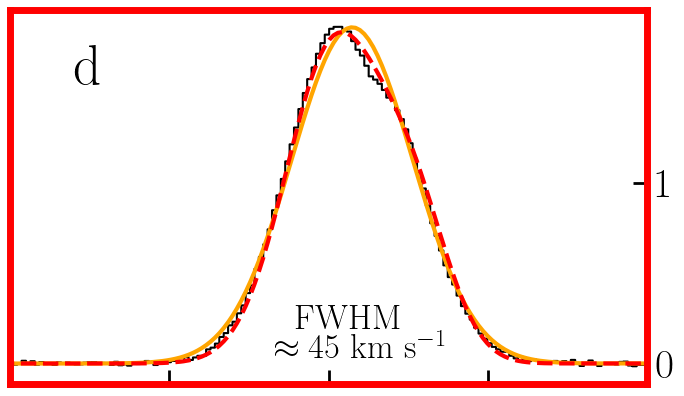}
	\label{subfig:spectrum_d}
	\end{subfigure}
	
	\begin{subfigure}{0.33\textwidth}
    \centering
	\includegraphics[width=\textwidth]{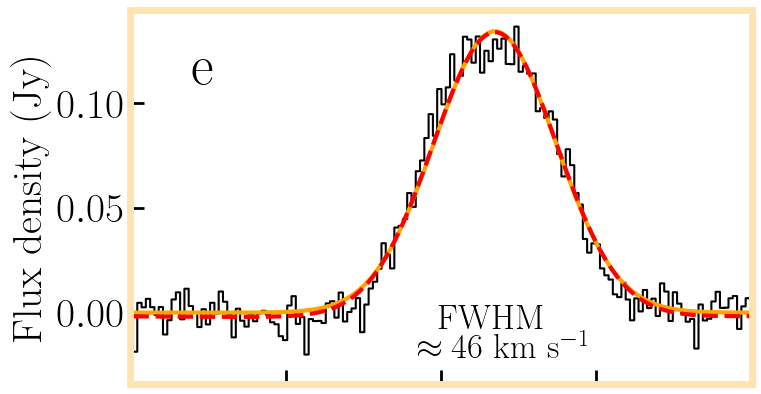}
	\label{subfig:spectrum_e}
	\end{subfigure} \hspace{2mm}
	\begin{subfigure}{0.30\textwidth}
    \centering
	\includegraphics[width=\textwidth]{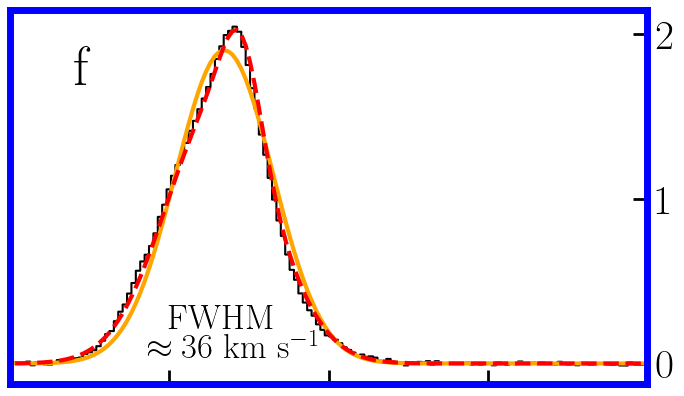}
	\label{subfig:spectrum_f}
	\end{subfigure}	
	
	\begin{subfigure}{0.34\textwidth}
    \centering
	\includegraphics[width=\textwidth]{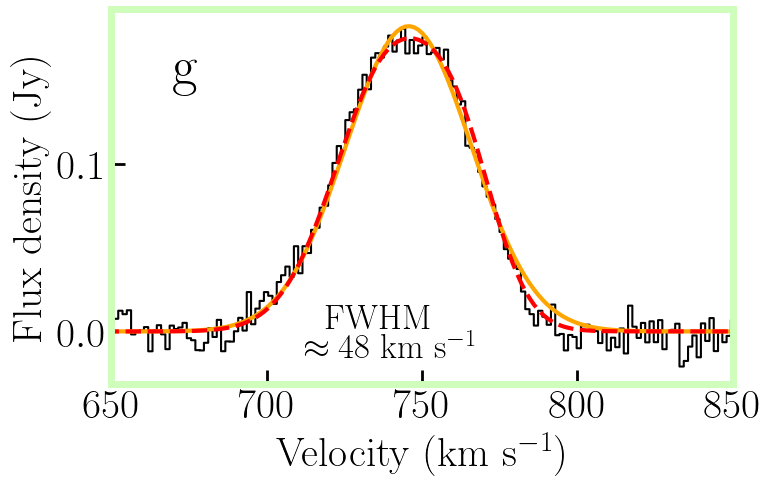}
	\label{subfig:spectrum_g}
	\end{subfigure} 
    \begin{subfigure}{0.32\textwidth}
    \centering
	\includegraphics[width=\textwidth]{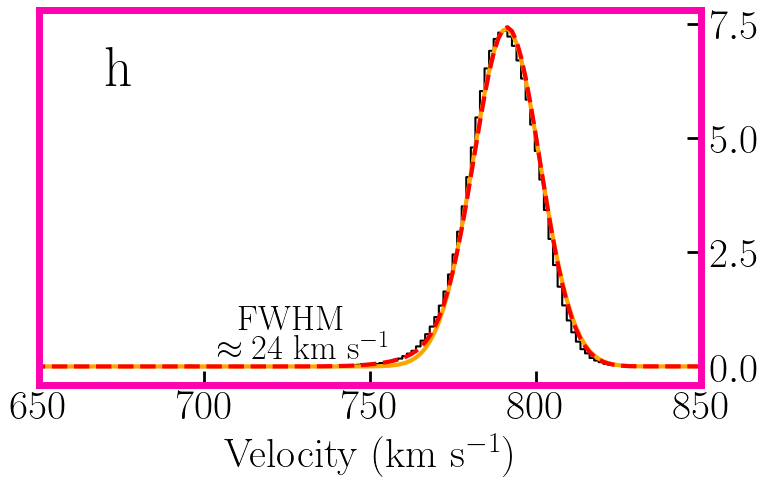}
	\label{subfig:spectrum_h}
	\end{subfigure}	   
	
	\begin{subfigure}{0.43\textwidth}
    \centering
	\includegraphics[width=\textwidth]{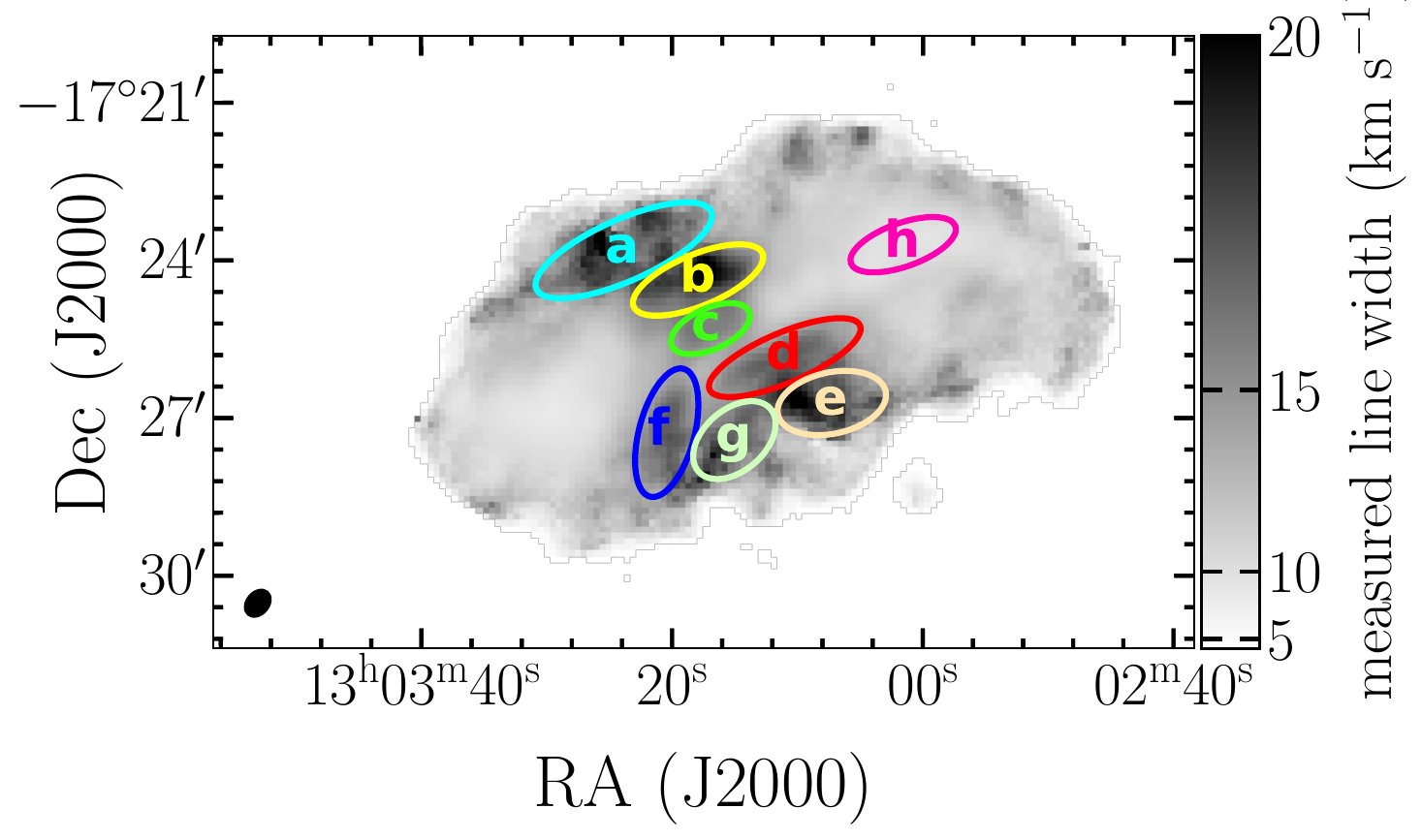}
	\label{subfig:spectrum_footprints}
	\end{subfigure}   
	
	\caption{\refrep{\HI\ spectra extracted from regions in UGCA~320 with increased observed line width. The regions from which spectra are extraced are indicated in the bottom panel, which shows the observed line width map with the beam-sized regions overlaid. The colour and letter associated which each region correspond to the colours of the frames of the sub-panels (marked a - h), and the letters indicated in their top-left corners. The moment two map is shown in greyscale with a logarithmic stretch to highlight the regions with increased values. Spectra a - g are extracted from regions with high measured line widths, and spectra h was extracted from the disc, serving as a control. Gaussian fits are overlaid for a single component (yellow solid line) and two components (red dashed lines). The full width at half maximum (FWHM) of the best-fit profile is indicated in each panel. Spectra a -- d and f are best fit with two Gaussian components, while a single component is sufficient for spectra e and g.}}
	\label{fig:spectra}
\end{figure*}

In the first part of this section we have identified several structures of anomalous gas in UGCA~320. Comparing Figure \ref{fig:anomalous_gas} to the moment two map (\refrep{right-hand panel of Figure \ref{fig:moment_maps} and bottom row of Figure \ref{fig:residuals}}), we can see that a substantial fraction of this gas is associated with regions with elevated measured line widths. For a highly inclined galaxy like UGCA~320, the highest measured line widths are usually distributed in a thinner strip associated with the mid-plane, reflecting \vthree{the increased rotational broadening along the line of sight}. To study the \HI\ in these regions in more detail, and investigate whether there is a correlation between the increased measured line widths and the anomalous gas, we take spectra of each coherent region in the moment two map. These regions, along with their extracted spectra, are shown in Figure \ref{fig:spectra}. 
The spectrum in the control region (panel \textit{h}) is Gaussian, with a FWHM of $\sim$24 km~s$^{-1}$. In areas with increased observed line widths some spectra are best described by wider Gaussians (\refrep{panels \textit{e} and \textit{g}}), whereas, \refrep{by eye,} others require multiple Gaussian components to be accurately described (\refrep{panels \textit{a} -- \textit{d} and \textit{f}}). This can be for a variety of reasons: regions \textit{a} and \textit{f} have skewed spectra, regions \textit{b} and \textit{d} have a clearly separate second component, and region \textit{c} has two wings on either side of the main component. The FWHM \refrep{(estimated using a single Gaussian component, for simplicity)} of the regions are similar, except regions \textit{a}, which covers a larger area (the effect goes away if the spectrum is extracted from an area with a size similar to those of the remaining regions), and \textit{f}, which is narrower than the other regions by $\sim$10 km~s$^{-1}$, but still wider than the control region by $\sim$10 km~s$^{-1}$. 

The presence of multiple Gaussian components, wings, and systematically broadened line profiles are consistent with the presence of an outflow. \vthree{The PVD along the minor axis, shown in panel \textbf{d} in \refrep{the top-right panel in Figure \ref{fig:pvd_slices_final_model}}, goes through the regions with elevated measured line width. Some radial motions are required to explain its shape (\S \ref{subsec:tirific}), which may be associated with the radial component of such an outflow.} Since UGCA~320 is a low-mass galaxy with no indication of harbouring an active galactic nucleus, such an outflow would most likely driven by star formation. To explore the possibility of a star formation driven outflow, we estimate the energy associated with the continuum presented in \S \ref{subsec:continuum}, and compare it with the \HI\ mass of the material displaced by the hypothetical outflow. To this end, we make the following assumptions:

\begin{itemize}
\item The detected radio continuum in the centre of UGCA~320 is non-thermal and entirely the result of recent star formation.
\item The proposed outflow has a conical geometry.
\item Only bulk motions associated with an outflow are responsible for the difference in line width between regions of high measured line width and the control region, everything else is constant.
\item The number of stars that explode as supernovae (SNe) per unit mass is 0.0068 M$_\odot^{-1}$ for a Salpeter initial mass function (IMF) \citep{Madau2014}, and thus
\item A Salpeter IMF (m$_\text{min} = 8 \text{M}_\odot$, m$_\text{max} = 40 \text{M}_\odot$ for SNe).
\item An individual supernova provides an energy of $10^{51}$ ergs.
\end{itemize}
\refrep{From \S \ref{subsec:continuum}, the integrated continuum flux at the basis of the potential outflow is $\sim$3~mJy. At a distance of $D = 6.03$ Mpc, this corresponds to a luminosity of $L_\nu \approx 10^{19}$ W Hz$^{-1}$.} Then, following \citet{Condon1992}, 

\begin{align}
\text{SFR} \left( \text{M}_\odot\ \text{yr}^{-1} \right) &= 0.75 \times 10^{-21}\ L_\text{1.4 GHz} \left( \text{W Hz}^{-1} \right) \\ 
&= 0.01 \text{ M}_\odot \text{ yr}^{-1}
\label{eqn:SFR}
\end{align}
\refrep{Assuming 0.0068 SNe per unit M$_\odot$ (see above), this means that $\sim 7 \times 10^{-5}$ SNe are created in the centre of UGCA~320 per year, collectively producing an energy of $7 \times 10^{46}$ ergs yr $^{-1}$. Assuming that 10\% of this energy is kinetic, this corresponds to a kinetic power of $\boldsymbol{\dot{E}}_{\text{out}} \approx \mathbf{2 \times 10^{39}}$ \textbf{erg s}$\mathbf{^{-1}}$, which is on the lower side of the distribution for local dwarf galaxies \citep{Romano2023}. We can then use the standard kinetic energy equation to derive the corresponding mass this hypothetical SNe-driven outflow is capable of displacing. For this we need the velocity of the outflow. Assuming Gaussian line shapes, we can derive this using the difference in width between the spectrum in the outflow regions and the control region: $\sigma_\text{bulk} = \sqrt{\sigma^2_\text{outflow} - \sigma^2_\text{control}}$. From Figure \ref{fig:spectra}, the average (mean) FWHM of the \HI\ line in the regions associated with the potential outflow is $\sim$48 km s $^{-1}$. Comparing this to the control region, which has a FWHM of $\sim$24 km s$^{-1}$, this implies a $\sigma_\text{bulk} \approx 18$ km s$^{-1}$. Because the galaxy is viewed almost edge-on, this velocity component along the line of sight is relatively small. Assuming that the maximum outflow velocity is perpendicular to the galaxy disc, we multiply this number by cos($i$), where $i \approx 80^\circ$ is the inclination angle of the galaxy. This gives a maximum outflow velocity of $v_\text{max} \approx 100$ km s$^{-1}$. From the kinetic energy equation, this means the outflow is able to displace a maximum mass of $\boldsymbol{\dot{M}}_{\text{out}} \mathbf{\approx 0.6\ \textbf{M}_\odot\ \textbf{yr}^{-1}}$. This is a reasonable outflow rate for a dwarf galaxy, although it is on the high side for the estimated SFR of UGCA~320 according to the scaling relation in \citet[][Figure 5]{Romano2023}. From the increased values in the moment two map (right-hand panel in Figure \ref{fig:moment_maps}) we estimate that $\sim30\%$ of the \HI\ mass of UGCA~320 is associated with the hypothetical outflow, which amounts to $\mathbf{M_\textbf{outflow} \sim 10^{8.5} \textbf{M}_\odot}$. This means that, with the current outflow rate, it would have taken at least $\mathbf{\sim 0.5}$ \textbf{Gyr} to displace the amount of anomalous \HI\ associated with the potential outflow. This would imply that we are observing long-lived signatures of an outflow alongside the relatively short-lived radio continuum. In that case, the outflow must have been stronger at earlier times or driven by multiple episodes rather than a single, steady event.}

\subsection{\vthree{Asymmetric extension \& peculiar velocities -- interaction with UGCA~319?}}
\label{subsec:interaction}

\begin{figure*}
  \centering
  \includegraphics[width=0.7\textwidth]{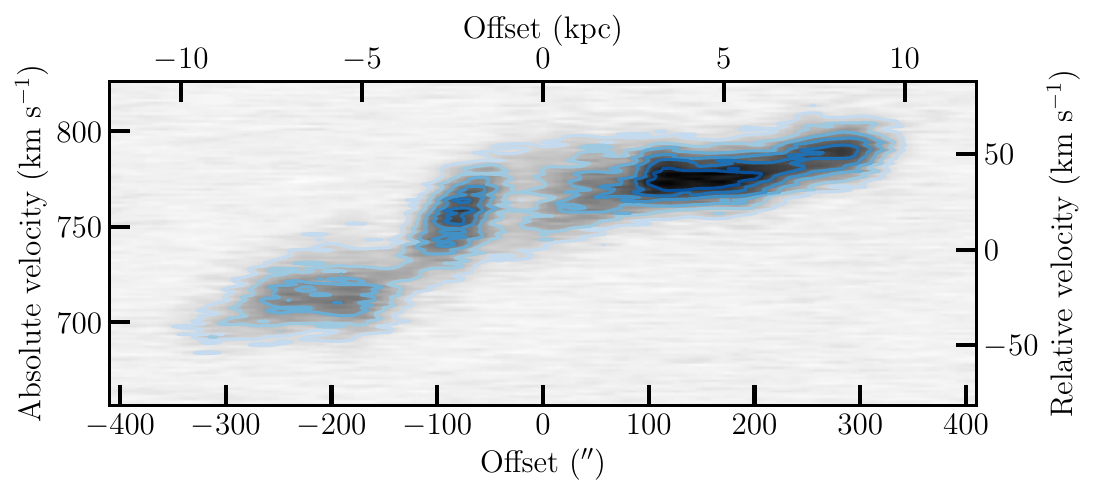}
  \caption{PVD of the low surface density \HI\ on the south side of UGCA~320 that is anomalous in velocity. The path of the PVD is shown in Figure \ref{fig:anomalous_gas}. Both on the receding and the approaching side there is a discontinuity with the gas in the centre of the galaxy. The anomalous gas on the approaching side is disconnected from that on the receding side both spatially and in velocity. \refrep{Therefore, we cannot confirm that both have the same origin.}}
  \label{fig:anomalous_pvd}
\end{figure*}

The other prominent \HI\ feature that is not reproduced by the tilted ring model is the low column density extended asymmetry on the \vthree{receding} side of the galaxy (Figure \ref{fig:anomalous_gas}). The level of asymmetry and location in the the outskirts of the \HI\ disc suggest an external origin. \vtwo{The warp and bend in the \HI\ disc and the shift in central coordinates and systemic velocity that were revealed by tilted ring modelling (see \S \ref{subsec:tirific}) are common signatures of tidal interaction (e.g. \citealt{Chemin2009, Eymeren2010, Kirby2012}). Since these interactions primarily affect the gas in the outer areas of the disc, this could result in a radially varying kinematic centre. Similarly, tidal interactions can result in a shift in the physical centre of the \HI\ disc with respect to the stellar disc, which can also be more pronounced in the outer rings.}

UGCA~320 is part of a small galaxy group, with two less massive and reasonably close companions (\S \ref{subsec:ugca_320}). \vthree{The more massive and nearest of the two}, UGCA~319, has been suggested by various authors to have interacted with UGCA~320 in the past. \vthree{For example,} \citet{Alabi2025} study the spectroscopic properties of UGCA~320 in detail using \vthree{optical} data from VLT/MUSE and SALT/RSS (see \S \ref{subsec:ugca_320}\vthree{)}. They observe a sharp transition in kinematic properties of the stars in the disc of UGCA~320 at $\sim 10 ^{\prime\prime}$ ($\sim$0.3 kpc), whereas the rotation amplitude of the ionised gas remains lower at the same radii. This kind of observation is a typical signature for a past tidal interaction, as stars respond to gravitational perturbations on longer timescales (see references in \citealt{Alabi2025}). \vthree{Furthermore, they find that this break in stellar kinematic properties does not correspond to a break in the stellar population properties, which further supports the hypothesis of a tidal interaction rather than e.g. a gas-rich minor merger or accretion.} They also identify an extended and asymmetric stellar region extending $\sim 100 ^{\prime \prime}$ in the north-west direction, along its major axis, which is additional evidence for a recent tidal interaction. Here, we see that UGCA~319 indeed has a very disturbed \HI\ morphology (Figure \ref{fig:optical}), which supports this hypothesis.

Carrying out an approximate calculation, we find that the tidal radius (the distance from the centre of a less massive system at which the gravitational influence of a more massive system becomes dominant) exceeds the extent of the \HI\ reservoir of UGCA~319 (Scannell et al., in prep.). We use the crossing time as an approximate timescale over which disturbances and irregularities in a system smooth out, and estimate this to be $\sim$100 Myr. Assuming UGCA~319 and UGCA~320 are roughly at the same distance along the line of sight (\S \ref{subsec:ugca_320}), thus adopting their projected distance as their total separation, this implies a relative velocity of $\sim$100 km~s$^{-1}$. \refrep{This is comparable to the rotation velocity of UGCA~320, which implies that it is well below its escape velocity. Thus, this is a reasonable velocity, and it is possible that the galaxies have undergone a close encounter in the past 100 Myr. Therefore, we suggest that a tidal interaction is the most plausible explanation for the extended features and asymmetries observed in the \HI\ disc of UGCA~320, which are not reproduced by the tilted ring model.}

\vthree{While a potential star formation driven outflow is estimated to have started $\sim$150 Myr -- 3 Gyr ago (\S \ref{subsec:spectra}), before any tidal interaction between UGCA~320 and UGCA~319 took place, it is plausible that it was boosted by such an interaction.} Evidence for mergers and tidal interactions enhancing outflow rates, through stimulating the funnelling of gas to the central regions of galaxies and increasing star formation rates, has been found both observationally \citep[e.g.][]{Barton2000, Lambas2003, Li2008, Haan2009, Alonso2004, Scudder2012} and in simulations \citep[e.g.][]{Noguchi1988, Barnes1991, Blumenthal2018, Li2025}. \vthree{Both the blue optical light from young stellar populations and radio continuum emission from SNe are expected to still be visible after 10 -- 100 Myr (e.g. \citealt{Bruzual2003, Hirashita2006}), which is in agreement with the estimated time since the possible tidal interaction.}

The final region in the galaxy that contains a significant amount of anomalous gas is the low column density gas on the \vthree{approaching} side of the galaxy. Above we discussed that UGCA~320 has likely undergone a tidal interaction with its nearest neighbour, UGCA~319. Besides the gas directly impacted by this interaction, tidal forces can also influence the gas on the opposite side of the galaxy. To test if this gas could be disturbed by the same mechanism, we take a PVD of one beam width through the anomalous gas on the south side of the galaxy, parallel to the major axis. This is shown in Figure \ref{fig:anomalous_pvd}, and the corresponding path is indicated with a dashed line in Figure \ref{fig:anomalous_gas}. The gas on the approaching side shows a clear discrepancy with the gas closer to the centre of the galaxy, as well as the gas on the receding side. \vthree{This means that we are not able to confirm that this gas was disturbed by the same tidal interaction as the anomalous gas on the redshifted side of the galaxy.}

\section{Summary}
\label{sec:discussion}
We have studied the \HI\ reservoir of the edge-on dwarf galaxy (log $\left( M_\star/\text{M}_\odot \right) = 7.91$) UGCA~320 in detail, using ultra-deep, high resolution observations from the MHONGOOSE programme on the MeerKAT telescope. UGCA~320 is a very blue, star forming galaxy, with a SFR \refrep{slightly above} the SFMS. It \vthree{is} the most massive galaxy of a small group of three, in which UGCA~319 is its nearest neighbour, and it is possibly associated with a larger, loose galaxy group around NGC 5068. UGCA~320 is asymmetric along the major axis, both in the optical and in \HI, exhibiting a significant extension towards the receding side of its disc. A careful analysis of the 3D structure of the \HI\ reservoir has revealed the following:

\begin{itemize}
\item The neutral gas reservoir in UGCA~320 extends to relatively large scale heights above the plane (between 2.5 and 4 kpc depending on the exact geometry), and shows a filamentary structure at larger scale heights. The \HI\ beyond the stellar plane makes up the $\sim$10\% lowest column density gas in the galaxy, with column densities $N_\text{\HI} \leq 10^{21}$ atoms cm$^{-2}$.

\item Detailed modelling of the \HI\ disc using a tilted ring approach reveals complex kinematics, particularly in the outer rings. Accurate modelling of the observed \HI\ disc requires separating the approaching and receding side of the disc, and additional features including an inclination warp, a change in position angle, radial motions, a varying systemic velocity \vthree{indicating additional bulk motions in the outer rings}, and a flare and/or \vthree{thick disc}.

\item Anomalies in the velocity field and asymmetric, broadened line profiles, often best described by \refrep{two (or even three)} Gaussian components, are consistent with the presence of an outflow. An estimate of the energetics of a potential star formation driven outflow, based on the 1.4 GHz continuum image, confirms that it is plausible that a star formation driven outflow has been going on in UGCA~320.

\item An extended asymmetric feature on the receding side of the disc is most likely explained by a past interaction with the galaxy's nearest neighbour UGCA~319. Such an interaction has been suggested in the literature. \vthree{According to our back-of-the-envelope calculations, it should have taken place in the last $\sim$100 Myr. Whether a similar albeit more subtle anomalous gas feature on the approaching side of the galaxy is caused by the same interaction is unclear.}
\end{itemize}

\vthree{In summary, while global measurements suggest that UGCA~320 is an ordinary galaxy, a more careful analysis of sensitive, resolved observations of its neutral gas reservoir reveals that it is actively undergoing evolution through interaction with its nearest neighbour, and likely harbours an ongoing outflow. This study highlights the importance of high-resolution, high-sensitivity observations in capturing the smaller scale physics that drive galaxy evolution, the role of environment in shaping galaxies, and the merit of studying dwarf galaxies.}

\section*{Acknowledgements}

NZ, DJP, and SK are/were supported through the South African Research Chairs Initiative of the Department of Science and Technology and National Research Foundation.

\vthree{PK is partially supported by the BMBF project 05A23PC1 for D-MeerKAT.}

\refrep{This work has received funding from the European Research Council (ERC) under the European Union’s Horizon 2020 research and innovation programme (grant agreement No 882793 `MeerGas').}

The MeerKAT telescope is operated by the South African Radio Astronomy Observatory, which is a facility of the National Research Foundation, an agency of the Department of Science and Innovation. 

We acknowledge the use of the ilifu cloud computing facility -- \url{www.ilifu.ac.za}, a partnership between the University of Cape Town, the University of the Western Cape, the University of Stellenbosch, Sol Plaatje University, the Cape Peninsula University of Technology, and the South African Radio Astronomy Observatory. The Ilifu facility is supported by contributions from the Inter-University Institute for Data Intensive Astronomy (IDIA -- a partnership between the University of Cape Town, the University of Pretoria, the University of the Western Cape and the South African Radio Astronomy Observatory), the Computational Biology division at UCT and the Data Intensive Research Initiative of South Africa (DIRISA).

(Part of) the data published here have been reduced using the CARACal pipeline, partially supported by ERC Starting grant number 679627 ``FORNAX'', MAECI Grant Number ZA18GR02, DST-NRF Grant Number 113121 as part of the ISARP Joint Research Scheme, and BMBF project 05A17PC2 for D-MeerKAT. Information about CARACal can be obtained online under the URL: \url{https://caracal.readthedocs.io}.

This work made use of Astropy:\footnote{http://www.astropy.org} a community-developed core Python package and an ecosystem of tools and resources for astronomy \citep{Astropy2013, Astropy2018, Astropy2022}.

This research made use of APLpy, an open-source plotting package for Python \citep{Robitaille2012}.

This work made use of SAOImage DS9 \citep{Joye2003}, TOPCAT \citep{Taylor2005}, Matplotlib \citep{Hunter2007}, Numpy \citep{Harris2020}, SciPy \citep{2020SciPy}, lifelines \citep{Davidson-Pilon2019}, and reproject: Python-based astronomical image reprojection \citep{Reproject2020}.

\section{Data Availability}
\vthree{The data used in this work are publicly available on the MHONGOOSE website: \url{https://mhongoose.astron.nl/}.}




\bibliographystyle{mnras}
\bibliography{References} 

@software{Heywood2020,
       author = {{Heywood}, Ian},
        title = "{oxkat: Semi-automated imaging of MeerKAT observations}",
 howpublished = {Astrophysics Source Code Library, record ascl:2009.003},
         year = 2020,
        month = sep,
          eid = {ascl:2009.003},
archivePrefix = {ascl},
       eprint = {2009.003},
       adsurl = {https://ui.adsabs.harvard.edu/abs/2020ascl.soft09003H}
}

@ARTICLE{English2017,
       author = {{English}, Jayanne},
        title = "{Canvas and cosmos: Visual art techniques applied to astronomy data}",
      journal = {International Journal of Modern Physics D},
         year = 2017,
        month = jan,
       volume = {26},
       number = {4},
          eid = {1730010},
        pages = {1730010},
          doi = {10.1142/S0218271817300105},
archivePrefix = {arXiv},
       eprint = {1703.04183},
 primaryClass = {astro-ph.IM},
       adsurl = {https://ui.adsabs.harvard.edu/abs/2017IJMPD..2630010E}
}

@ARTICLE{Blok2008,
       author = {{de Blok}, W.~J.~G. and {Walter}, F. and {Brinks}, E. and {Trachternach}, C. and {Oh}, S.-H. and {Kennicutt}, Jr., R.~C.},
        title = "{High-Resolution Rotation Curves and Galaxy Mass Models from THINGS}",
      journal = {\aj},
         year = 2008,
        month = dec,
       volume = {136},
       number = {6},
        pages = {2648-2719},
          doi = {10.1088/0004-6256/136/6/2648},
archivePrefix = {arXiv},
       eprint = {0810.2100},
 primaryClass = {astro-ph},
       adsurl = {https://ui.adsabs.harvard.edu/abs/2008AJ....136.2648D}
}

@ARTICLE{Battaglia2006,
       author = {{Battaglia}, G. and {Fraternali}, F. and {Oosterloo}, T. and {Sancisi}, R.},
        title = "{ion\{H\}\{i\} study of the warped spiral galaxy NGC 5055: a disk/dark matter halo offset?}",
      journal = {\aap},
         year = 2006,
        month = feb,
       volume = {447},
       number = {1},
        pages = {49-62},
          doi = {10.1051/0004-6361:20053210},
archivePrefix = {arXiv},
       eprint = {astro-ph/0509382},
 primaryClass = {astro-ph},
       adsurl = {https://ui.adsabs.harvard.edu/abs/2006A&A...447...49B}
}

@ARTICLE{Jovanovic2017,
       author = {{Jovanovi{\'c}}, Milena},
        title = "{Two regimes of galaxy dynamics: mass models of NGC 5055 and DDO 154}",
      journal = {\mnras},
         year = 2017,
        month = aug,
       volume = {469},
       number = {3},
        pages = {3564-3575},
          doi = {10.1093/mnras/stx1009},
       adsurl = {https://ui.adsabs.harvard.edu/abs/2017MNRAS.469.3564J}
}

@ARTICLE{Flaugher2015,
       author = {{Flaugher}, B. and {Diehl}, H.~T. and {Honscheid}, K. and {Abbott}, T.~M.~C. and {Alvarez}, O. and {Angstadt}, R. and {Annis}, J.~T. and {Antonik}, M. and {Ballester}, O. and {Beaufore}, L. and {Bernstein}, G.~M. and {Bernstein}, R.~A. and {Bigelow}, B. and {Bonati}, M. and {Boprie}, D. and {Brooks}, D. and {Buckley-Geer}, E.~J. and {Campa}, J. and {Cardiel-Sas}, L. and {Castander}, F.~J. and {Castilla}, J. and {Cease}, H. and {Cela-Ruiz}, J.~M. and {Chappa}, S. and {Chi}, E. and {Cooper}, C. and {da Costa}, L.~N. and {Dede}, E. and {Derylo}, G. and {DePoy}, D.~L. and {de Vicente}, J. and {Doel}, P. and {Drlica-Wagner}, A. and {Eiting}, J. and {Elliott}, A.~E. and {Emes}, J. and {Estrada}, J. and {Fausti Neto}, A. and {Finley}, D.~A. and {Flores}, R. and {Frieman}, J. and {Gerdes}, D. and {Gladders}, M.~D. and {Gregory}, B. and {Gutierrez}, G.~R. and {Hao}, J. and {Holland}, S.~E. and {Holm}, S. and {Huffman}, D. and {Jackson}, C. and {James}, D.~J. and {Jonas}, M. and {Karcher}, A. and {Karliner}, I. and {Kent}, S. and {Kessler}, R. and {Kozlovsky}, M. and {Kron}, R.~G. and {Kubik}, D. and {Kuehn}, K. and {Kuhlmann}, S. and {Kuk}, K. and {Lahav}, O. and {Lathrop}, A. and {Lee}, J. and {Levi}, M.~E. and {Lewis}, P. and {Li}, T.~S. and {Mandrichenko}, I. and {Marshall}, J.~L. and {Martinez}, G. and {Merritt}, K.~W. and {Miquel}, R. and {Mu{\~n}oz}, F. and {Neilsen}, E.~H. and {Nichol}, R.~C. and {Nord}, B. and {Ogando}, R. and {Olsen}, J. and {Palaio}, N. and {Patton}, K. and {Peoples}, J. and {Plazas}, A.~A. and {Rauch}, J. and {Reil}, K. and {Rheault}, J.-P. and {Roe}, N.~A. and {Rogers}, H. and {Roodman}, A. and {Sanchez}, E. and {Scarpine}, V. and {Schindler}, R.~H. and {Schmidt}, R. and {Schmitt}, R. and {Schubnell}, M. and {Schultz}, K. and {Schurter}, P. and {Scott}, L. and {Serrano}, S. and {Shaw}, T.~M. and {Smith}, R.~C. and {Soares-Santos}, M. and {Stefanik}, A. and {Stuermer}, W. and {Suchyta}, E. and {Sypniewski}, A. and {Tarle}, G. and {Thaler}, J. and {Tighe}, R. and {Tran}, C. and {Tucker}, D. and {Walker}, A.~R. and {Wang}, G. and {Watson}, M. and {Weaverdyck}, C. and {Wester}, W. and {Woods}, R. and {Yanny}, B. and {DES Collaboration}},
        title = "{The Dark Energy Camera}",
      journal = {\aj},
         year = 2015,
        month = nov,
       volume = {150},
       number = {5},
          eid = {150},
        pages = {150},
          doi = {10.1088/0004-6256/150/5/150},
archivePrefix = {arXiv},
       eprint = {1504.02900},
 primaryClass = {astro-ph.IM},
       adsurl = {https://ui.adsabs.harvard.edu/abs/2015AJ....150..150F}
}

@ARTICLE{Alabi2025,
       author = {{Alabi}, Adebusola B. and {Loubser}, S. Ilani and {Mogotsi}, Moses K. and {Zabel}, N.},
        title = "{Stars and ionized gas in UGCA 320: a nearby gas-rich, dwarf irregular galaxy}",
      journal = {\mnras},
         year = 2025,
        month = nov,
       volume = {543},
       number = {4},
        pages = {3613-3627},
          doi = {10.1093/mnras/staf1553},
archivePrefix = {arXiv},
       eprint = {2509.20359},
 primaryClass = {astro-ph.GA},
       adsurl = {https://ui.adsabs.harvard.edu/abs/2025MNRAS.543.3613A}
}

@ARTICLE{Romano2023,
       author = {{Romano}, M. and {Nanni}, A. and {Donevski}, D. and {Ginolfi}, M. and {Jones}, G.~C. and {Shivaei}, I. and {Junais} and {Salak}, D. and {Sawant}, P.},
        title = "{Star-formation-driven outflows in local dwarf galaxies as revealed from [CII] observations by Herschel}",
      journal = {\aap},
         year = 2023,
        month = sep,
       volume = {677},
          eid = {A44},
        pages = {A44},
          doi = {10.1051/0004-6361/202346143},
archivePrefix = {arXiv},
       eprint = {2306.10433},
 primaryClass = {astro-ph.GA},
       adsurl = {https://ui.adsabs.harvard.edu/abs/2023A&A...677A..44R}
}

@ARTICLE{Hirashita2006,
       author = {{Hirashita}, H. and {Hunt}, L.~K.},
        title = "{Time evolution of the radio continuum of young starbursts: the importance of synchrotron emission}",
      journal = {\aap},
         year = 2006,
        month = dec,
       volume = {460},
       number = {1},
        pages = {67-81},
          doi = {10.1051/0004-6361:20065629},
archivePrefix = {arXiv},
       eprint = {astro-ph/0609733},
 primaryClass = {astro-ph},
       adsurl = {https://ui.adsabs.harvard.edu/abs/2006A&A...460...67H}
}

@ARTICLE{Bruzual2003,
       author = {{Bruzual}, G. and {Charlot}, S.},
        title = "{Stellar population synthesis at the resolution of 2003}",
      journal = {\mnras},
         year = 2003,
        month = oct,
       volume = {344},
       number = {4},
        pages = {1000-1028},
          doi = {10.1046/j.1365-8711.2003.06897.x},
archivePrefix = {arXiv},
       eprint = {astro-ph/0309134},
 primaryClass = {astro-ph},
       adsurl = {https://ui.adsabs.harvard.edu/abs/2003MNRAS.344.1000B}
}

@ARTICLE{Kamphuis2013,
       author = {{Kamphuis}, P. and {Rand}, R.~J. and {J{\'o}zsa}, G.~I.~G. and {Zschaechner}, L.~K. and {Heald}, G.~H. and {Patterson}, M.~T. and {Gentile}, G. and {Walterbos}, R.~A.~M. and {Serra}, P. and {de Blok}, W.~J.~G.},
        title = "{HALOGAS observations of NGC 5023 and UGC 2082: modelling of non-cylindrically symmetric gas distributions in edge-on galaxies}",
      journal = {\mnras},
         year = 2013,
        month = sep,
       volume = {434},
       number = {3},
        pages = {2069-2093},
          doi = {10.1093/mnras/stt1153},
archivePrefix = {arXiv},
       eprint = {1306.5312},
 primaryClass = {astro-ph.GA},
       adsurl = {https://ui.adsabs.harvard.edu/abs/2013MNRAS.434.2069K}
}

@ARTICLE{Fraternali2008,
       author = {{Fraternali}, F. and {Binney}, J.~J.},
        title = "{Accretion of gas on to nearby spiral galaxies}",
      journal = {\mnras},
         year = 2008,
        month = may,
       volume = {386},
       number = {2},
        pages = {935-944},
          doi = {10.1111/j.1365-2966.2008.13071.x},
archivePrefix = {arXiv},
       eprint = {0802.0496},
 primaryClass = {astro-ph},
       adsurl = {https://ui.adsabs.harvard.edu/abs/2008MNRAS.386..935F}
}

@ARTICLE{Bregman1980,
       author = {{Bregman}, J.~N.},
        title = "{The galactic fountain of high-velocity clouds.}",
      journal = {\apj},
         year = 1980,
        month = mar,
       volume = {236},
        pages = {577-591},
          doi = {10.1086/157776},
       adsurl = {https://ui.adsabs.harvard.edu/abs/1980ApJ...236..577B}
}

@ARTICLE{Shapiro1976,
       author = {{Shapiro}, Paul R. and {Field}, George B.},
        title = "{Consequences of a New Hot Component of the Interstellar Medium}",
      journal = {\apj},
         year = 1976,
        month = may,
       volume = {205},
        pages = {762-765},
          doi = {10.1086/154332},
       adsurl = {https://ui.adsabs.harvard.edu/abs/1976ApJ...205..762S}
}

@ARTICLE{Kamphuis2022,
       author = {{Kamphuis}, P. and {J{\"u}tte}, E. and {Heald}, G.~H. and {Herrera Ruiz}, N. and {J{\'o}zsa}, G.~I.~G. and {de Blok}, W.~J.~G. and {Serra}, P. and {Marasco}, A. and {Dettmar}, R.-J. and {Pingel}, N.~M. and {Oosterloo}, T. and {Rand}, R.~J. and {Walterbos}, R.~A.~M. and {van der Hulst}, J.~M.},
        title = "{HALOGAS: Strong constraints on the neutral gas reservoir and accretion rate in nearby spiral galaxies}",
      journal = {\aap},
         year = 2022,
        month = dec,
       volume = {668},
          eid = {A182},
        pages = {A182},
          doi = {10.1051/0004-6361/202140704},
archivePrefix = {arXiv},
       eprint = {2210.09383},
 primaryClass = {astro-ph.GA},
       adsurl = {https://ui.adsabs.harvard.edu/abs/2022A&A...668A.182K}
}

@ARTICLE{Marasco2025,
       author = {{Marasco}, A. and {de Blok}, W.~J.~G. and {Maccagni}, F.~M. and {Fraternali}, F. and {Oman}, K.~A. and {Oosterloo}, T. and {Combes}, F. and {McGaugh}, S.~S. and {Kamphuis}, P. and {Spekkens}, K. and {Kleiner}, D. and {Veronese}, S. and {Amram}, P. and {Chemin}, L. and {Brinks}, E.},
        title = "{HI within and around observed and simulated galaxy discs: Comparing MeerKAT observations with mock data from TNG50 and FIRE-2}",
      journal = {\aap},
         year = 2025,
        month = may,
       volume = {697},
          eid = {A86},
        pages = {A86},
          doi = {10.1051/0004-6361/202453172},
archivePrefix = {arXiv},
       eprint = {2503.03818},
 primaryClass = {astro-ph.GA},
       adsurl = {https://ui.adsabs.harvard.edu/abs/2025A&A...697A..86M}
}

@ARTICLE{Tamburro2009,
       author = {{Tamburro}, D. and {Rix}, H. -W. and {Leroy}, A.~K. and {Mac Low}, M. -M. and {Walter}, F. and {Kennicutt}, R.~C. and {Brinks}, E. and {de Blok}, W.~J.~G.},
        title = "{What is Driving the H I Velocity Dispersion?}",
      journal = {\aj},
         year = 2009,
        month = may,
       volume = {137},
       number = {5},
        pages = {4424-4435},
          doi = {10.1088/0004-6256/137/5/4424},
archivePrefix = {arXiv},
       eprint = {0903.0183},
 primaryClass = {astro-ph.GA},
       adsurl = {https://ui.adsabs.harvard.edu/abs/2009AJ....137.4424T}
}

@ARTICLE{Blumenthal2018,
       author = {{Blumenthal}, Kelly A. and {Barnes}, Joshua E.},
        title = "{Go with the Flow: Understanding inflow mechanisms in galaxy collisions}",
      journal = {\mnras},
         year = 2018,
        month = sep,
       volume = {479},
       number = {3},
        pages = {3952-3965},
          doi = {10.1093/mnras/sty1605},
archivePrefix = {arXiv},
       eprint = {1806.05132},
 primaryClass = {astro-ph.GA},
       adsurl = {https://ui.adsabs.harvard.edu/abs/2018MNRAS.479.3952B}
}

@ARTICLE{Li2025,
       author = {{Li}, Fei and {Rahman}, Mubdi and {Murray}, Norman and {Kere{\v{s}}}, Du{\v{s}}an and {Wetzel}, Andrew and {Faucher-Gigu{\`e}re}, Claude-Andr{\'e} and {Hopkins}, Philip F. and {Moreno}, Jorge},
        title = "{The Effect of Galaxy Interactions on Starbursts in Milky Way-mass Galaxies in FIRE Simulations}",
      journal = {\apj},
         year = 2025,
        month = jan,
       volume = {979},
       number = {1},
          eid = {7},
        pages = {7},
          doi = {10.3847/1538-4357/ad94ef},
archivePrefix = {arXiv},
       eprint = {2411.10373},
 primaryClass = {astro-ph.GA},
       adsurl = {https://ui.adsabs.harvard.edu/abs/2025ApJ...979....7L}
}

@ARTICLE{Lambas2003,
       author = {{Lambas}, Diego G. and {Tissera}, Patricia B. and {Alonso}, M. Sol and {Coldwell}, Georgina},
        title = "{Galaxy pairs in the 2dF survey - I. Effects of interactions on star formation in the field}",
      journal = {\mnras},
         year = 2003,
        month = dec,
       volume = {346},
       number = {4},
        pages = {1189-1196},
          doi = {10.1111/j.1365-2966.2003.07179.x},
archivePrefix = {arXiv},
       eprint = {astro-ph/0212222},
 primaryClass = {astro-ph},
       adsurl = {https://ui.adsabs.harvard.edu/abs/2003MNRAS.346.1189L}
}

@ARTICLE{Barton2000,
       author = {{Barton}, Elizabeth J. and {Geller}, Margaret J. and {Kenyon}, Scott J.},
        title = "{Tidally Triggered Star Formation in Close Pairs of Galaxies}",
      journal = {\apj},
         year = 2000,
        month = feb,
       volume = {530},
       number = {2},
        pages = {660-679},
          doi = {10.1086/308392},
archivePrefix = {arXiv},
       eprint = {astro-ph/9909217},
 primaryClass = {astro-ph},
       adsurl = {https://ui.adsabs.harvard.edu/abs/2000ApJ...530..660B}
}

@ARTICLE{Alonso2004,
       author = {{Alonso}, M. Sol and {Tissera}, Patricia B. and {Coldwell}, Georgina and {Lambas}, Diego G.},
        title = "{Galaxy pairs in the 2dF survey - II. Effects of interactions on star formation in groups and clusters}",
      journal = {\mnras},
         year = 2004,
        month = aug,
       volume = {352},
       number = {3},
        pages = {1081-1088},
          doi = {10.1111/j.1365-2966.2004.08002.x},
archivePrefix = {arXiv},
       eprint = {astro-ph/0401455},
 primaryClass = {astro-ph},
       adsurl = {https://ui.adsabs.harvard.edu/abs/2004MNRAS.352.1081A}
}

@ARTICLE{Scudder2012,
       author = {{Scudder}, Jillian M. and {Ellison}, Sara L. and {Mendel}, J. Trevor},
        title = "{The dependence of galaxy group star formation rates and metallicities on large-scale environment}",
      journal = {\mnras},
         year = 2012,
        month = jul,
       volume = {423},
       number = {3},
        pages = {2690-2704},
          doi = {10.1111/j.1365-2966.2012.21080.x},
archivePrefix = {arXiv},
       eprint = {1204.2828},
 primaryClass = {astro-ph.CO},
       adsurl = {https://ui.adsabs.harvard.edu/abs/2012MNRAS.423.2690S}
}

@ARTICLE{Haan2009,
       author = {{Haan}, Sebastian and {Schinnerer}, Eva and {Emsellem}, Eric and {Garc{\'\i}a-Burillo}, Santiago and {Combes}, Francoise and {Mundell}, Carole G. and {Rix}, Hans-Walter},
        title = "{Dynamical Evolution of AGN Host Galaxies{\textemdash}Gas In/Out-Flow Rates in Seven NUGA Galaxies}",
      journal = {\apj},
         year = 2009,
        month = feb,
       volume = {692},
       number = {2},
        pages = {1623-1661},
          doi = {10.1088/0004-637X/692/2/1623},
archivePrefix = {arXiv},
       eprint = {0811.1988},
 primaryClass = {astro-ph},
       adsurl = {https://ui.adsabs.harvard.edu/abs/2009ApJ...692.1623H}
}

@ARTICLE{Noguchi1988,
       author = {{Noguchi}, M.},
        title = "{Gas dynamics in interacting disc galaxies}",
      journal = {\aap},
         year = 1988,
        month = sep,
       volume = {203},
        pages = {259-272},
       adsurl = {https://ui.adsabs.harvard.edu/abs/1988A&A...203..259N}
}

@ARTICLE{Barnes1991,
       author = {{Barnes}, Joshua E. and {Hernquist}, Lars E.},
        title = "{Fueling Starburst Galaxies with Gas-rich Mergers}",
      journal = {\apjl},
         year = 1991,
        month = apr,
       volume = {370},
        pages = {L65},
          doi = {10.1086/185978},
       adsurl = {https://ui.adsabs.harvard.edu/abs/1991ApJ...370L..65B}
}

@ARTICLE{Li2008,
       author = {{Li}, Cheng and {Kauffmann}, Guinevere and {Heckman}, Timothy M. and {Jing}, Y.~P. and {White}, Simon D.~M.},
        title = "{Interaction-induced star formation in a complete sample of {}10$^{5}$ nearby star-forming galaxies}",
      journal = {\mnras},
         year = 2008,
        month = apr,
       volume = {385},
       number = {4},
        pages = {1903-1914},
          doi = {10.1111/j.1365-2966.2008.13000.x},
archivePrefix = {arXiv},
       eprint = {0711.3792},
 primaryClass = {astro-ph},
       adsurl = {https://ui.adsabs.harvard.edu/abs/2008MNRAS.385.1903L}
}

@ARTICLE{Eymeren2010,
       author = {{van Eymeren}, Janine and {Koribalski}, B{\"a}rbel S. and {L{\'o}pez-S{\'a}nchez}, {\'A}ngel R. and {Dettmar}, Ralf-J{\"u}rgen and {Bomans}, Dominik J.},
        title = "{A kinematic study of the neutral and ionized gas in the irregular dwarf galaxies IC4662 and NGC5408}",
      journal = {\mnras},
         year = 2010,
        month = sep,
       volume = {407},
       number = {1},
        pages = {113-132},
          doi = {10.1111/j.1365-2966.2010.16923.x},
archivePrefix = {arXiv},
       eprint = {1004.4757},
 primaryClass = {astro-ph.CO},
       adsurl = {https://ui.adsabs.harvard.edu/abs/2010MNRAS.407..113V}
}

@ARTICLE{Chemin2009,
       author = {{Chemin}, Laurent and {Carignan}, Claude and {Foster}, Tyler},
        title = "{H I Kinematics and Dynamics of Messier 31}",
      journal = {\apj},
         year = 2009,
        month = nov,
       volume = {705},
       number = {2},
        pages = {1395-1415},
          doi = {10.1088/0004-637X/705/2/1395},
archivePrefix = {arXiv},
       eprint = {0909.3846},
 primaryClass = {astro-ph.CO},
       adsurl = {https://ui.adsabs.harvard.edu/abs/2009ApJ...705.1395C}
}

@ARTICLE{Kirby2012,
       author = {{Kirby}, Emma M. and {Koribalski}, B{\"a}rbel and {Jerjen}, Helmut and {L{\'o}pez-S{\'a}nchez}, {\'A}ngel},
        title = "{The Local Volume H I Survey: galaxy kinematics}",
      journal = {\mnras},
         year = 2012,
        month = mar,
       volume = {420},
       number = {4},
        pages = {2924-2943},
          doi = {10.1111/j.1365-2966.2011.20103.x},
archivePrefix = {arXiv},
       eprint = {1202.0354},
 primaryClass = {astro-ph.CO},
       adsurl = {https://ui.adsabs.harvard.edu/abs/2012MNRAS.420.2924K}
}

@ARTICLE{Condon1992,
       author = {{Condon}, J.~J.},
        title = "{Radio emission from normal galaxies.}",
      journal = {\araa},
         year = 1992,
        month = jan,
       volume = {30},
        pages = {575-611},
          doi = {10.1146/annurev.aa.30.090192.003043},
       adsurl = {https://ui.adsabs.harvard.edu/abs/1992ARA&A..30..575C}
}

@ARTICLE{Madau2014,
       author = {{Madau}, Piero and {Dickinson}, Mark},
        title = "{Cosmic Star-Formation History}",
      journal = {\araa},
         year = 2014,
        month = aug,
       volume = {52},
        pages = {415-486},
          doi = {10.1146/annurev-astro-081811-125615},
archivePrefix = {arXiv},
       eprint = {1403.0007},
 primaryClass = {astro-ph.CO},
       adsurl = {https://ui.adsabs.harvard.edu/abs/2014ARA&A..52..415M}
}

@ARTICLE{Roberts1978A,
       author = {{Roberts}, M.~S.},
        title = "{Twenty-one centimeter line widths of galaxies.}",
      journal = {\aj},
         year = 1978,
        month = sep,
       volume = {83},
        pages = {1026-1035},
          doi = {10.1086/112287},
       adsurl = {https://ui.adsabs.harvard.edu/abs/1978AJ.....83.1026R}
}

@ARTICLE{Maccagni2024,
       author = {{Maccagni}, F.~M. and {de Blok}, W.~J.~G. and {Mancera Pi{\~n}a}, P.~E. and {Ragusa}, R. and {Iodice}, E. and {Spavone}, M. and {McGaugh}, S. and {Oman}, K.~A. and {Oosterloo}, T.~A. and {Koribalski}, B.~S. and {Kim}, M. and {Adams}, E.~A.~K. and {Amram}, P. and {Bosma}, A. and {Bigiel}, F. and {Brinks}, E. and {Chemin}, L. and {Combes}, F. and {Gibson}, B. and {Healy}, J. and {Holwerda}, B.~W. and {J{\'o}zsa}, G.~I.~G. and {Kamphuis}, P. and {Kleiner}, D. and {Kurapati}, S. and {Marasco}, A. and {Spekkens}, K. and {Veronese}, S. and {Walter}, F. and {Zabel}, N. and {Zijlstra}, A.},
        title = "{MHONGOOSE discovery of a gas-rich low surface brightness galaxy in the Dorado group}",
      journal = {\aap},
         year = 2024,
        month = oct,
       volume = {690},
          eid = {A69},
        pages = {A69},
          doi = {10.1051/0004-6361/202449441},
archivePrefix = {arXiv},
       eprint = {2405.17000},
 primaryClass = {astro-ph.GA},
       adsurl = {https://ui.adsabs.harvard.edu/abs/2024A&A...690A..69M}
}

@ARTICLE{Veronese2025,
       author = {{Veronese}, S. and {de Blok}, W.~J.~G. and {Healy}, J. and {Kleiner}, D. and {Marasco}, A. and {Maccagni}, F.~M. and {Kamphuis}, P. and {Brinks}, E. and {Holwerda}, B.~W. and {Zabel}, N. and {Chemin}, L. and {Adams}, E.~A.~K. and {Kurapati}, S. and {Sorgho}, A. and {Spekkens}, K. and {Combes}, F. and {Pisano}, D.~J. and {Walter}, F. and {Amram}, P. and {Bigiel}, F. and {Wong}, O.~I. and {Athanassoula}, E.},
        title = "{Searching for HI around MHONGOOSE galaxies via spectral stacking}",
      journal = {\aap},
         year = 2025,
        month = jan,
       volume = {693},
          eid = {A97},
        pages = {A97},
          doi = {10.1051/0004-6361/202452085},
archivePrefix = {arXiv},
       eprint = {2411.11584},
 primaryClass = {astro-ph.GA},
       adsurl = {https://ui.adsabs.harvard.edu/abs/2025A&A...693A..97V}
}

@ARTICLE{Kurapati2025,
       author = {{Kurapati}, Sushma and {Pisano}, D.~J. and {de Blok}, W.~J.~G. and {Kamphuis}, Peter and {Zabel}, Nikki and {de Villiers}, Mikhail and {Healy}, Julia and {Maccagni}, Filippo M. and {Kleiner}, Dane and {Adams}, Elizabeth A.~K. and {Amram}, Philippe and {Athanassoula}, E. and {Bigiel}, Frank and {Bosma}, Albert and {Brinks}, Elias and {Chemin}, Laurent and {Combes}, Francoise and {Dettmar}, Ralf-J{\"u}rgen and {J{\'o}zsa}, Gyula and {Koribalski}, Baerbel and {Marasco}, Antonino and {Meurer}, Gerhardt and {Mogotsi}, Moses and {Mohapatra}, Abhisek and {Rajohnson}, Sambatriniaina H.~A. and {Schinnerer}, Eva and {Sorgho}, Amidou and {Spekkens}, Kristine and {Verdes-Montenegro}, Lourdes and {Veronese}, Simone and {Walter}, Fabian},
        title = "{Uncovering extraplanar gas in UGCA 250 with the Ultra-deep MHONGOOSE Survey}",
      journal = {\mnras},
         year = 2025,
        month = apr,
       volume = {538},
       number = {2},
        pages = {1272-1287},
          doi = {10.1093/mnras/staf387},
archivePrefix = {arXiv},
       eprint = {2503.03483},
 primaryClass = {astro-ph.GA},
       adsurl = {https://ui.adsabs.harvard.edu/abs/2025MNRAS.538.1272K}
}

@ARTICLE{Oosterloo2007,
       author = {{Oosterloo}, Tom and {Fraternali}, Filippo and {Sancisi}, Renzo},
        title = "{The Cold Gaseous Halo of NGC 891}",
      journal = {\aj},
         year = 2007,
        month = sep,
       volume = {134},
       number = {3},
        pages = {1019},
          doi = {10.1086/520332},
archivePrefix = {arXiv},
       eprint = {0705.4034},
 primaryClass = {astro-ph},
       adsurl = {https://ui.adsabs.harvard.edu/abs/2007AJ....134.1019O}
}

@ARTICLE{Vargas2017,
       author = {{Vargas}, Carlos J. and {Heald}, George and {Walterbos}, Ren{\'e} A.~M. and {Fraternali}, Filippo and {Patterson}, Maria T. and {Rand}, Richard J. and {J{\'o}zsa}, Gyula I.~G. and {Gentile}, Gianfranco and {Serra}, Paolo},
        title = "{HALOGAS Observations of NGC 4559: Anomalous and Extraplanar H I and its Relation to Star Formation}",
      journal = {\apj},
         year = 2017,
        month = apr,
       volume = {839},
       number = {2},
          eid = {118},
        pages = {118},
          doi = {10.3847/1538-4357/aa692c},
       adsurl = {https://ui.adsabs.harvard.edu/abs/2017ApJ...839..118V}
}

@ARTICLE{Gentile2013,
       author = {{Gentile}, G. and {J{\'o}zsa}, G.~I.~G. and {Serra}, P. and {Heald}, G.~H. and {de Blok}, W.~J.~G. and {Fraternali}, F. and {Patterson}, M.~T. and {Walterbos}, R.~A.~M. and {Oosterloo}, T.},
        title = "{HALOGAS: Extraplanar gas in NGC 3198}",
      journal = {\aap},
         year = 2013,
        month = jun,
       volume = {554},
          eid = {A125},
        pages = {A125},
          doi = {10.1051/0004-6361/201321116},
archivePrefix = {arXiv},
       eprint = {1304.4232},
 primaryClass = {astro-ph.CO},
       adsurl = {https://ui.adsabs.harvard.edu/abs/2013A&A...554A.125G}
}

@ARTICLE{Swaters1997,
       author = {{Swaters}, R.~A. and {Sancisi}, R. and {van der Hulst}, J.~M.},
        title = "{The H I Halo of NGC 891}",
      journal = {\apj},
         year = 1997,
        month = dec,
       volume = {491},
       number = {1},
        pages = {140-145},
          doi = {10.1086/304958},
archivePrefix = {arXiv},
       eprint = {astro-ph/9707150},
 primaryClass = {astro-ph},
       adsurl = {https://ui.adsabs.harvard.edu/abs/1997ApJ...491..140S}
}

@ARTICLE{Marasco2019,
       author = {{Marasco}, A. and {Fraternali}, F. and {Heald}, G. and {de Blok}, W.~J.~G. and {Oosterloo}, T. and {Kamphuis}, P. and {J{\'o}zsa}, G.~I.~G. and {Vargas}, C.~J. and {Winkel}, B. and {Walterbos}, R.~A.~M. and {Dettmar}, R.~J. and {Juẗte}, E.},
        title = "{HALOGAS: the properties of extraplanar HI in disc galaxies}",
      journal = {\aap},
         year = 2019,
        month = nov,
       volume = {631},
          eid = {A50},
        pages = {A50},
          doi = {10.1051/0004-6361/201936338},
archivePrefix = {arXiv},
       eprint = {1909.04048},
 primaryClass = {astro-ph.GA},
       adsurl = {https://ui.adsabs.harvard.edu/abs/2019A&A...631A..50M}
}

@ARTICLE{Lucero2015,
       author = {{Lucero}, D.~M. and {Carignan}, C. and {Elson}, E.~C. and {Randriamampandry}, T.~H. and {Jarrett}, T.~H. and {Oosterloo}, T.~A. and {Heald}, G.~H.},
        title = "{H I observations of the nearest starburst galaxy NGC 253 with the SKA precursor KAT-7}",
      journal = {\mnras},
         year = 2015,
        month = jul,
       volume = {450},
       number = {4},
        pages = {3935-3951},
          doi = {10.1093/mnras/stv856},
       adsurl = {https://ui.adsabs.harvard.edu/abs/2015MNRAS.450.3935L}
}

@ARTICLE{Ramesh2023,
       author = {{Ramesh}, Rahul and {Nelson}, Dylan and {Pillepich}, Annalisa},
        title = "{The circumgalactic medium of Milky Way-like galaxies in the TNG50 simulation - I: halo gas properties and the role of SMBH feedback}",
      journal = {\mnras},
         year = 2023,
        month = jan,
       volume = {518},
       number = {4},
        pages = {5754-5777},
          doi = {10.1093/mnras/stac3524},
archivePrefix = {arXiv},
       eprint = {2211.00020},
 primaryClass = {astro-ph.GA},
       adsurl = {https://ui.adsabs.harvard.edu/abs/2023MNRAS.518.5754R}
}

@ARTICLE{Voort2019,
       author = {{van de Voort}, Freeke and {Springel}, Volker and {Mandelker}, Nir and {van den Bosch}, Frank C. and {Pakmor}, R{\"u}diger},
        title = "{Cosmological simulations of the circumgalactic medium with 1 kpc resolution: enhanced H I column densities}",
      journal = {\mnras},
         year = 2019,
        month = jan,
       volume = {482},
       number = {1},
        pages = {L85-L89},
          doi = {10.1093/mnrasl/sly190},
archivePrefix = {arXiv},
       eprint = {1808.04369},
 primaryClass = {astro-ph.GA},
       adsurl = {https://ui.adsabs.harvard.edu/abs/2019MNRAS.482L..85V}
}

@ARTICLE{Keres2009,
       author = {{Kere{\v{s}}}, Du{\v{s}}an and {Katz}, Neal and {Fardal}, Mark and {Dav{\'e}}, Romeel and {Weinberg}, David H.},
        title = "{Galaxies in a simulated {\ensuremath{\Lambda}}CDM Universe - I. Cold mode and hot cores}",
      journal = {\mnras},
         year = 2009,
        month = may,
       volume = {395},
       number = {1},
        pages = {160-179},
          doi = {10.1111/j.1365-2966.2009.14541.x},
archivePrefix = {arXiv},
       eprint = {0809.1430},
 primaryClass = {astro-ph},
       adsurl = {https://ui.adsabs.harvard.edu/abs/2009MNRAS.395..160K}
}

@ARTICLE{Brooks2009,
       author = {{Brooks}, A.~M. and {Governato}, F. and {Quinn}, T. and {Brook}, C.~B. and {Wadsley}, J.},
        title = "{The Role of Cold Flows in the Assembly of Galaxy Disks}",
      journal = {\apj},
         year = 2009,
        month = mar,
       volume = {694},
       number = {1},
        pages = {396-410},
          doi = {10.1088/0004-637X/694/1/396},
archivePrefix = {arXiv},
       eprint = {0812.0007},
 primaryClass = {astro-ph},
       adsurl = {https://ui.adsabs.harvard.edu/abs/2009ApJ...694..396B}
}

@ARTICLE{Voort2011,
       author = {{van de Voort}, Freeke and {Schaye}, Joop and {Booth}, C.~M. and {Haas}, Marcel R. and {Dalla Vecchia}, Claudio},
        title = "{The rates and modes of gas accretion on to galaxies and their gaseous haloes}",
      journal = {\mnras},
         year = 2011,
        month = jul,
       volume = {414},
       number = {3},
        pages = {2458-2478},
          doi = {10.1111/j.1365-2966.2011.18565.x},
archivePrefix = {arXiv},
       eprint = {1011.2491},
 primaryClass = {astro-ph.CO},
       adsurl = {https://ui.adsabs.harvard.edu/abs/2011MNRAS.414.2458V}
}

@ARTICLE{DiTeodoro2014,
       author = {{Di Teodoro}, E.~M. and {Fraternali}, F.},
        title = "{Gas accretion from minor mergers in local spiral galaxies}",
      journal = {\aap},
         year = 2014,
        month = jul,
       volume = {567},
          eid = {A68},
        pages = {A68},
          doi = {10.1051/0004-6361/201423596},
archivePrefix = {arXiv},
       eprint = {1406.0856},
 primaryClass = {astro-ph.GA},
       adsurl = {https://ui.adsabs.harvard.edu/abs/2014A&A...567A..68D}
}

@ARTICLE{Sancisi2008,
       author = {{Sancisi}, Renzo and {Fraternali}, Filippo and {Oosterloo}, Tom and {van der Hulst}, Thijs},
        title = "{Cold gas accretion in galaxies}",
      journal = {\aapr},
         year = 2008,
        month = jun,
       volume = {15},
       number = {3},
        pages = {189-223},
          doi = {10.1007/s00159-008-0010-0},
archivePrefix = {arXiv},
       eprint = {0803.0109},
 primaryClass = {astro-ph},
       adsurl = {https://ui.adsabs.harvard.edu/abs/2008A&ARv..15..189S}
}

@ARTICLE{Tully2015,
       author = {{Tully}, R. Brent},
        title = "{Galaxy Groups: A 2MASS Catalog}",
      journal = {\aj},
         year = 2015,
        month = may,
       volume = {149},
       number = {5},
          eid = {171},
        pages = {171},
          doi = {10.1088/0004-6256/149/5/171},
archivePrefix = {arXiv},
       eprint = {1503.03134},
 primaryClass = {astro-ph.CO},
       adsurl = {https://ui.adsabs.harvard.edu/abs/2015AJ....149..171T}
}

@ARTICLE{Kourkchi2017,
       author = {{Kourkchi}, Ehsan and {Tully}, R. Brent},
        title = "{Galaxy Groups Within 3500 km s$^{-1}$}",
      journal = {\apj},
         year = 2017,
        month = jul,
       volume = {843},
       number = {1},
          eid = {16},
        pages = {16},
          doi = {10.3847/1538-4357/aa76db},
archivePrefix = {arXiv},
       eprint = {1705.08068},
 primaryClass = {astro-ph.GA},
       adsurl = {https://ui.adsabs.harvard.edu/abs/2017ApJ...843...16K}
}

@ARTICLE{Garcia1993,
       author = {{Garcia}, A.~M.},
        title = "{General study of group membership. II. Determination of nearby groups.}",
      journal = {\aaps},
         year = 1993,
        month = jul,
       volume = {100},
        pages = {47-90},
       adsurl = {https://ui.adsabs.harvard.edu/abs/1993A&AS..100...47G}
}

@ARTICLE{Healy2024,
       author = {{Healy}, J. and {de Blok}, W.~J.~G. and {Maccagni}, F.~M. and {Amram}, P. and {Chemin}, L. and {Combes}, F. and {Holwerda}, B.~W. and {Kamphuis}, P. and {Pisano}, D.~J. and {Schinnerer}, E. and {Spekkens}, K. and {Verdes-Montenegro}, L. and {Walter}, F. and {Adams}, E.~A.~K. and {Gibson}, B.~K. and {Kleiner}, D. and {Veronese}, S. and {Zabel}, N. and {English}, J. and {Carignan}, C.},
        title = "{Possible origins of anomalous H I gas around MHONGOOSE galaxy, NGC 5068}",
      journal = {\aap},
         year = 2024,
        month = jul,
       volume = {687},
          eid = {A254},
        pages = {A254},
          doi = {10.1051/0004-6361/202347475},
archivePrefix = {arXiv},
       eprint = {2402.13749},
 primaryClass = {astro-ph.GA},
       adsurl = {https://ui.adsabs.harvard.edu/abs/2024A&A...687A.254H}
}

@ARTICLE{Kamphuis2015,
       author = {{Kamphuis}, P. and {J{\'o}zsa}, G.~I.~G. and {Oh}, S. -. H. and {Spekkens}, K. and {Urbancic}, N. and {Serra}, P. and {Koribalski}, B.~S. and {Dettmar}, R. -J.},
        title = "{Automated kinematic modelling of warped galaxy discs in large H I surveys: 3D tilted-ring fitting of H I emission cubes}",
      journal = {\mnras},
         year = 2015,
        month = sep,
       volume = {452},
       number = {3},
        pages = {3139-3158},
          doi = {10.1093/mnras/stv1480},
archivePrefix = {arXiv},
       eprint = {1507.00413},
 primaryClass = {astro-ph.GA},
       adsurl = {https://ui.adsabs.harvard.edu/abs/2015MNRAS.452.3139K}
}

@ARTICLE{Kuijken2011,
       author = {{Kuijken}, K.},
        title = "{OmegaCAM: ESO's Newest Imager}",
      journal = {The Messenger},
         year = 2011,
        month = dec,
       volume = {146},
        pages = {8-11},
       adsurl = {https://ui.adsabs.harvard.edu/abs/2011Msngr.146....8K}
}

@ARTICLE{Kuijken2002,
       author = {{Kuijken}, K. and {Bender}, R. and {Cappellaro}, E. and {Muschielok}, B. and {Baruffolo}, A. and {Cascone}, E. and {Iwert}, O. and {Mitsch}, W. and {Nicklas}, H. and {Valentijn}, E.~A. and {Baade}, D. and {Begeman}, K.~G. and {Bortolussi}, A. and {Boxhoorn}, D. and {Christen}, F. and {Deul}, E.~R. and {Geimer}, C. and {Greggio}, L. and {Harke}, R. and {H{\"a}fner}, R. and {Hess}, G. and {Hess}, H. -J. and {Hopp}, U. and {Ilijevski}, I. and {Klink}, G. and {Kravcar}, H. and {Lizon}, J.~L. and {Magagna}, C.~E. and {M{\"u}ller}, Ph. and {Niemeczek}, R. and {de Pizzol}, L. and {Poschmann}, H. and {Reif}, K. and {Rengelink}, R. and {Reyes}, J. and {Silber}, A. and {Wellem}, W.},
        title = "{OmegaCAM: the 16k{\texttimes}16k CCD camera for the VLT survey telescope}",
      journal = {The Messenger},
         year = 2002,
        month = dec,
       volume = {110},
        pages = {15-18},
       adsurl = {https://ui.adsabs.harvard.edu/abs/2002Msngr.110...15K}
}

@ARTICLE{Arnaboldi1998,
       author = {{Arnaboldi}, M. and {Capaccioli}, M. and {Mancini}, D. and {Rafanelli}, P. and {Scaramella}, R. and {Sedmak}, G. and {Vettolani}, G.~P.},
        title = "{VST: VLT Survey Telescope.}",
      journal = {The Messenger},
         year = 1998,
        month = sep,
       volume = {93},
        pages = {30-35},
       adsurl = {https://ui.adsabs.harvard.edu/abs/1998Msngr..93...30A}
}

@ARTICLE{Pisano2011,
       author = {{Pisano}, D.~J. and {Barnes}, David G. and {Staveley-Smith}, Lister and {Gibson}, Brad K. and {Kilborn}, Virginia A. and {Freeman}, Ken C.},
        title = "{An H I Survey of Six Local Group Analogs. II. H I Properties of Group Galaxies}",
      journal = {\apjs},
         year = 2011,
        month = dec,
       volume = {197},
       number = {2},
          eid = {28},
        pages = {28},
          doi = {10.1088/0067-0049/197/2/28},
archivePrefix = {arXiv},
       eprint = {1110.3431},
 primaryClass = {astro-ph.CO},
       adsurl = {https://ui.adsabs.harvard.edu/abs/2011ApJS..197...28P}
}

@ARTICLE{Westmeier2021,
       author = {{Westmeier}, T. and {Kitaeff}, S. and {Pallot}, D. and {Serra}, P. and {van der Hulst}, J.~M. and {Jurek}, R.~J. and {Elagali}, A. and {For}, B. -Q. and {Kleiner}, D. and {Koribalski}, B.~S. and {Lee-Waddell}, K. and {Mould}, J.~R. and {Reynolds}, T.~N. and {Rhee}, J. and {Staveley-Smith}, L.},
        title = "{SOFIA 2 - an automated, parallel H I source finding pipeline for the WALLABY survey}",
      journal = {\mnras},
         year = 2021,
        month = sep,
       volume = {506},
       number = {3},
        pages = {3962-3976},
          doi = {10.1093/mnras/stab1881},
archivePrefix = {arXiv},
       eprint = {2106.15789},
 primaryClass = {astro-ph.IM},
       adsurl = {https://ui.adsabs.harvard.edu/abs/2021MNRAS.506.3962W}
}

@ARTICLE{Offringa2014,
       author = {{Offringa}, A.~R. and {McKinley}, B. and {Hurley-Walker}, N. and {Briggs}, F.~H. and {Wayth}, R.~B. and {Kaplan}, D.~L. and {Bell}, M.~E. and {Feng}, L. and {Neben}, A.~R. and {Hughes}, J.~D. and {Rhee}, J. and {Murphy}, T. and {Bhat}, N.~D.~R. and {Bernardi}, G. and {Bowman}, J.~D. and {Cappallo}, R.~J. and {Corey}, B.~E. and {Deshpande}, A.~A. and {Emrich}, D. and {Ewall-Wice}, A. and {Gaensler}, B.~M. and {Goeke}, R. and {Greenhill}, L.~J. and {Hazelton}, B.~J. and {Hindson}, L. and {Johnston-Hollitt}, M. and {Jacobs}, D.~C. and {Kasper}, J.~C. and {Kratzenberg}, E. and {Lenc}, E. and {Lonsdale}, C.~J. and {Lynch}, M.~J. and {McWhirter}, S.~R. and {Mitchell}, D.~A. and {Morales}, M.~F. and {Morgan}, E. and {Kudryavtseva}, N. and {Oberoi}, D. and {Ord}, S.~M. and {Pindor}, B. and {Procopio}, P. and {Prabu}, T. and {Riding}, J. and {Roshi}, D.~A. and {Shankar}, N. Udaya and {Srivani}, K.~S. and {Subrahmanyan}, R. and {Tingay}, S.~J. and {Waterson}, M. and {Webster}, R.~L. and {Whitney}, A.~R. and {Williams}, A. and {Williams}, C.~L.},
        title = "{WSCLEAN: an implementation of a fast, generic wide-field imager for radio astronomy}",
      journal = {\mnras},
         year = 2014,
        month = oct,
       volume = {444},
       number = {1},
        pages = {606-619},
          doi = {10.1093/mnras/stu1368},
archivePrefix = {arXiv},
       eprint = {1407.1943},
 primaryClass = {astro-ph.IM},
       adsurl = {https://ui.adsabs.harvard.edu/abs/2014MNRAS.444..606O}
}

@software{Jozsa2020,
       author = {{J{\'o}zsa}, Gyula I.~G. and {White}, Sarah V. and {Thorat}, Kshitij and {Smirnov}, Oleg M. and {Serra}, Paolo and {Ramatsoku}, Mpati and {Ramaila}, Athanaseus J.~T. and {Perkins}, Simon J. and {Moln{\'a}r}, D{\'a}niel Cs. and {Makhathini}, Sphesihle and {Maccagni}, Filippo M. and {Kleiner}, Dane and {Kamphuis}, Peter and {Hugo}, Benjamin V. and {de Blok}, W.~J.~G. and {Andati}, Lexy A.~L.},
        title = "{CARACal: Containerized Automated Radio Astronomy Calibration pipeline}",
 howpublished = {Astrophysics Source Code Library, record ascl:2006.014},
         year = 2020,
        month = jun,
          eid = {ascl:2006.014},
       adsurl = {https://ui.adsabs.harvard.edu/abs/2020ascl.soft06014J}
}

@ARTICLE{Karachentsev2017,
       author = {{Karachentsev}, I.~D. and {Makarova}, L.~N. and {Tully}, R.~B. and {Rizzi}, L. and {Karachentseva}, V.~E. and {Shaya}, E.~J.},
        title = "{DDO 161 and UGCA 319: an isolated pair of nearby dwarf galaxies}",
      journal = {\mnras},
         year = 2017,
        month = jul,
       volume = {469},
       number = {1},
        pages = {L113-L117},
          doi = {10.1093/mnrasl/slx061},
archivePrefix = {arXiv},
       eprint = {1704.07648},
 primaryClass = {astro-ph.GA},
       adsurl = {https://ui.adsabs.harvard.edu/abs/2017MNRAS.469L.113K}
}

@ARTICLE{Blok2024,
       author = {{de Blok}, W.~J.~G. and {Healy}, J. and {Maccagni}, F.~M. and {Pisano}, D.~J. and {Bosma}, A. and {English}, J. and {Jarrett}, T. and {Marasco}, A. and {Meurer}, G.~R. and {Veronese}, S. and {Bigiel}, F. and {Chemin}, L. and {Fraternali}, F. and {Holwerda}, B.~W. and {Kamphuis}, P. and {Kl{\"o}ckner}, H.~R. and {Kleiner}, D. and {Leroy}, A.~K. and {Mogotsi}, M. and {Oman}, K.~A. and {Schinnerer}, E. and {Verdes-Montenegro}, L. and {Westmeier}, T. and {Wong}, O.~I. and {Zabel}, N. and {Amram}, P. and {Carignan}, C. and {Combes}, F. and {Brinks}, E. and {Dettmar}, R.~J. and {Gibson}, B.~K. and {Jozsa}, G.~I.~G. and {Koribalski}, B.~S. and {McGaugh}, S.~S. and {Oosterloo}, T.~A. and {Spekkens}, K. and {Schr{\"o}der}, A.~C. and {Adams}, E.~A.~K. and {Athanassoula}, E. and {Bershady}, M.~A. and {Beswick}, R.~J. and {Blyth}, S. and {Elson}, E.~C. and {Frank}, B.~S. and {Heald}, G. and {Henning}, P.~A. and {Kurapati}, S. and {Loubser}, S.~I. and {Lucero}, D. and {Meyer}, M. and {Namumba}, B. and {Oh}, S. -H. and {Sardone}, A. and {Sheth}, K. and {Smith}, M.~W.~L. and {Sorgho}, A. and {Walter}, F. and {Williams}, T. and {Woudt}, P.~A. and {Zijlstra}, A.},
        title = "{MHONGOOSE: A MeerKAT nearby galaxy H I survey}",
      journal = {\aap},
         year = 2024,
        month = aug,
       volume = {688},
          eid = {A109},
        pages = {A109},
          doi = {10.1051/0004-6361/202348297},
archivePrefix = {arXiv},
       eprint = {2404.01774},
 primaryClass = {astro-ph.GA},
       adsurl = {https://ui.adsabs.harvard.edu/abs/2024A&A...688A.109D}
}

@ARTICLE{Jozsa2007,
       author = {{J{\'o}zsa}, G.~I.~G. and {Kenn}, F. and {Klein}, U. and {Oosterloo}, T.~A.},
        title = "{Kinematic modelling of disk galaxies. I. A new method to fit tilted rings to data cubes}",
      journal = {\aap},
         year = 2007,
        month = jun,
       volume = {468},
       number = {2},
        pages = {731-774},
          doi = {10.1051/0004-6361:20066164},
archivePrefix = {arXiv},
       eprint = {astro-ph/0703207},
 primaryClass = {astro-ph},
       adsurl = {https://ui.adsabs.harvard.edu/abs/2007A&A...468..731J}
}

@INPROCEEDINGS{Taylor2005,
       author = {{Taylor}, M.~B.},
        title = "{TOPCAT \& STIL: Starlink Table/VOTable Processing Software}",
    booktitle = {Astronomical Data Analysis Software and Systems XIV},
         year = 2005,
       editor = {{Shopbell}, P. and {Britton}, M. and {Ebert}, R.},
       series = {Astronomical Society of the Pacific Conference Series},
       volume = {347},
        month = dec,
        pages = {29},
       adsurl = {https://ui.adsabs.harvard.edu/abs/2005ASPC..347...29T}
}

@INPROCEEDINGS{Joye2003,
       author = {{Joye}, W.~A. and {Mandel}, E.},
        title = "{New Features of SAOImage DS9}",
    booktitle = {Astronomical Data Analysis Software and Systems XII},
         year = 2003,
       editor = {{Payne}, H.~E. and {Jedrzejewski}, R.~I. and {Hook}, R.~N.},
       series = {Astronomical Society of the Pacific Conference Series},
       volume = {295},
        month = jan,
        pages = {489},
       adsurl = {https://ui.adsabs.harvard.edu/abs/2003ASPC..295..489J}
}

@software{Reproject2020,
       author = {{Robitaille}, Thomas and {Deil}, Christoph and {Ginsburg}, Adam},
        title = "{reproject: Python-based astronomical image reprojection}",
 howpublished = {Astrophysics Source Code Library, record ascl:2011.023},
         year = 2020,
        month = nov,
          eid = {ascl:2011.023},
       adsurl = {https://ui.adsabs.harvard.edu/abs/2020ascl.soft11023R}
}

@ARTICLE{Astropy2022,
       author = {{Astropy Collaboration} and {Price-Whelan}, Adrian M. and {Lim}, Pey Lian and {Earl}, Nicholas and {Starkman}, Nathaniel and {Bradley}, Larry and {Shupe}, David L. and {Patil}, Aarya A. and {Corrales}, Lia and {Brasseur}, C.~E. and {N{\"o}the}, Maximilian and {Donath}, Axel and {Tollerud}, Erik and {Morris}, Brett M. and {Ginsburg}, Adam and {Vaher}, Eero and {Weaver}, Benjamin A. and {Tocknell}, James and {Jamieson}, William and {van Kerkwijk}, Marten H. and {Robitaille}, Thomas P. and {Merry}, Bruce and {Bachetti}, Matteo and {G{\"u}nther}, H. Moritz and {Aldcroft}, Thomas L. and {Alvarado-Montes}, Jaime A. and {Archibald}, Anne M. and {B{\'o}di}, Attila and {Bapat}, Shreyas and {Barentsen}, Geert and {Baz{\'a}n}, Juanjo and {Biswas}, Manish and {Boquien}, M{\'e}d{\'e}ric and {Burke}, D.~J. and {Cara}, Daria and {Cara}, Mihai and {Conroy}, Kyle E. and {Conseil}, Simon and {Craig}, Matthew W. and {Cross}, Robert M. and {Cruz}, Kelle L. and {D'Eugenio}, Francesco and {Dencheva}, Nadia and {Devillepoix}, Hadrien A.~R. and {Dietrich}, J{\"o}rg P. and {Eigenbrot}, Arthur Davis and {Erben}, Thomas and {Ferreira}, Leonardo and {Foreman-Mackey}, Daniel and {Fox}, Ryan and {Freij}, Nabil and {Garg}, Suyog and {Geda}, Robel and {Glattly}, Lauren and {Gondhalekar}, Yash and {Gordon}, Karl D. and {Grant}, David and {Greenfield}, Perry and {Groener}, Austen M. and {Guest}, Steve and {Gurovich}, Sebastian and {Handberg}, Rasmus and {Hart}, Akeem and {Hatfield-Dodds}, Zac and {Homeier}, Derek and {Hosseinzadeh}, Griffin and {Jenness}, Tim and {Jones}, Craig K. and {Joseph}, Prajwel and {Kalmbach}, J. Bryce and {Karamehmetoglu}, Emir and {Ka{\l}uszy{\'n}ski}, Miko{\l}aj and {Kelley}, Michael S.~P. and {Kern}, Nicholas and {Kerzendorf}, Wolfgang E. and {Koch}, Eric W. and {Kulumani}, Shankar and {Lee}, Antony and {Ly}, Chun and {Ma}, Zhiyuan and {MacBride}, Conor and {Maljaars}, Jakob M. and {Muna}, Demitri and {Murphy}, N.~A. and {Norman}, Henrik and {O'Steen}, Richard and {Oman}, Kyle A. and {Pacifici}, Camilla and {Pascual}, Sergio and {Pascual-Granado}, J. and {Patil}, Rohit R. and {Perren}, Gabriel I. and {Pickering}, Timothy E. and {Rastogi}, Tanuj and {Roulston}, Benjamin R. and {Ryan}, Daniel F. and {Rykoff}, Eli S. and {Sabater}, Jose and {Sakurikar}, Parikshit and {Salgado}, Jes{\'u}s and {Sanghi}, Aniket and {Saunders}, Nicholas and {Savchenko}, Volodymyr and {Schwardt}, Ludwig and {Seifert-Eckert}, Michael and {Shih}, Albert Y. and {Jain}, Anany Shrey and {Shukla}, Gyanendra and {Sick}, Jonathan and {Simpson}, Chris and {Singanamalla}, Sudheesh and {Singer}, Leo P. and {Singhal}, Jaladh and {Sinha}, Manodeep and {Sip{\H{o}}cz}, Brigitta M. and {Spitler}, Lee R. and {Stansby}, David and {Streicher}, Ole and {{\v{S}}umak}, Jani and {Swinbank}, John D. and {Taranu}, Dan S. and {Tewary}, Nikita and {Tremblay}, Grant R. and {de Val-Borro}, Miguel and {Van Kooten}, Samuel J. and {Vasovi{\'c}}, Zlatan and {Verma}, Shresth and {de Miranda Cardoso}, Jos{\'e} Vin{\'\i}cius and {Williams}, Peter K.~G. and {Wilson}, Tom J. and {Winkel}, Benjamin and {Wood-Vasey}, W.~M. and {Xue}, Rui and {Yoachim}, Peter and {Zhang}, Chen and {Zonca}, Andrea and {Astropy Project Contributors}},
        title = "{The Astropy Project: Sustaining and Growing a Community-oriented Open-source Project and the Latest Major Release (v5.0) of the Core Package}",
      journal = {\apj},
         year = 2022,
        month = aug,
       volume = {935},
       number = {2},
          eid = {167},
        pages = {167},
          doi = {10.3847/1538-4357/ac7c74},
archivePrefix = {arXiv},
       eprint = {2206.14220},
 primaryClass = {astro-ph.IM},
       adsurl = {https://ui.adsabs.harvard.edu/abs/2022ApJ...935..167A}
}

@ARTICLE{Serra2024,
       author = {{Serra}, P. and {Oosterloo}, T.~A. and {Kamphuis}, P. and {Jozsa}, G.~I.~G. and {de Blok}, W.~J.~G. and {Bryan}, G.~L. and {van Gorkom}, J.~H. and {Iodice}, E. and {Kleiner}, D. and {Loni}, A. and {Loubser}, S.~I. and {Maccagni}, F.~M. and {Molnar}, D. and {Peletier}, R. and {Pisano}, D.~J. and {Ramatsoku}, M. and {Smith}, M.~W.~L. and {Verheijen}, M.~A.~W. and {Zabel}, N.},
        title = "{The MeerKAT Fornax Survey. III. Ram-pressure stripping of the tidally interacting galaxy NGC 1427A in the Fornax cluster}",
      journal = {arXiv e-prints},
         year = 2024,
        month = jul,
          eid = {arXiv:2407.09082},
        pages = {arXiv:2407.09082},
          doi = {10.48550/arXiv.2407.09082},
archivePrefix = {arXiv},
       eprint = {2407.09082},
 primaryClass = {astro-ph.GA},
       adsurl = {https://ui.adsabs.harvard.edu/abs/2024arXiv240709082S}
}

@ARTICLE{Jarrett2023,
       author = {{Jarrett}, T.~H. and {Cluver}, M.~E. and {Taylor}, Edward N. and {Bellstedt}, Sabine and {Robotham}, A.~S.~G. and {Yao}, H.~F.~M.},
        title = "{A New Wide-field Infrared Survey Explorer Calibration of Stellar Mass}",
      journal = {\apj},
         year = 2023,
        month = apr,
       volume = {946},
       number = {2},
          eid = {95},
        pages = {95},
          doi = {10.3847/1538-4357/acb68f},
       adsurl = {https://ui.adsabs.harvard.edu/abs/2023ApJ...946...95J}
}

@ARTICLE{Catinella2018,
       author = {{Catinella}, Barbara and {Saintonge}, Am{\'e}lie and {Janowiecki}, Steven and {Cortese}, Luca and {Dav{\'e}}, Romeel and {Lemonias}, Jenna J. and {Cooper}, Andrew P. and {Schiminovich}, David and {Hummels}, Cameron B. and {Fabello}, Silvia and {Ger{\'e}b}, Katinka and {Kilborn}, Virginia and {Wang}, Jing},
        title = "{xGASS: total cold gas scaling relations and molecular-to-atomic gas ratios of galaxies in the local Universe}",
      journal = {\mnras},
         year = 2018,
        month = may,
       volume = {476},
       number = {1},
        pages = {875-895},
          doi = {10.1093/mnras/sty089},
archivePrefix = {arXiv},
       eprint = {1802.02373},
 primaryClass = {astro-ph.GA},
       adsurl = {https://ui.adsabs.harvard.edu/abs/2018MNRAS.476..875C}
}

@ARTICLE{Walter2008,
       author = {{Walter}, Fabian and {Brinks}, Elias and {de Blok}, W.~J.~G. and {Bigiel}, Frank and {Kennicutt}, Robert C., Jr. and {Thornley}, Michele D. and {Leroy}, Adam},
        title = "{THINGS: The H I Nearby Galaxy Survey}",
      journal = {\aj},
         year = 2008,
        month = dec,
       volume = {136},
       number = {6},
        pages = {2563-2647},
          doi = {10.1088/0004-6256/136/6/2563},
archivePrefix = {arXiv},
       eprint = {0810.2125},
 primaryClass = {astro-ph},
       adsurl = {https://ui.adsabs.harvard.edu/abs/2008AJ....136.2563W}
}

@article{Davidson-Pilon2019,
  doi = {10.21105/joss.01317},
  url = {https://doi.org/10.21105/joss.01317},
  year = {2019},
  publisher = {The Open Journal},
  volume = {4},
  number = {40},
  pages = {1317},
  author = {Cameron Davidson-Pilon},
  title = {lifelines: survival analysis in Python},
  journal = {Journal of Open Source Software}
}

@ARTICLE{2020SciPy,
  author  = {Virtanen, Pauli and Gommers, Ralf and Oliphant, Travis E. and
            Haberland, Matt and Reddy, Tyler and Cournapeau, David and
            Burovski, Evgeni and Peterson, Pearu and Weckesser, Warren and
            Bright, Jonathan and {van der Walt}, St{\'e}fan J. and
            Brett, Matthew and Wilson, Joshua and Millman, K. Jarrod and
            Mayorov, Nikolay and Nelson, Andrew R. J. and Jones, Eric and
            Kern, Robert and Larson, Eric and Carey, C J and
            Polat, {\.I}lhan and Feng, Yu and Moore, Eric W. and
            {VanderPlas}, Jake and Laxalde, Denis and Perktold, Josef and
            Cimrman, Robert and Henriksen, Ian and Quintero, E. A. and
            Harris, Charles R. and Archibald, Anne M. and
            Ribeiro, Ant{\^o}nio H. and Pedregosa, Fabian and
            {van Mulbregt}, Paul and {SciPy 1.0 Contributors}},
  title   = {{{SciPy} 1.0: Fundamental Algorithms for Scientific
            Computing in Python}},
  journal = {Nature Methods},
  year    = {2020},
  volume  = {17},
  pages   = {261--272},
  adsurl  = {https://rdcu.be/b08Wh},
  doi     = {10.1038/s41592-019-0686-2},
}

@Article{Harris2020,
 title         = {Array programming with {NumPy}},
 author        = {Charles R. Harris and K. Jarrod Millman and St{\'{e}}fan J.
                 van der Walt and Ralf Gommers and Pauli Virtanen and David
                 Cournapeau and Eric Wieser and Julian Taylor and Sebastian
                 Berg and Nathaniel J. Smith and Robert Kern and Matti Picus
                 and Stephan Hoyer and Marten H. van Kerkwijk and Matthew
                 Brett and Allan Haldane and Jaime Fern{\'{a}}ndez del
                 R{\'{i}}o and Mark Wiebe and Pearu Peterson and Pierre
                 G{\'{e}}rard-Marchant and Kevin Sheppard and Tyler Reddy and
                 Warren Weckesser and Hameer Abbasi and Christoph Gohlke and
                 Travis E. Oliphant},
 year          = {2020},
 month         = sep,
 journal       = {Nature},
 volume        = {585},
 number        = {7825},
 pages         = {357--362},
 doi           = {10.1038/s41586-020-2649-2},
 publisher     = {Springer Science and Business Media {LLC}},
 url           = {https://doi.org/10.1038/s41586-020-2649-2}
}

@Article{Hunter2007,
  Author    = {Hunter, J. D.},
  Title     = {Matplotlib: A 2D graphics environment},
  Journal   = {Computing in Science \& Engineering},
  Volume    = {9},
  Number    = {3},
  Pages     = {90--95},
  publisher = {IEEE COMPUTER SOC},
  doi       = {10.1109/MCSE.2007.55},
  year      = 2007
}

@ARTICLE{Leroy2019,
       author = {{Leroy}, Adam K. and {Sandstrom}, Karin M. and {Lang}, Dustin and {Lewis}, Alexia and {Salim}, Samir and {Behrens}, Erica A. and {Chastenet}, J{\'e}r{\'e}my and {Chiang}, I. -Da and {Gallagher}, Molly J. and {Kessler}, Sarah and {Utomo}, Dyas},
        title = "{A z = 0 Multiwavelength Galaxy Synthesis. I. A WISE and GALEX Atlas of Local Galaxies}",
      journal = {\apjs},
         year = 2019,
        month = oct,
       volume = {244},
       number = {2},
          eid = {24},
        pages = {24},
          doi = {10.3847/1538-4365/ab3925},
archivePrefix = {arXiv},
       eprint = {1910.13470},
 primaryClass = {astro-ph.GA},
       adsurl = {https://ui.adsabs.harvard.edu/abs/2019ApJS..244...24L}
}

@MISC{Robitaille2012,
       author = {{Robitaille}, Thomas and {Bressert}, Eli},
        title = "{APLpy: Astronomical Plotting Library in Python}",
         year = 2012,
        month = aug,
          eid = {ascl:1208.017},
        pages = {ascl:1208.017},
archivePrefix = {ascl},
       eprint = {1208.017},
       adsurl = {https://ui.adsabs.harvard.edu/abs/2012ascl.soft08017R}
}

@article{Leroy2013,
	Archiveprefix = {arXiv},
	Arxivid = {arXiv:1301.2328v1},
	Author = {Leroy, Adam K. and Walter, Fabian and Sandstrom, Karin and Schruba, Andreas and Munoz-Mateos, Juan-Carlos and Bigiel, Frank and Bolatto, Alberto and Brinks, Elias and de Blok, W. J. G. and Meidt, Sharon and Rix, Hans-Walter and Rosolowsky, Erik and Schinnerer, Eva and Schuster, Karl-Friedrich and Usero, Antonio},
	Doi = {10.1088/0004-6256/146/2/19},
	Eprint = {arXiv:1301.2328v1},
	Issn = {0004-6256},
	Journal = {\aj},
	Number = {2},
	Pages = {19},
	Title = {{Molecular Gas and Star Formation in Nearby Disk Galaxies}},
	Url = {http://adsabs.harvard.edu/abs/2013AJ....146...19L{\%}5Cnhttp://stacks.iop.org/1538-3881/146/i=2/a=19},
	Volume = {146},
	Year = {2013}}

@ARTICLE{Astropy2018,
   author = {{Astropy Collaboration} and {Price-Whelan}, A.~M. and {Sip{\H o}cz}, B.~M. and 
	{G{\"u}nther}, H.~M. and {Lim}, P.~L. and {Crawford}, S.~M. and 
	{Conseil}, S. and {Shupe}, D.~L. and {Craig}, M.~W. and {Dencheva}, N. and 
	{Ginsburg}, A. and {VanderPlas}, J.~T. and {Bradley}, L.~D. and 
	{P{\'e}rez-Su{\'a}rez}, D. and {de Val-Borro}, M. and {Aldcroft}, T.~L. and 
	{Cruz}, K.~L. and {Robitaille}, T.~P. and {Tollerud}, E.~J. and 
	{Ardelean}, C. and {Babej}, T. and {Bach}, Y.~P. and {Bachetti}, M. and 
	{Bakanov}, A.~V. and {Bamford}, S.~P. and {Barentsen}, G. and 
	{Barmby}, P. and {Baumbach}, A. and {Berry}, K.~L. and {Biscani}, F. and 
	{Boquien}, M. and {Bostroem}, K.~A. and {Bouma}, L.~G. and {Brammer}, G.~B. and 
	{Bray}, E.~M. and {Breytenbach}, H. and {Buddelmeijer}, H. and 
	{Burke}, D.~J. and {Calderone}, G. and {Cano Rodr{\'{\i}}guez}, J.~L. and 
	{Cara}, M. and {Cardoso}, J.~V.~M. and {Cheedella}, S. and {Copin}, Y. and 
	{Corrales}, L. and {Crichton}, D. and {D'Avella}, D. and {Deil}, C. and 
	{Depagne}, {\'E}. and {Dietrich}, J.~P. and {Donath}, A. and 
	{Droettboom}, M. and {Earl}, N. and {Erben}, T. and {Fabbro}, S. and 
	{Ferreira}, L.~A. and {Finethy}, T. and {Fox}, R.~T. and {Garrison}, L.~H. and 
	{Gibbons}, S.~L.~J. and {Goldstein}, D.~A. and {Gommers}, R. and 
	{Greco}, J.~P. and {Greenfield}, P. and {Groener}, A.~M. and 
	{Grollier}, F. and {Hagen}, A. and {Hirst}, P. and {Homeier}, D. and 
	{Horton}, A.~J. and {Hosseinzadeh}, G. and {Hu}, L. and {Hunkeler}, J.~S. and 
	{Ivezi{\'c}}, {\v Z}. and {Jain}, A. and {Jenness}, T. and {Kanarek}, G. and 
	{Kendrew}, S. and {Kern}, N.~S. and {Kerzendorf}, W.~E. and 
	{Khvalko}, A. and {King}, J. and {Kirkby}, D. and {Kulkarni}, A.~M. and 
	{Kumar}, A. and {Lee}, A. and {Lenz}, D. and {Littlefair}, S.~P. and 
	{Ma}, Z. and {Macleod}, D.~M. and {Mastropietro}, M. and {McCully}, C. and 
	{Montagnac}, S. and {Morris}, B.~M. and {Mueller}, M. and {Mumford}, S.~J. and 
	{Muna}, D. and {Murphy}, N.~A. and {Nelson}, S. and {Nguyen}, G.~H. and 
	{Ninan}, J.~P. and {N{\"o}the}, M. and {Ogaz}, S. and {Oh}, S. and 
	{Parejko}, J.~K. and {Parley}, N. and {Pascual}, S. and {Patil}, R. and 
	{Patil}, A.~A. and {Plunkett}, A.~L. and {Prochaska}, J.~X. and 
	{Rastogi}, T. and {Reddy Janga}, V. and {Sabater}, J. and {Sakurikar}, P. and 
	{Seifert}, M. and {Sherbert}, L.~E. and {Sherwood-Taylor}, H. and 
	{Shih}, A.~Y. and {Sick}, J. and {Silbiger}, M.~T. and {Singanamalla}, S. and 
	{Singer}, L.~P. and {Sladen}, P.~H. and {Sooley}, K.~A. and 
	{Sornarajah}, S. and {Streicher}, O. and {Teuben}, P. and {Thomas}, S.~W. and 
	{Tremblay}, G.~R. and {Turner}, J.~E.~H. and {Terr{\'o}n}, V. and 
	{van Kerkwijk}, M.~H. and {de la Vega}, A. and {Watkins}, L.~L. and 
	{Weaver}, B.~A. and {Whitmore}, J.~B. and {Woillez}, J. and 
	{Zabalza}, V. and {Astropy Contributors}},
    title = "{The Astropy Project: Building an Open-science Project and Status of the v2.0 Core Package}",
  journal = {\aj},
archivePrefix = "arXiv",
   eprint = {1801.02634},
 primaryClass = "astro-ph.IM",
     year = 2018,
    month = sep,
   volume = 156,
      eid = {123},
    pages = {123},
      doi = {10.3847/1538-3881/aabc4f},
   adsurl = {https://ui.adsabs.harvard.edu/abs/2018AJ....156..123A}
}

@ARTICLE{Astropy2013,
   author = {{Astropy Collaboration} and {Robitaille}, T.~P. and {Tollerud}, E.~J. and 
	{Greenfield}, P. and {Droettboom}, M. and {Bray}, E. and {Aldcroft}, T. and 
	{Davis}, M. and {Ginsburg}, A. and {Price-Whelan}, A.~M. and 
	{Kerzendorf}, W.~E. and {Conley}, A. and {Crighton}, N. and 
	{Barbary}, K. and {Muna}, D. and {Ferguson}, H. and {Grollier}, F. and 
	{Parikh}, M.~M. and {Nair}, P.~H. and {Unther}, H.~M. and {Deil}, C. and 
	{Woillez}, J. and {Conseil}, S. and {Kramer}, R. and {Turner}, J.~E.~H. and 
	{Singer}, L. and {Fox}, R. and {Weaver}, B.~A. and {Zabalza}, V. and 
	{Edwards}, Z.~I. and {Azalee Bostroem}, K. and {Burke}, D.~J. and 
	{Casey}, A.~R. and {Crawford}, S.~M. and {Dencheva}, N. and 
	{Ely}, J. and {Jenness}, T. and {Labrie}, K. and {Lim}, P.~L. and 
	{Pierfederici}, F. and {Pontzen}, A. and {Ptak}, A. and {Refsdal}, B. and 
	{Servillat}, M. and {Streicher}, O.},
    title = "{Astropy: A community Python package for astronomy}",
  journal = {\aap},
archivePrefix = "arXiv",
   eprint = {1307.6212},
 primaryClass = "astro-ph.IM",
     year = 2013,
    month = oct,
   volume = 558,
      eid = {A33},
    pages = {A33},
      doi = {10.1051/0004-6361/201322068},
   adsurl = {https://ui.adsabs.harvard.edu/abs/2013A%26A...558A..33A}
}

@ARTICLE{Kennicutt1998,
   author = {{Kennicutt}, Jr., R.~C.},
    title = "{The Global Schmidt Law in Star-forming Galaxies}",
  journal = {\apj},
   eprint = {astro-ph/9712213},
     year = 1998,
    month = may,
   volume = 498,
    pages = {541-552},
      doi = {10.1086/305588},
   adsurl = {https://ui.adsabs.harvard.edu/abs/1998ApJ...498..541K}
}

@ARTICLE{Bigiel2008,
   author = {{Bigiel}, F. and {Leroy}, A. and {Walter}, F. and {Brinks}, E. and 
	{de Blok}, W.~J.~G. and {Madore}, B. and {Thornley}, M.~D.},
    title = "{The Star Formation Law in Nearby Galaxies on Sub-Kpc Scales}",
  journal = {\aj},
archivePrefix = "arXiv",
   eprint = {0810.2541},
     year = 2008,
    month = dec,
   volume = 136,
    pages = {2846-2871},
      doi = {10.1088/0004-6256/136/6/2846},
   adsurl = {https://ui.adsabs.harvard.edu/abs/2008AJ....136.2846B}
}

@ARTICLE{Hunter2012,
   author = {{Hunter}, D.~A. and {Ficut-Vicas}, D. and {Ashley}, T. and {Brinks}, E. and 
	{Cigan}, P. and {Elmegreen}, B.~G. and {Heesen}, V. and {Herrmann}, K.~A. and 
	{Johnson}, M. and {Oh}, S.-H. and {Rupen}, M.~P. and {Schruba}, A. and 
	{Simpson}, C.~E. and {Walter}, F. and {Westpfahl}, D.~J. and 
	{Young}, L.~M. and {Zhang}, H.-X.},
    title = "{Little Things}",
  journal = {\aj},
archivePrefix = "arXiv",
   eprint = {1208.5834},
     year = 2012,
    month = nov,
   volume = 144,
      eid = {134},
    pages = {134},
      doi = {10.1088/0004-6256/144/5/134},
   adsurl = {http://adsabs.harvard.edu/abs/2012AJ....144..134H}
}

@ARTICLE{Wright2010,
   author = {{Wright}, E.~L. and {Eisenhardt}, P.~R.~M. and {Mainzer}, A.~K. and 
	{Ressler}, M.~E. and {Cutri}, R.~M. and {Jarrett}, T. and {Kirkpatrick}, J.~D. and 
	{Padgett}, D. and {McMillan}, R.~S. and {Skrutskie}, M. and 
	{Stanford}, S.~A. and {Cohen}, M. and {Walker}, R.~G. and {Mather}, J.~C. and 
	{Leisawitz}, D. and {Gautier}, III, T.~N. and {McLean}, I. and 
	{Benford}, D. and {Lonsdale}, C.~J. and {Blain}, A. and {Mendez}, B. and 
	{Irace}, W.~R. and {Duval}, V. and {Liu}, F. and {Royer}, D. and 
	{Heinrichsen}, I. and {Howard}, J. and {Shannon}, M. and {Kendall}, M. and 
	{Walsh}, A.~L. and {Larsen}, M. and {Cardon}, J.~G. and {Schick}, S. and 
	{Schwalm}, M. and {Abid}, M. and {Fabinsky}, B. and {Naes}, L. and 
	{Tsai}, C.-W.},
    title = "{The Wide-field Infrared Survey Explorer (WISE): Mission Description and Initial On-orbit Performance}",
  journal = {\aj},
archivePrefix = "arXiv",
   eprint = {1008.0031},
 primaryClass = "astro-ph.IM",
     year = 2010,
    month = dec,
   volume = 140,
    pages = {1868-1881},
      doi = {10.1088/0004-6256/140/6/1868},
   adsurl = {http://adsabs.harvard.edu/abs/2010AJ....140.1868W}
}

@ARTICLE{Elbaz2007,
   author = {{Elbaz}, D. and {Daddi}, E. and {Le Borgne}, D. and {Dickinson}, M. and 
	{Alexander}, D.~M. and {Chary}, R.-R. and {Starck}, J.-L. and 
	{Brandt}, W.~N. and {Kitzbichler}, M. and {MacDonald}, E. and 
	{Nonino}, M. and {Popesso}, P. and {Stern}, D. and {Vanzella}, E.
	},
    title = "{The reversal of the star formation-density relation in the distant universe}",
  journal = {\aap},
   eprint = {astro-ph/0703653},
     year = 2007,
    month = jun,
   volume = 468,
    pages = {33-48},
      doi = {10.1051/0004-6361:20077525},
   adsurl = {http://adsabs.harvard.edu/abs/2007A%26A...468...33E}
}

@ARTICLE{Barnes2001,
   author = {{Barnes}, D.~G. and {Staveley-Smith}, L. and {de Blok}, W.~J.~G. and 
	{Oosterloo}, T. and {Stewart}, I.~M. and {Wright}, A.~E. and 
	{Banks}, G.~D. and {Bhathal}, R. and {Boyce}, P.~J. and {Calabretta}, M.~R. and 
	{Disney}, M.~J. and {Drinkwater}, M.~J. and {Ekers}, R.~D. and 
	{Freeman}, K.~C. and {Gibson}, B.~K. and {Green}, A.~J. and 
	{Haynes}, R.~F. and {te Lintel Hekkert}, P. and {Henning}, P.~A. and 
	{Jerjen}, H. and {Juraszek}, S. and {Kesteven}, M.~J. and {Kilborn}, V.~A. and 
	{Knezek}, P.~M. and {Koribalski}, B. and {Kraan-Korteweg}, R.~C. and 
	{Malin}, D.~F. and {Marquarding}, M. and {Minchin}, R.~F. and 
	{Mould}, J.~R. and {Price}, R.~M. and {Putman}, M.~E. and {Ryder}, S.~D. and 
	{Sadler}, E.~M. and {Schr{\"o}der}, A. and {Stootman}, F. and 
	{Webster}, R.~L. and {Wilson}, W.~E. and {Ye}, T.},
    title = "{The HI Parkes All Sky Survey: southern observations, calibration and robust imaging}",
  journal = {\mnras},
     year = 2001,
    month = apr,
   volume = 322,
    pages = {486-498},
      doi = {10.1046/j.1365-8711.2001.04102.x},
   adsurl = {http://adsabs.harvard.edu/abs/2001MNRAS.322..486B}
}

@ARTICLE{Saintonge2017,
   author = {{Saintonge}, A. and {Catinella}, B. and {Tacconi}, L.~J. and 
	{Kauffmann}, G. and {Genzel}, R. and {Cortese}, L. and {Dav{\'e}}, R. and 
	{Fletcher}, T.~J. and {Graci{\'a}-Carpio}, J. and {Kramer}, C. and 
	{Heckman}, T.~M. and {Janowiecki}, S. and {Lutz}, K. and {Rosario}, D. and 
	{Schiminovich}, D. and {Schuster}, K. and {Wang}, J. and {Wuyts}, S. and 
	{Borthakur}, S. and {Lamperti}, I. and {Roberts-Borsani}, G.~W.
	},
    title = "{xCOLD GASS: The Complete IRAM 30 m Legacy Survey of Molecular Gas for Galaxy Evolution Studies}",
  journal = {\apjs},
archivePrefix = "arXiv",
   eprint = {1710.02157},
     year = 2017,
    month = dec,
   volume = 233,
      eid = {22},
    pages = {22},
      doi = {10.3847/1538-4365/aa97e0},
   adsurl = {http://adsabs.harvard.edu/abs/2017ApJS..233...22S}
}




\newpage

\appendix

\section{\vthree{Moment maps of UGCA~320 at all six MHONGOOSE resolutions}}
\label{app:moment_maps}

\vthree{We present the moment maps for UGCA~320 at all six MHONGOOSE resolutions in Figure \ref{fig:moment_maps_multi_res}. Details of these cubes can be found in \ref{sec:observations}, and the resolution and column density sensitivity are quoted under each panel. }

\begin{figure*}
    \begin{subfigure}{0.315\textwidth}
    \centering
	\includegraphics[width=\textwidth]{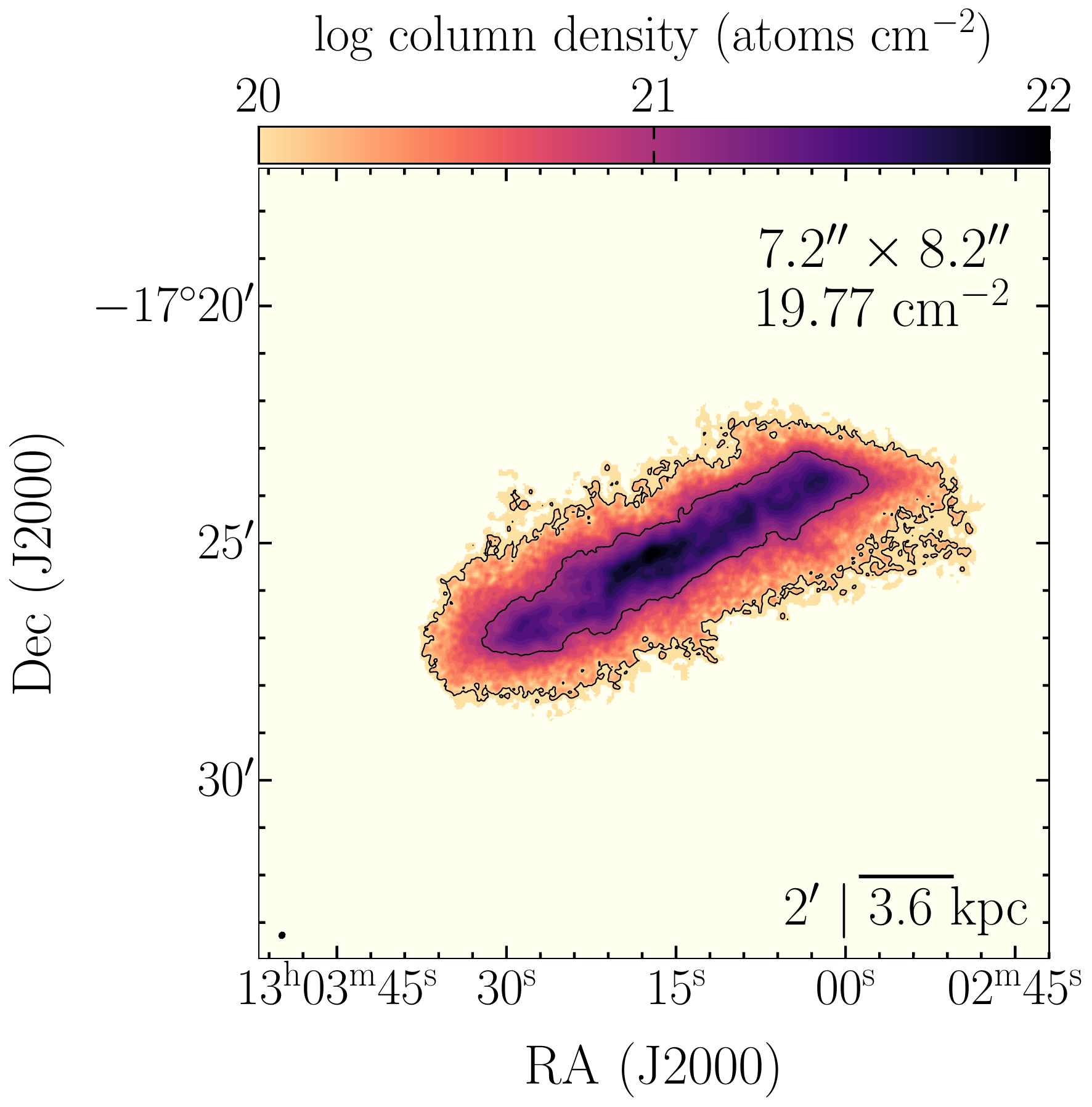}
	\label{subfig:mom0_r00_t00}
	\end{subfigure}
    \begin{subfigure}{0.25\textwidth}
    \centering
	\includegraphics[width=\textwidth]{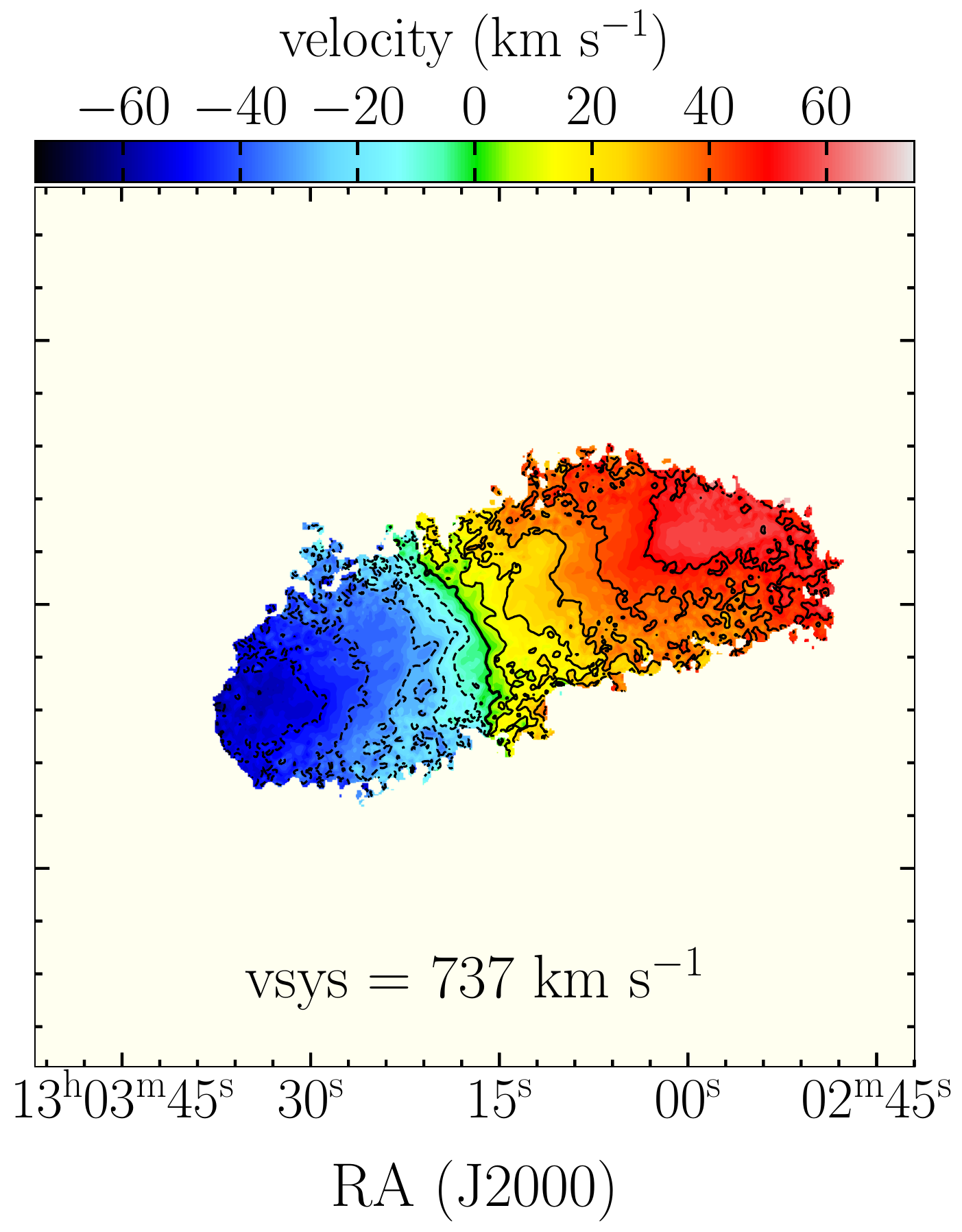}
	\label{subfig:mom1_r00_t00}
	\end{subfigure}	
	\begin{subfigure}{0.29\textwidth}
    \centering
	\includegraphics[width=\textwidth]{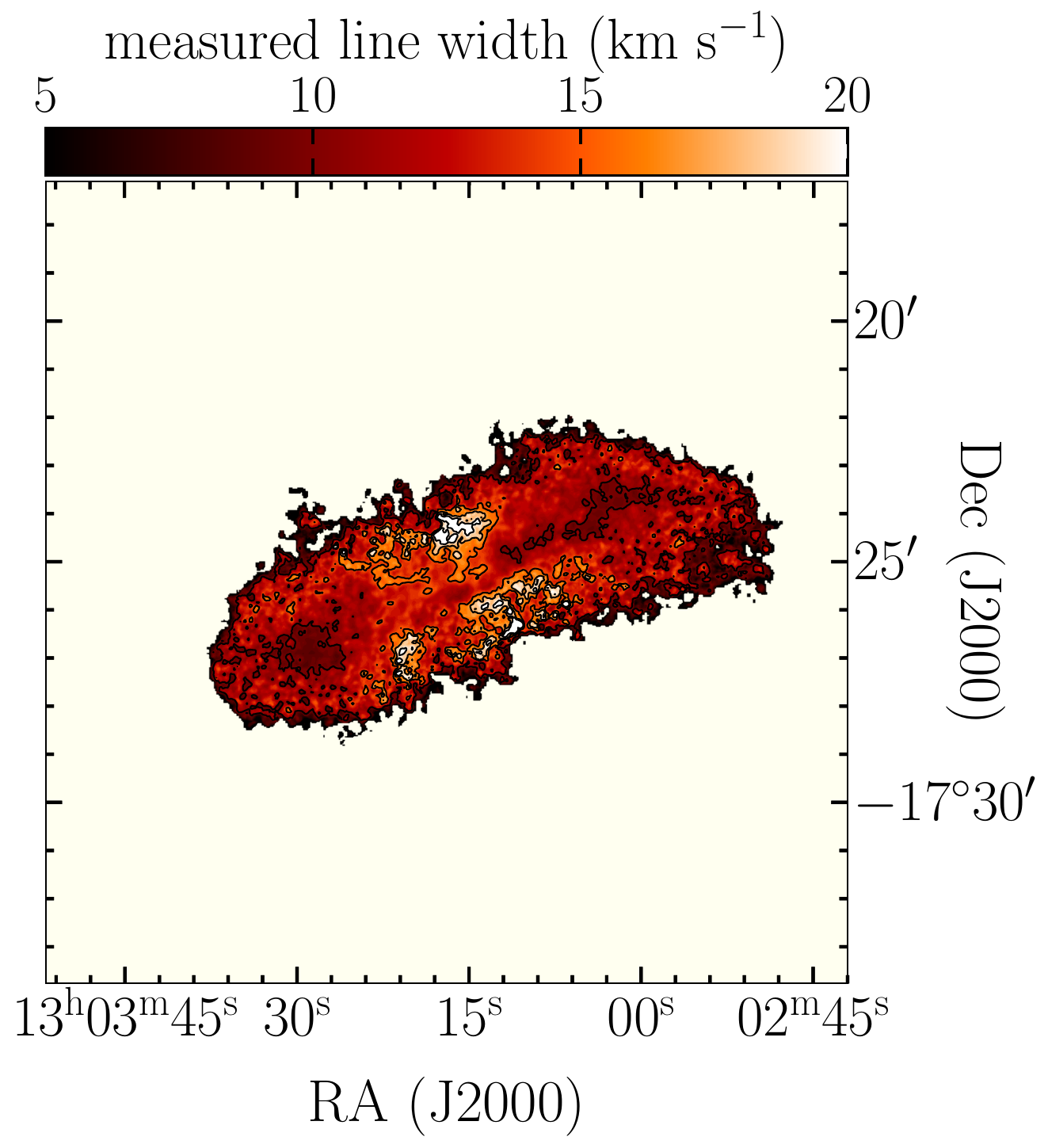}
	\label{subfig:mom2_r00_t00}
	\end{subfigure}
\end{figure*}

\begin{figure*} \ContinuedFloat
    \begin{subfigure}{0.32\textwidth}
    \centering
	\includegraphics[width=\textwidth]{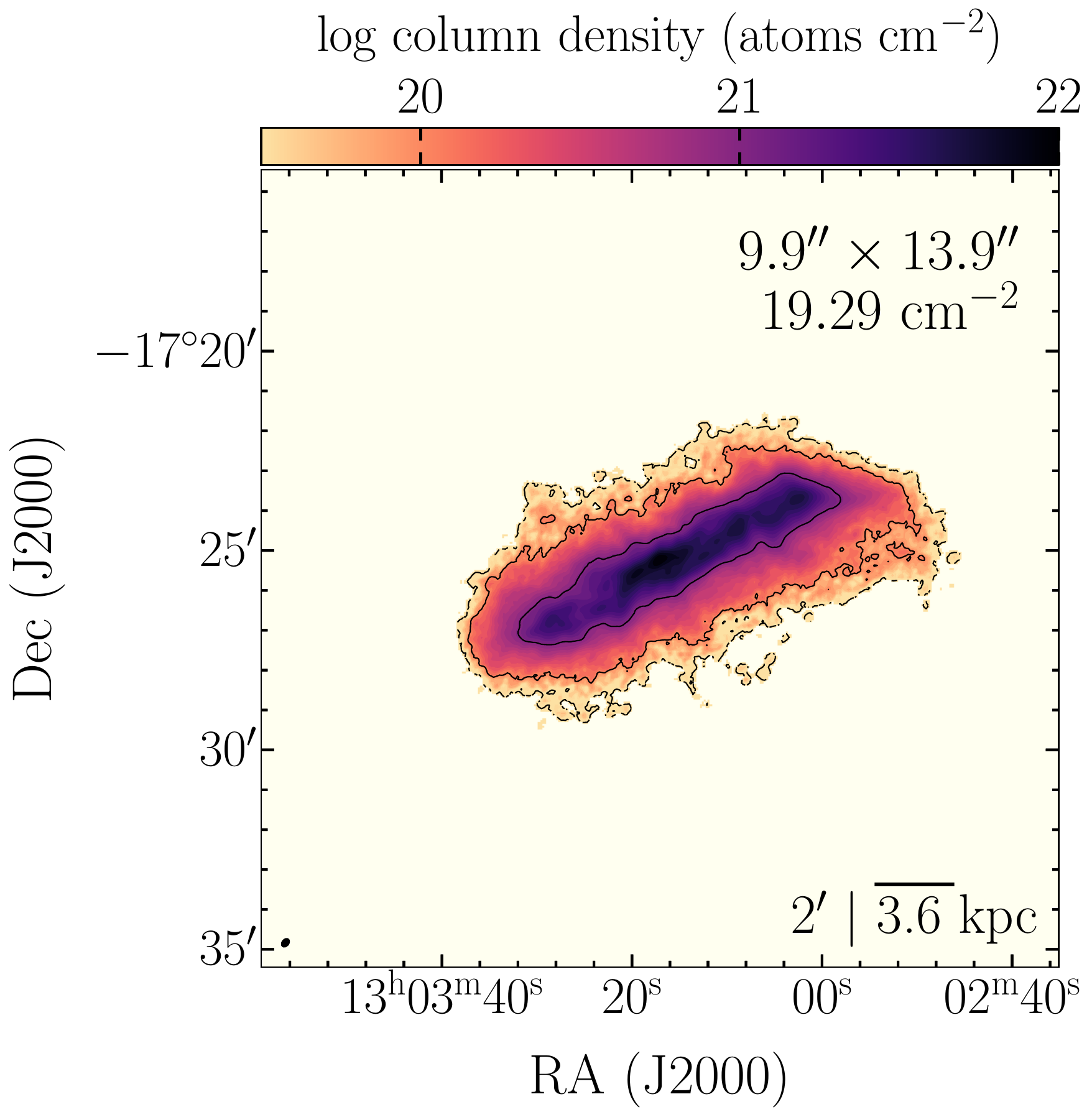}
	\label{subfig:mom0_r05_t00}
	\end{subfigure}
    \begin{subfigure}{0.245\textwidth}
    \centering
	\includegraphics[width=\textwidth]{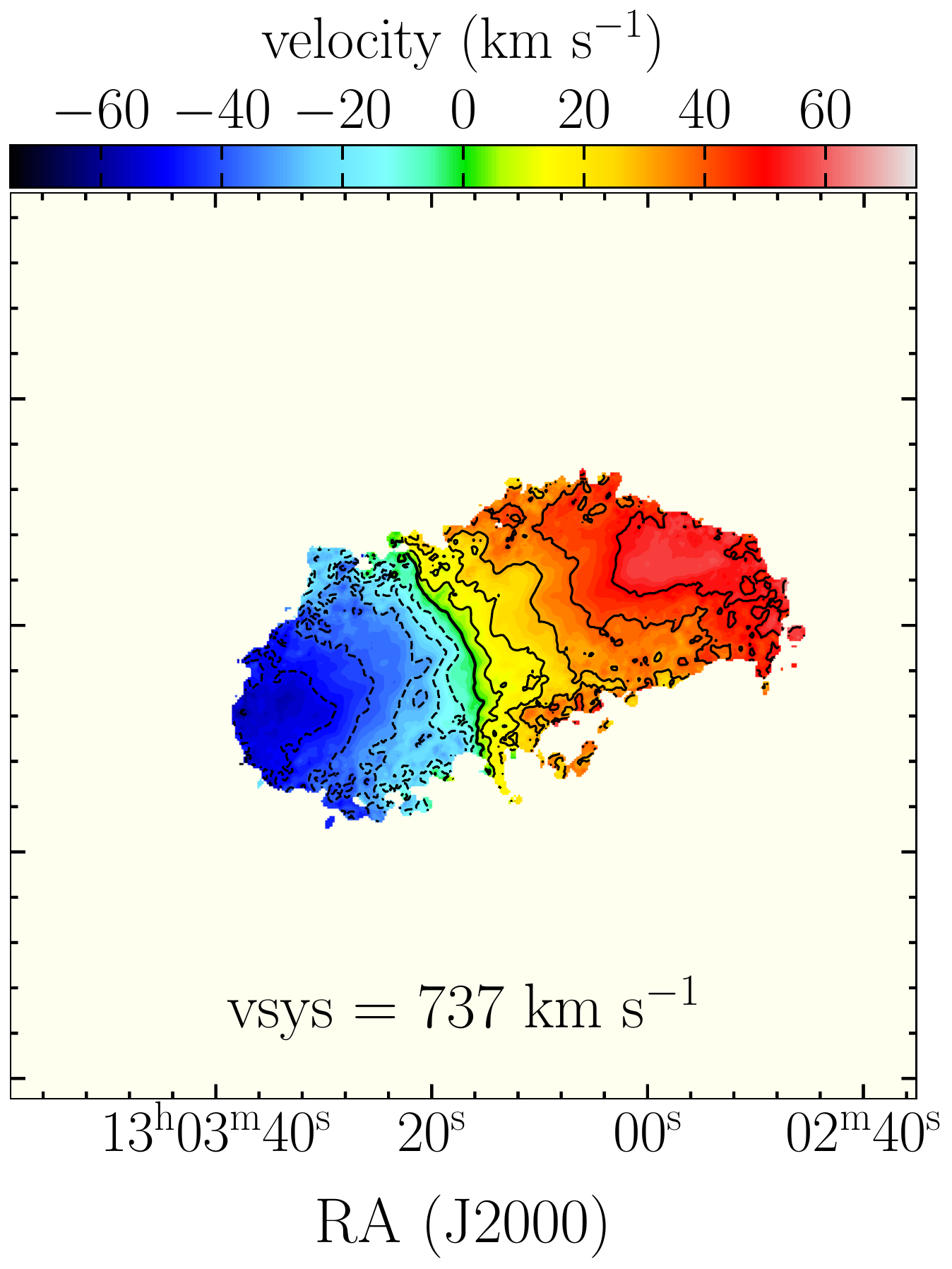}
	\label{subfig:mom1_r05_t00}
	\end{subfigure}	
	\begin{subfigure}{0.29\textwidth}
    \centering
	\includegraphics[width=\textwidth]{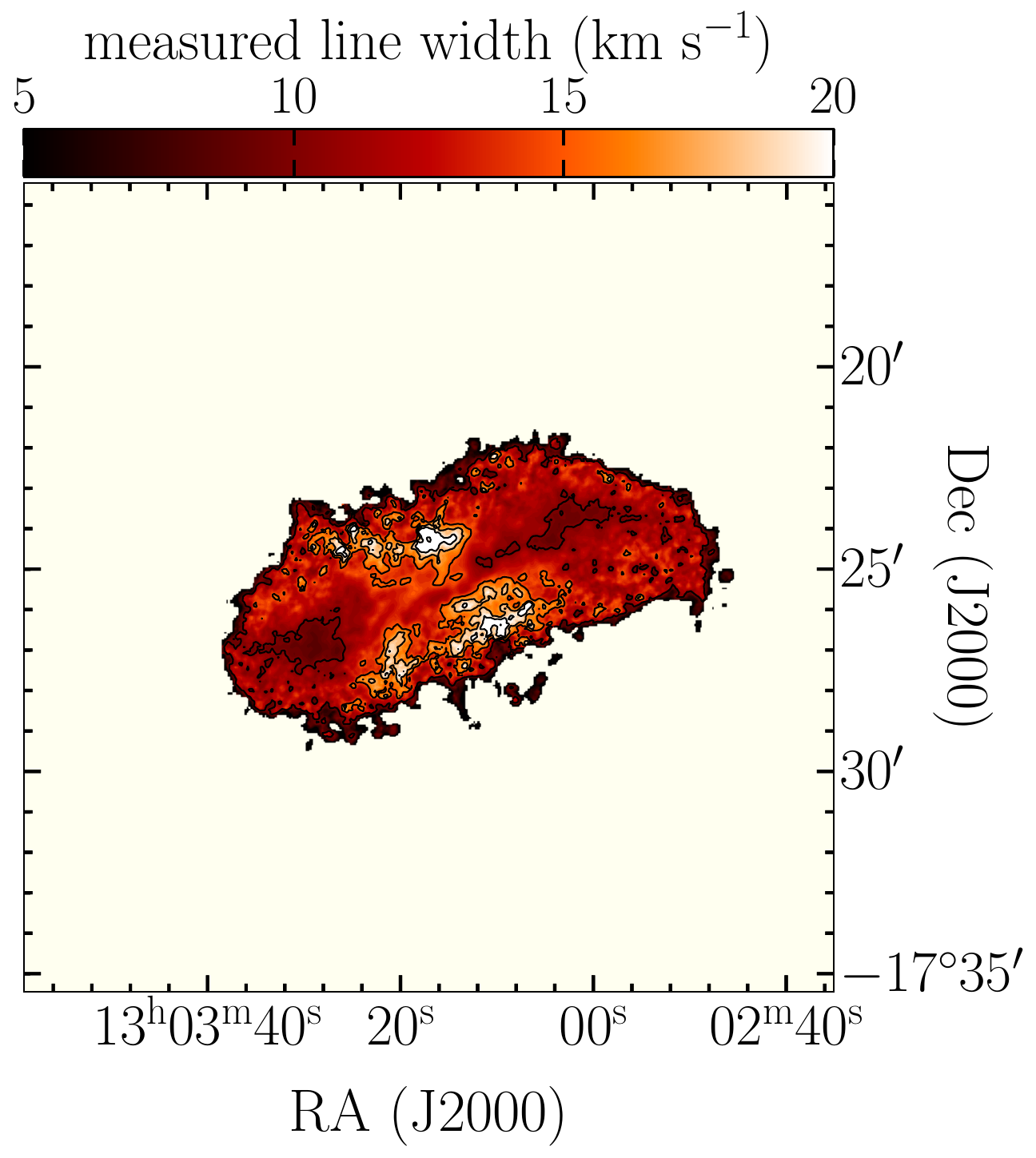}
	\label{subfig:mom2_r05_t00}
	\end{subfigure}
\end{figure*}

\begin{figure*} \ContinuedFloat
    \begin{subfigure}{0.32\textwidth}
    \centering
	\includegraphics[width=\textwidth]{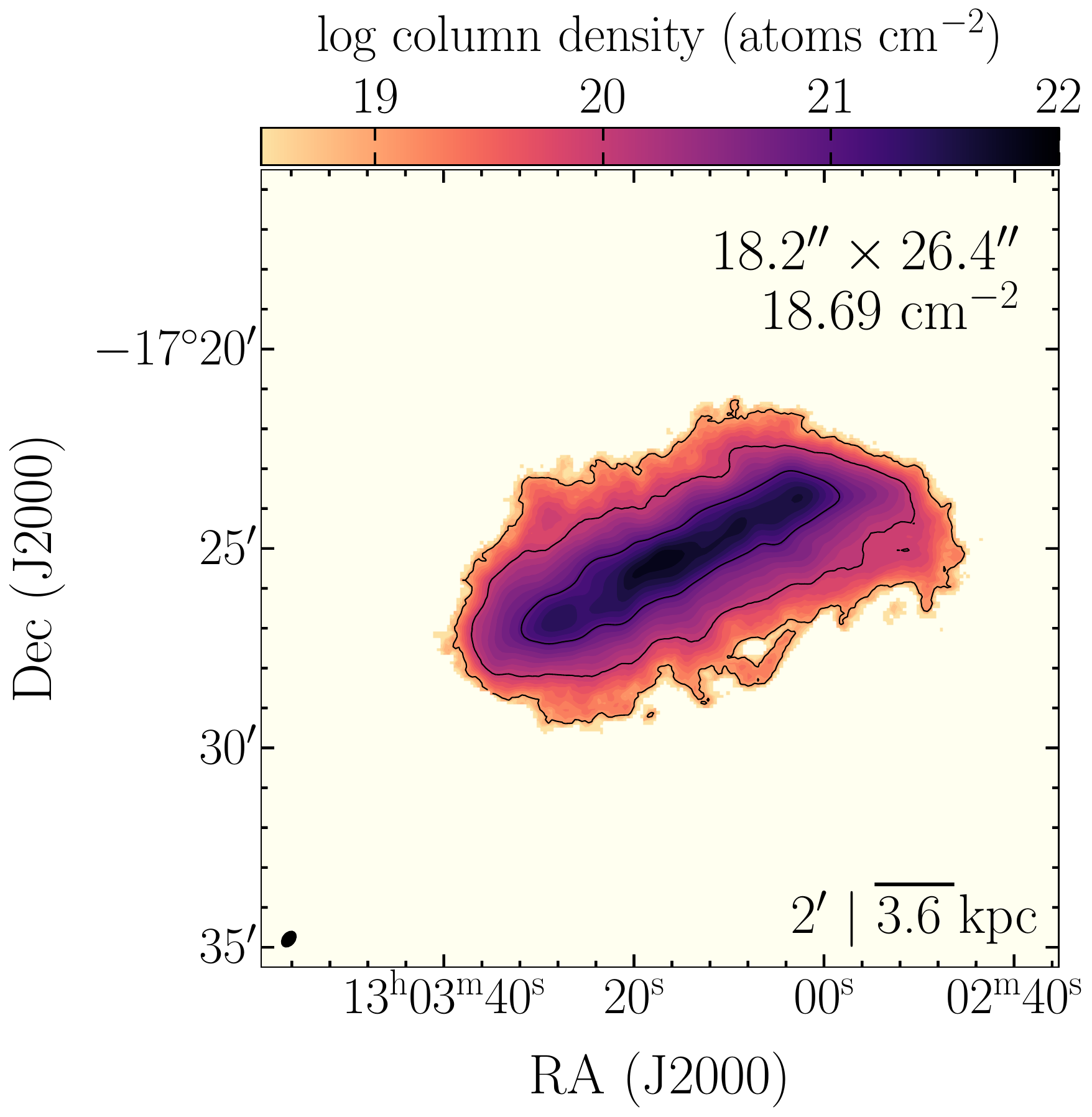}
	\label{subfig:mom0_r10_t00}
	\end{subfigure}
    \begin{subfigure}{0.245\textwidth}
    \centering
	\includegraphics[width=\textwidth]{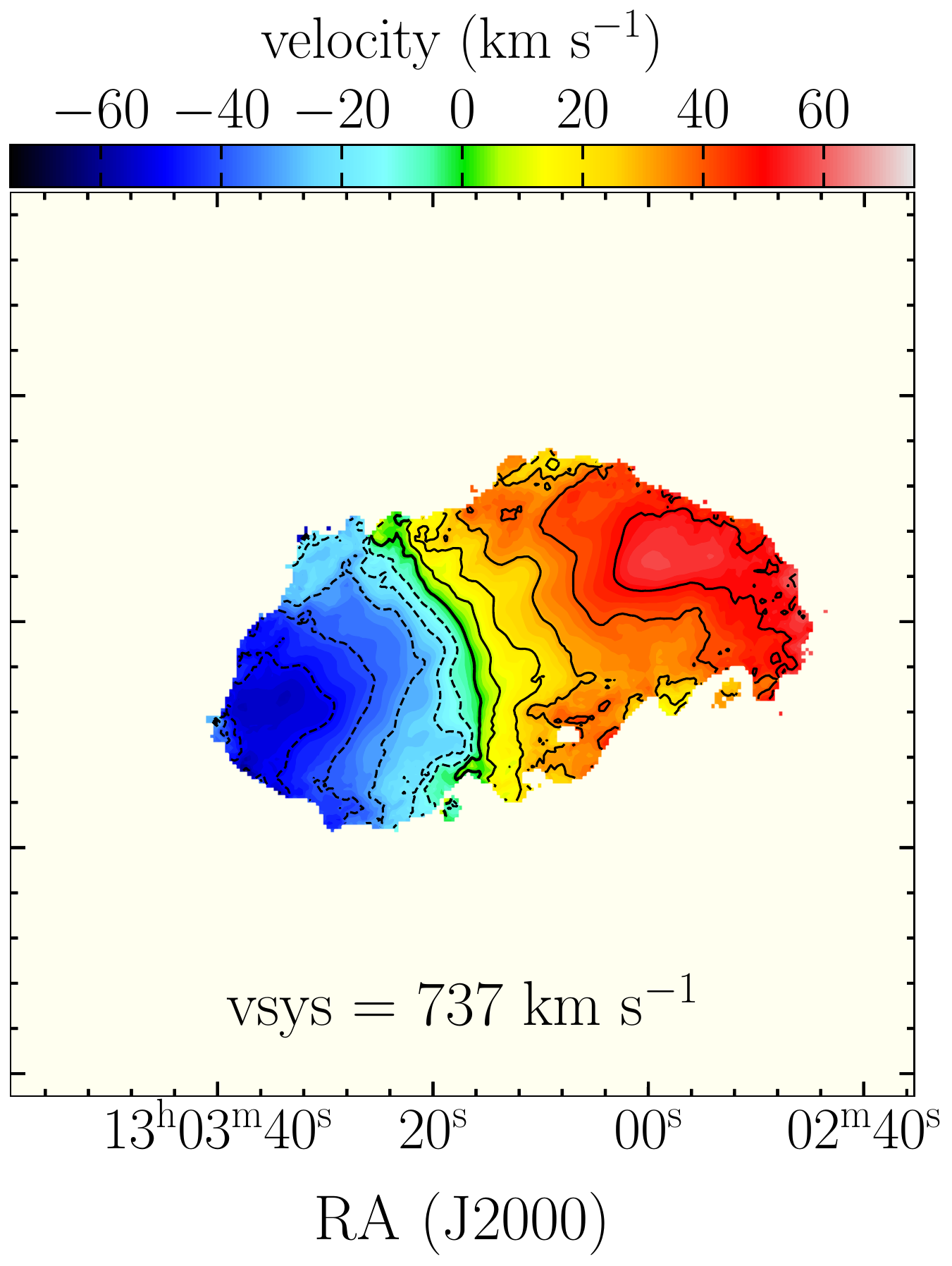}
	\label{subfig:mom1_r10_t00}
	\end{subfigure}	
	\begin{subfigure}{0.29\textwidth}
    \centering
	\includegraphics[width=\textwidth]{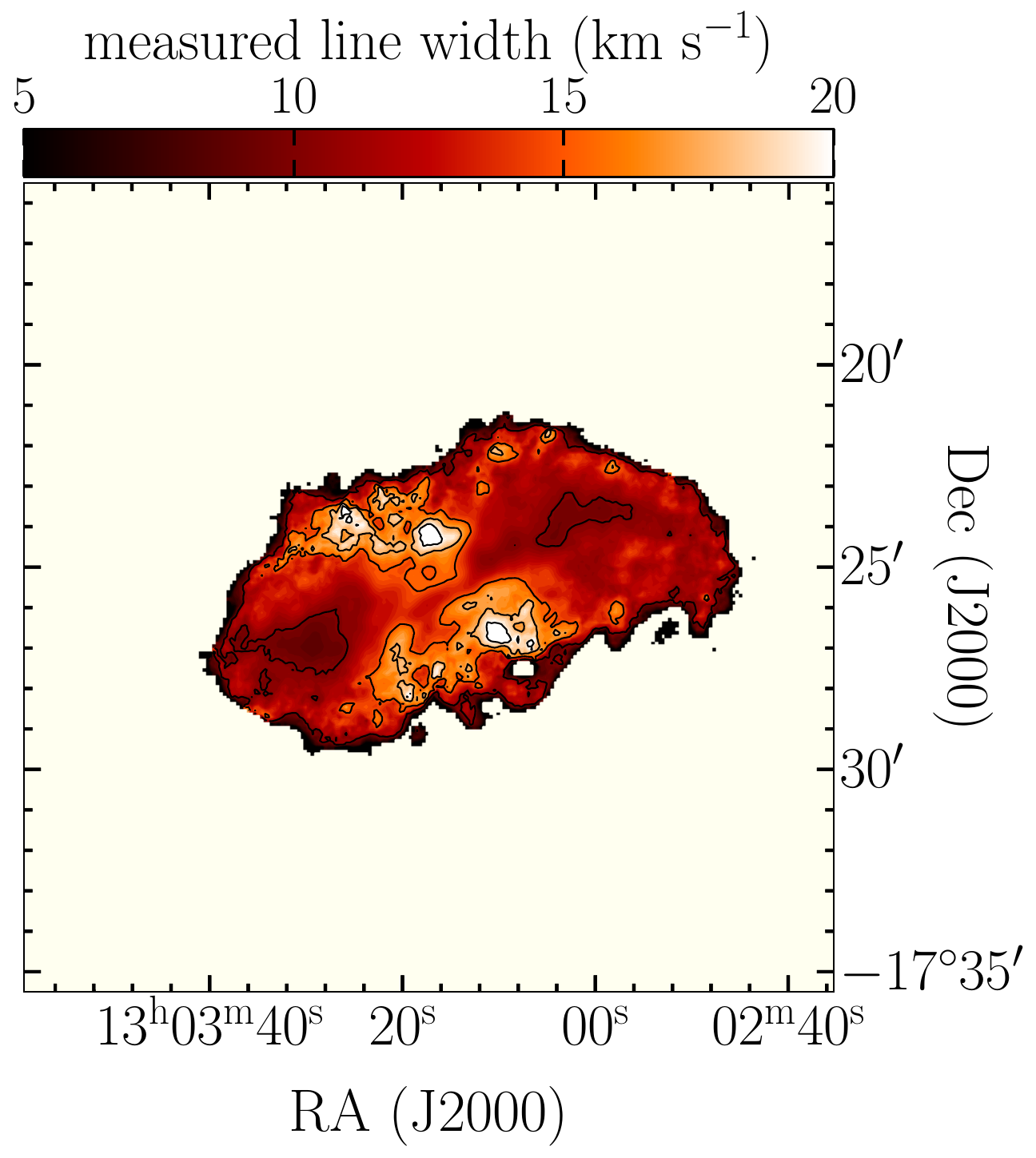}
	\label{subfig:mom2_r10_t00}
	\end{subfigure}
\end{figure*}

\begin{figure*} \ContinuedFloat
    \begin{subfigure}{0.32\textwidth}
    \centering
	\includegraphics[width=\textwidth]{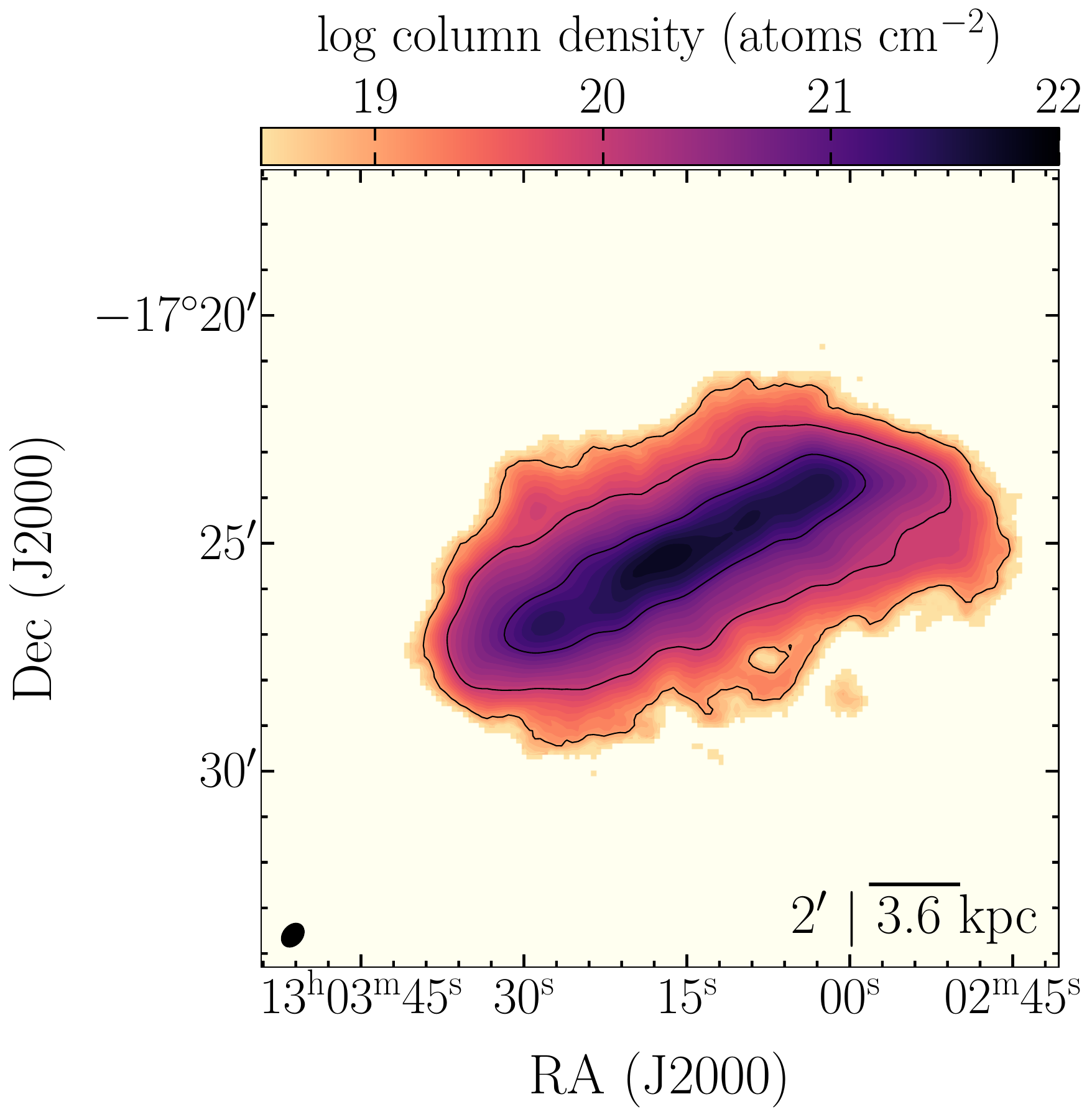}
    \hspace{5mm}
	\end{subfigure}
    \begin{subfigure}{0.245\textwidth}
    \centering
	\includegraphics[width=\textwidth]{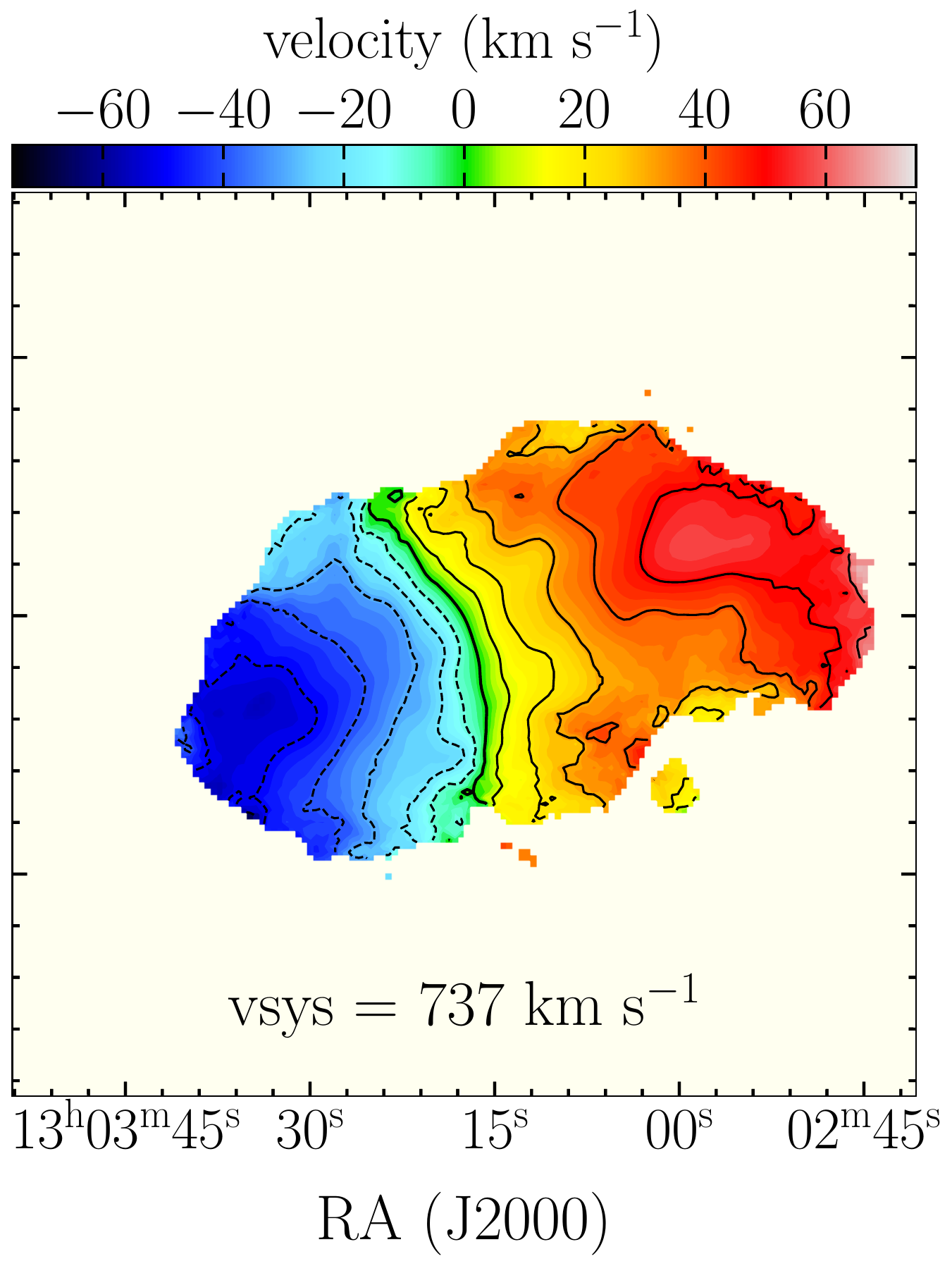}
	\label{subfig:mom1_r15_t00}
	\end{subfigure}	
	\begin{subfigure}{0.29\textwidth}
    \centering
	\includegraphics[width=\textwidth]{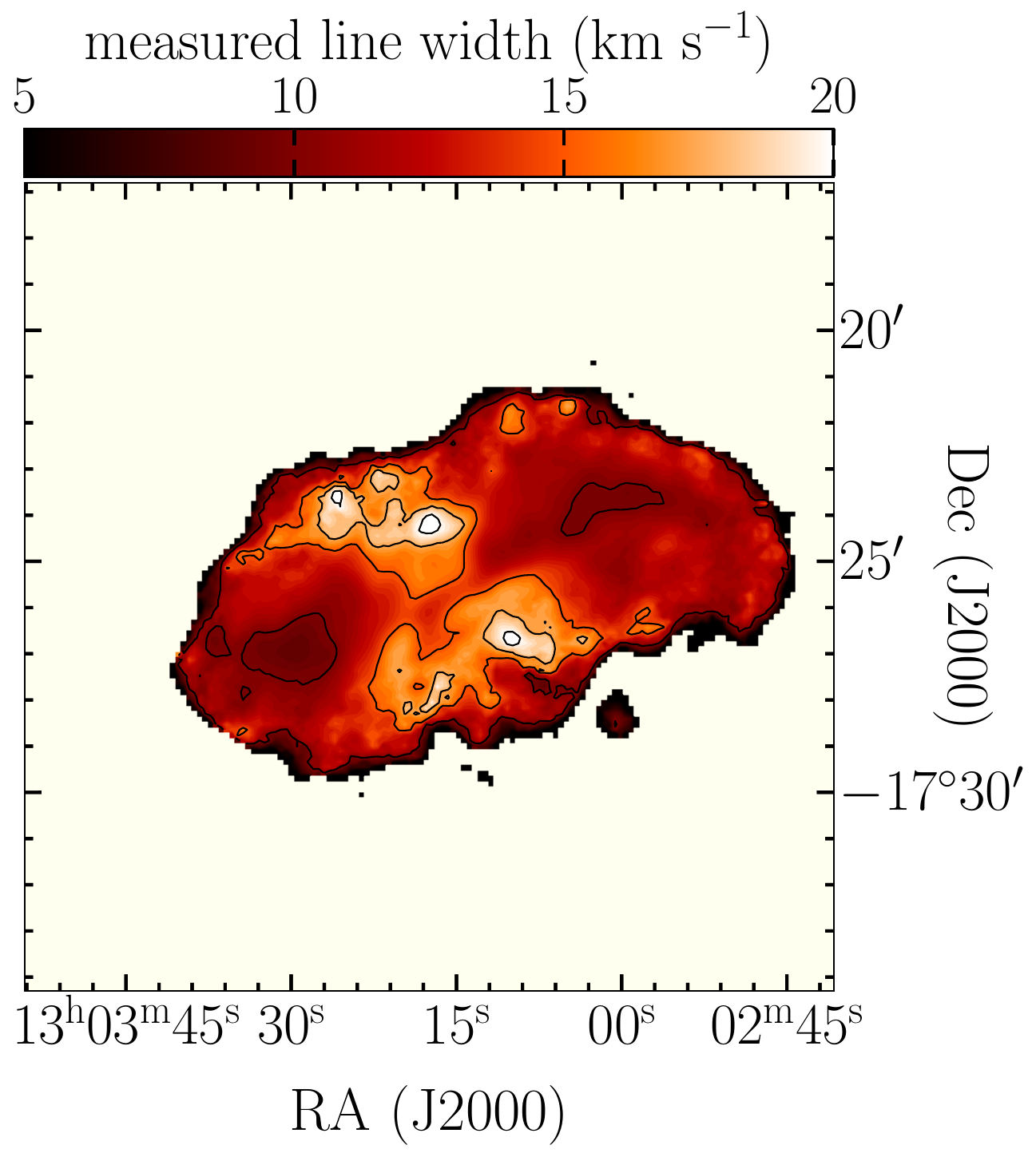}
	\label{subfig:mom2_r15_t00}
	\end{subfigure}
\end{figure*}

\begin{figure*} \ContinuedFloat
    \begin{subfigure}{0.3\textwidth}
    \centering
	\includegraphics[width=\textwidth]{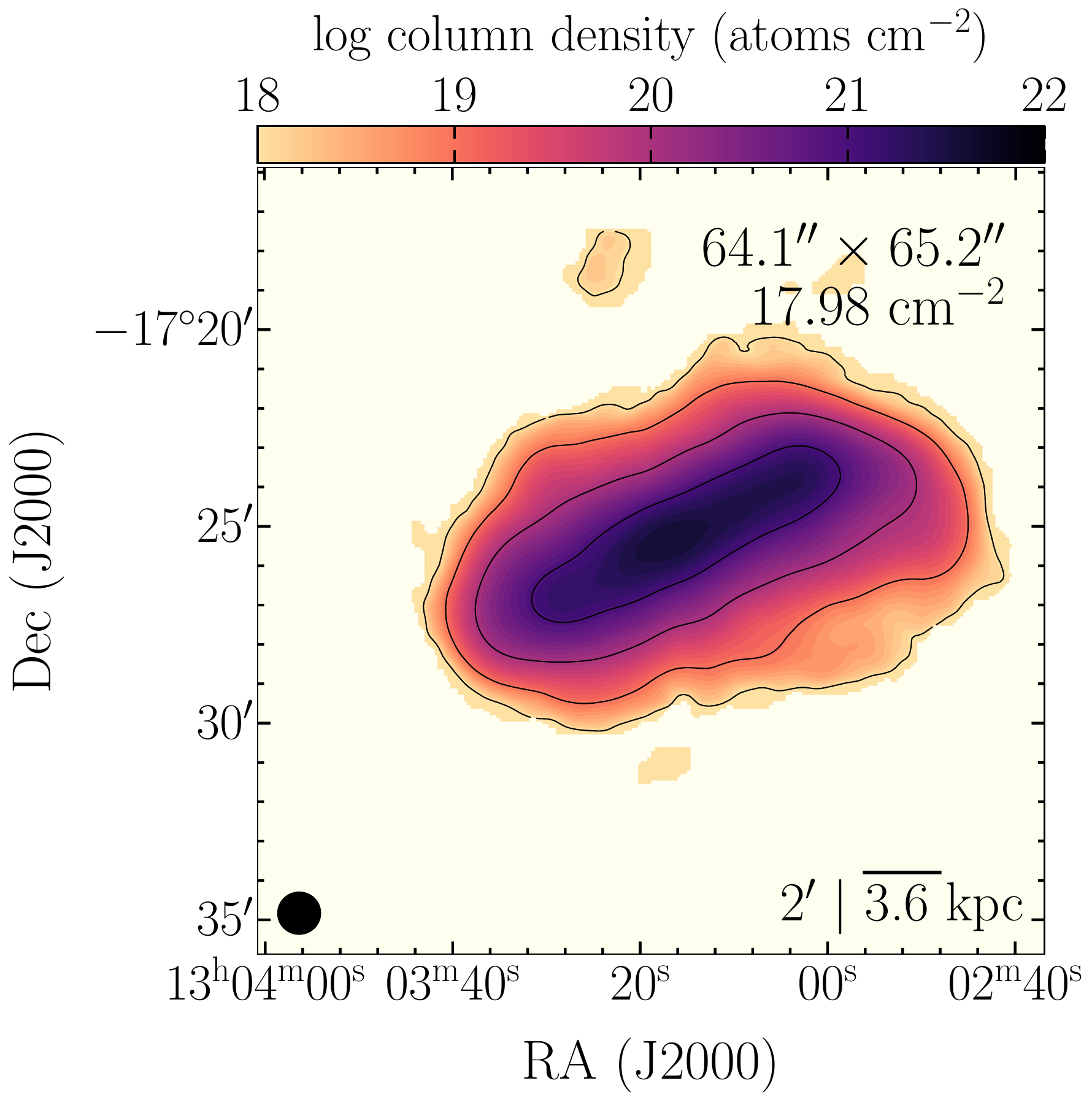}
	\label{subfig:mom0_r05_t60}
	\end{subfigure}
    \begin{subfigure}{0.255\textwidth}
    \centering
	\includegraphics[width=\textwidth]{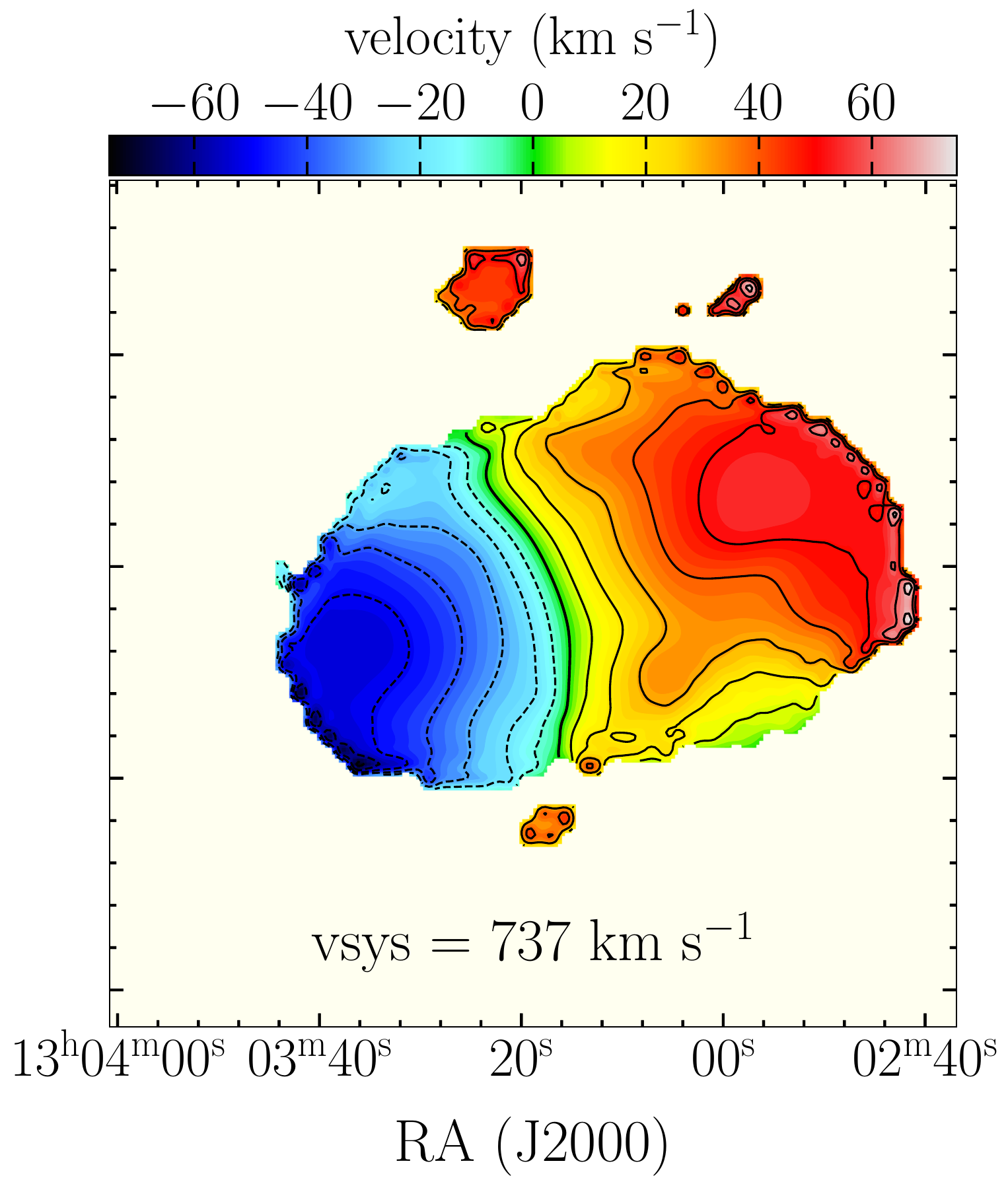}
	\label{subfig:mom1_r05_t60}
	\end{subfigure}	
	\begin{subfigure}{0.295\textwidth}
    \centering
	\includegraphics[width=\textwidth]{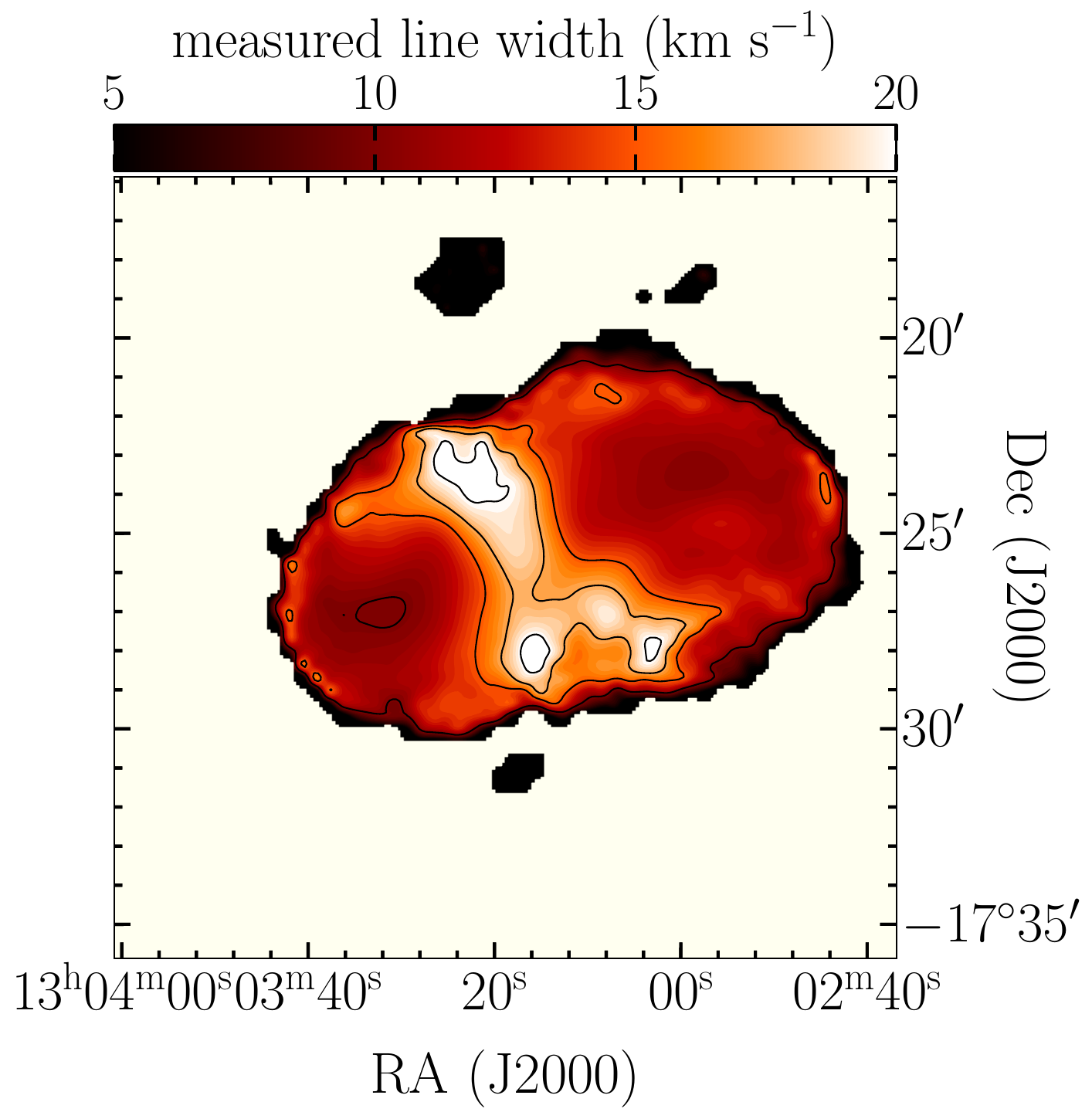}
	\label{subfig:mom2_r05_t60}
	\end{subfigure}
\end{figure*}

\begin{figure*} \ContinuedFloat
    \begin{subfigure}{0.315\textwidth}
    \centering
	\includegraphics[width=\textwidth]{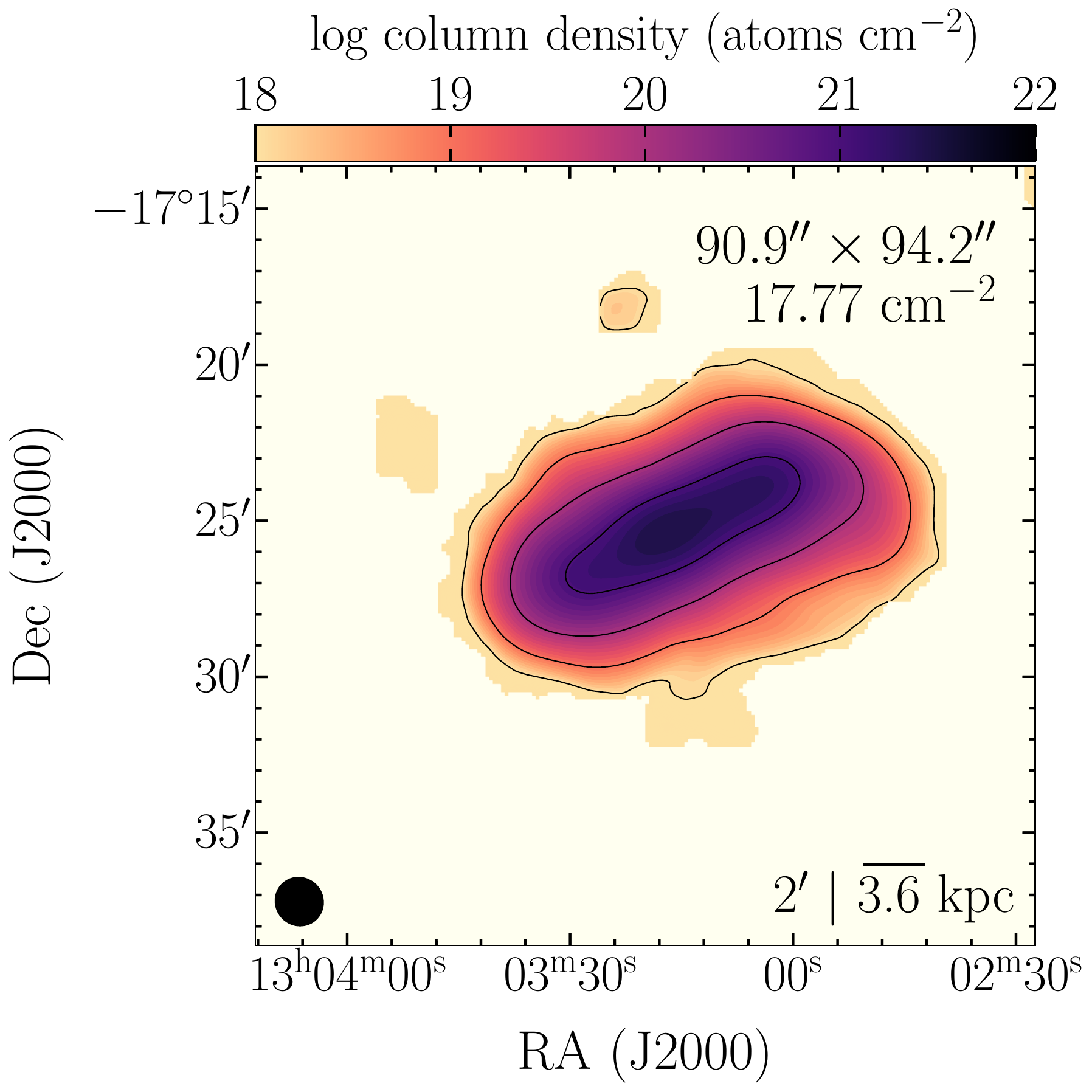}
	\label{subfig:mom0_r10_t90}
	\end{subfigure}
    \begin{subfigure}{0.25\textwidth}
    \centering
	\includegraphics[width=\textwidth]{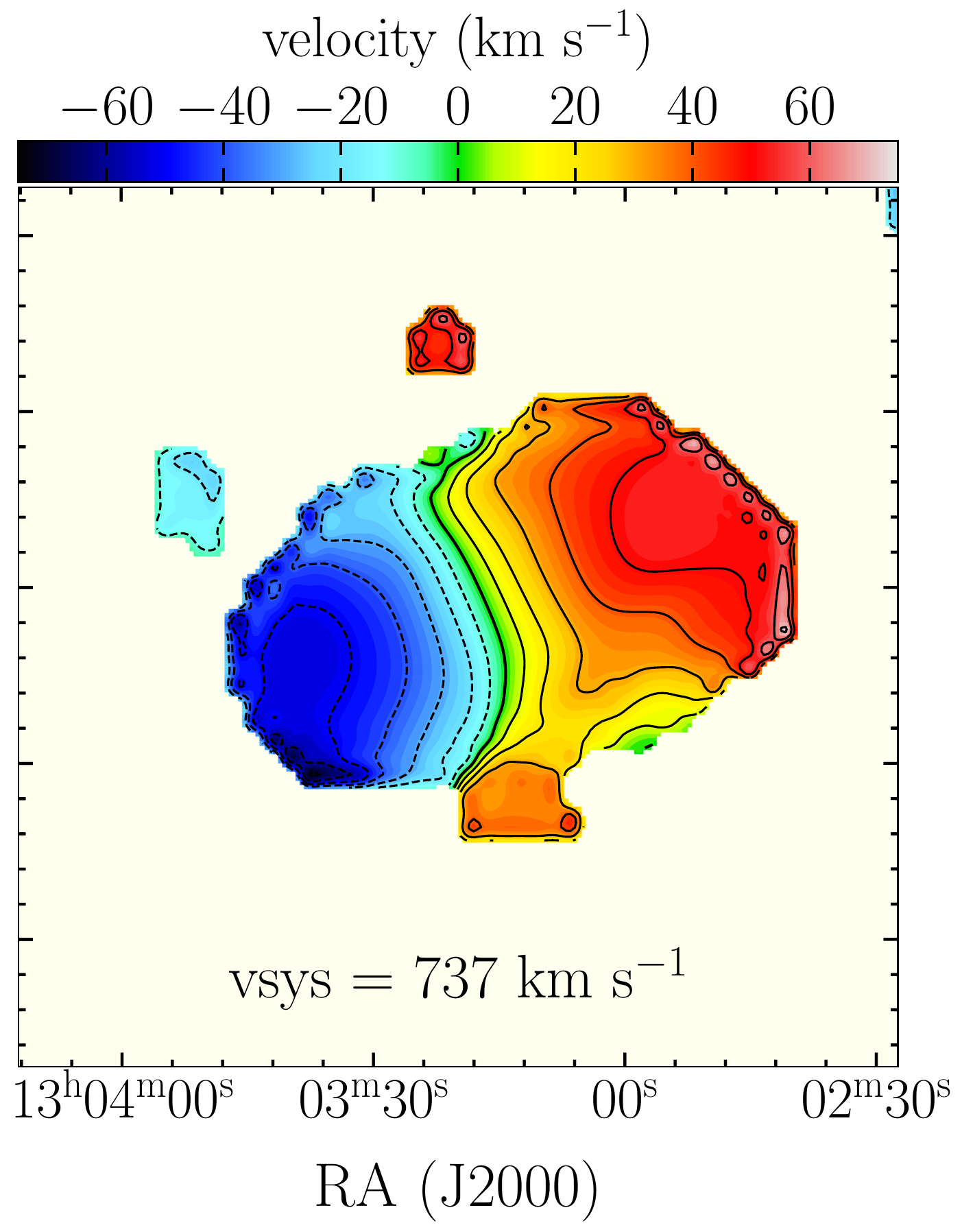}
	\label{subfig:mom1_r10_t90}
	\end{subfigure}	
	\begin{subfigure}{0.285\textwidth}
    \centering
	\includegraphics[width=\textwidth]{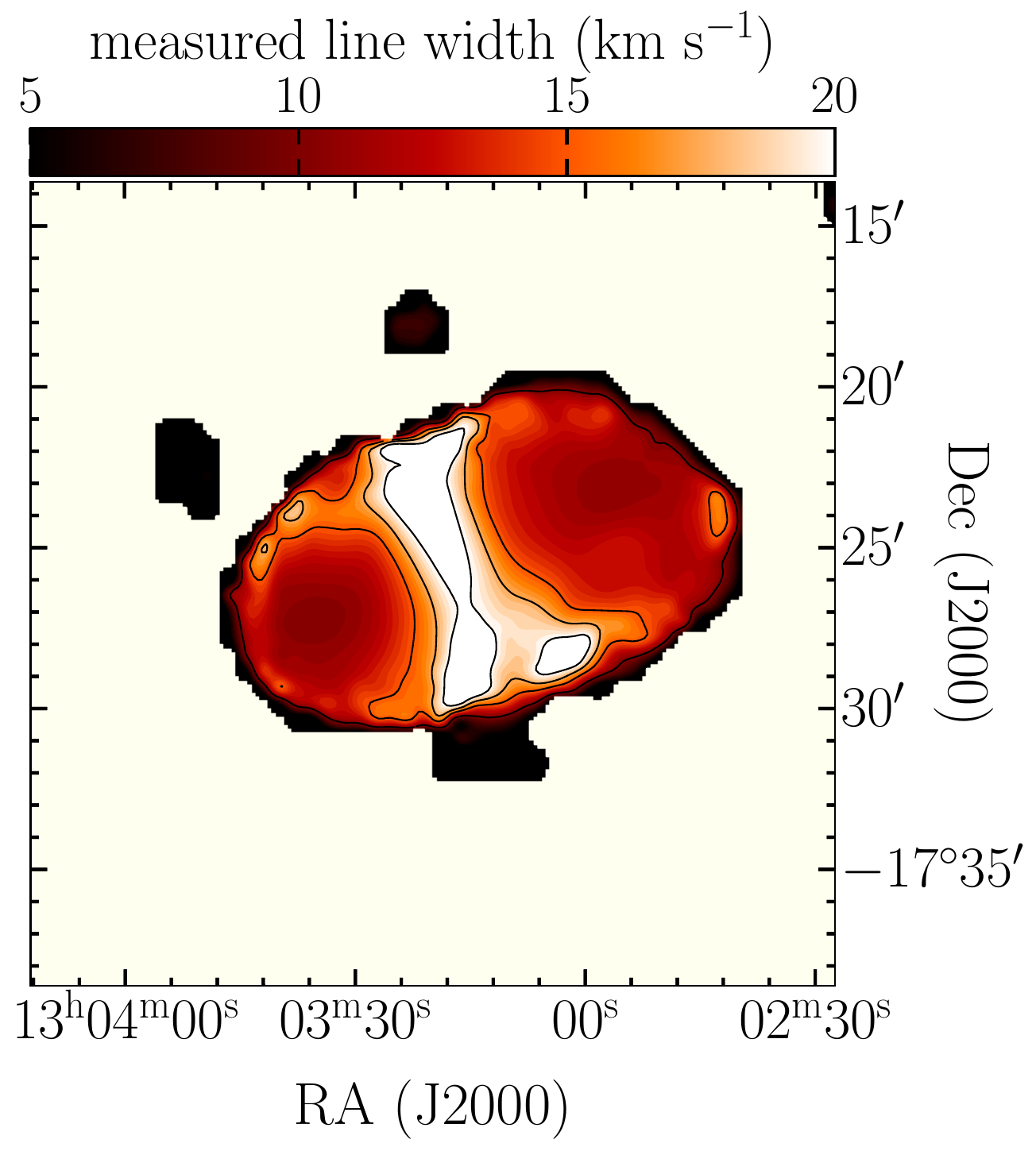}
	\label{subfig:mom2_r10_t90}
	\end{subfigure}
	\caption{\refrep{Moment zero, one, and two maps (from left to right) for UGCA~320 at all six MHONGOOSE resolutions (\texttt{r00\_t00}, \texttt{r05\_t00}, \texttt{r10\_t00}, \texttt{r15\_t00}, \texttt{r05\_t60}, \texttt{r10\_t90}, respectively, from top to bottom). The beams are shown in the lower-left corners of the moment zero maps, and a scale bar is shown in their lower right corners. In the top-right corners of the moment zero maps the size of the beam is quoted, along with the $3 \sigma$ column density of the \HI\ over 16 km~s$^{-1}$, in atoms cm$^{-2}$. The systemic velocity is indicated in each moment one map. The fourth row, which corresponds to \texttt{r15\_t00}, is identical to the corresponding panels in Figure \ref{fig:moment_maps}, and are repeated for completeness.}}
	\label{fig:moment_maps_multi_res}
\end{figure*}

\section{\vthree{Alternative tilted ring models}}

\vthree{In Figure \ref{fig:alternative_models} we present alternative TiRiFiC tilted ring models that we did not select as our final model. Most of these exclude a parameter that is key to recreating the data, to demonstrate its significance, or otherwise highlight an peculiarity in the \HI\ reservoir of UGCA~320.}

\begin{figure*}

	\begin{subfigure}{0.4\textwidth}
    \centering
	\includegraphics[width=\textwidth]{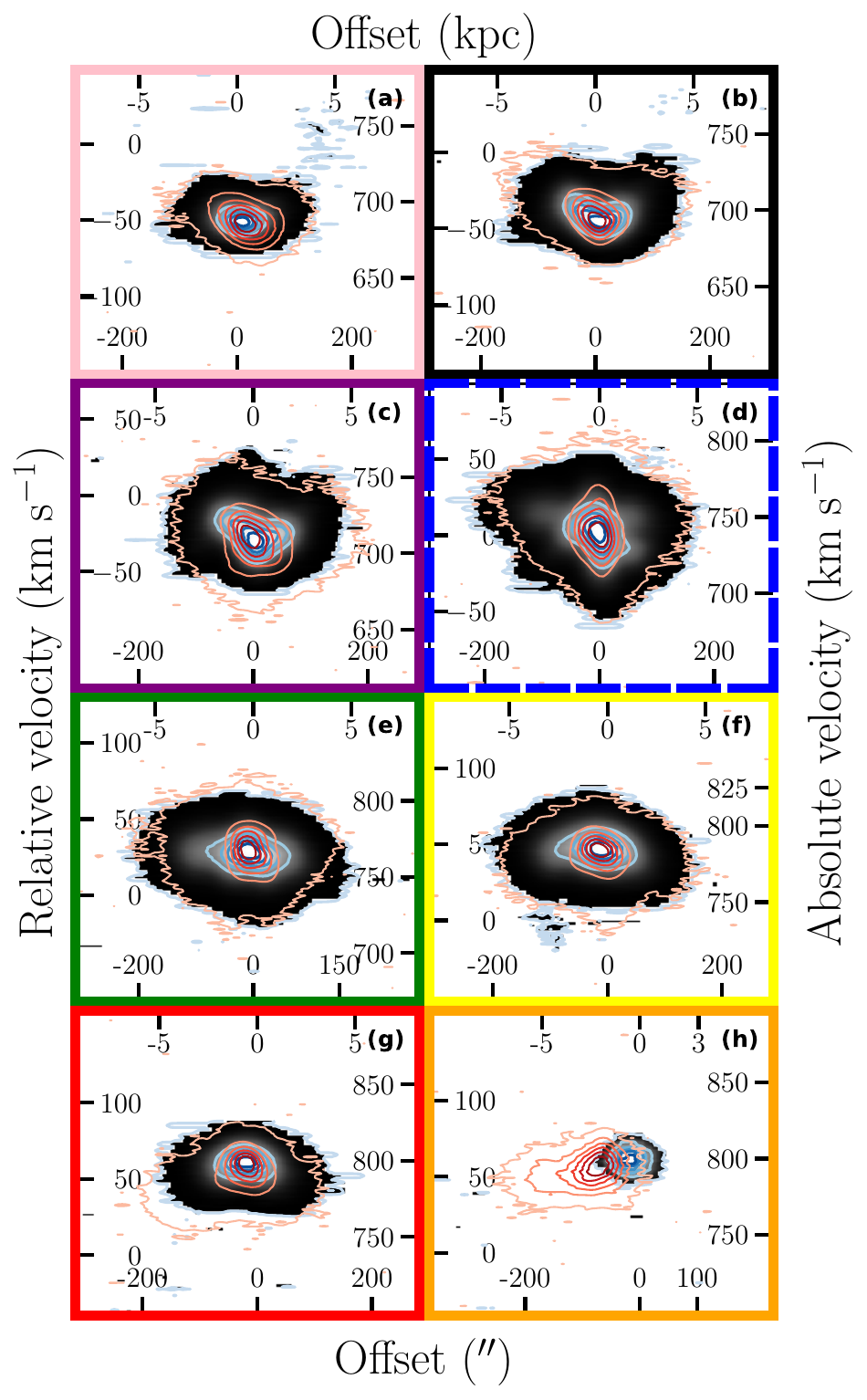}
	\caption{\refrep{PVD slices for the final model with the change in position angle removed.}}
	\label{subfig:pvd_no_pa}
	\end{subfigure}
    \begin{subfigure}{0.4\textwidth}
    \centering
	\includegraphics[width=\textwidth]{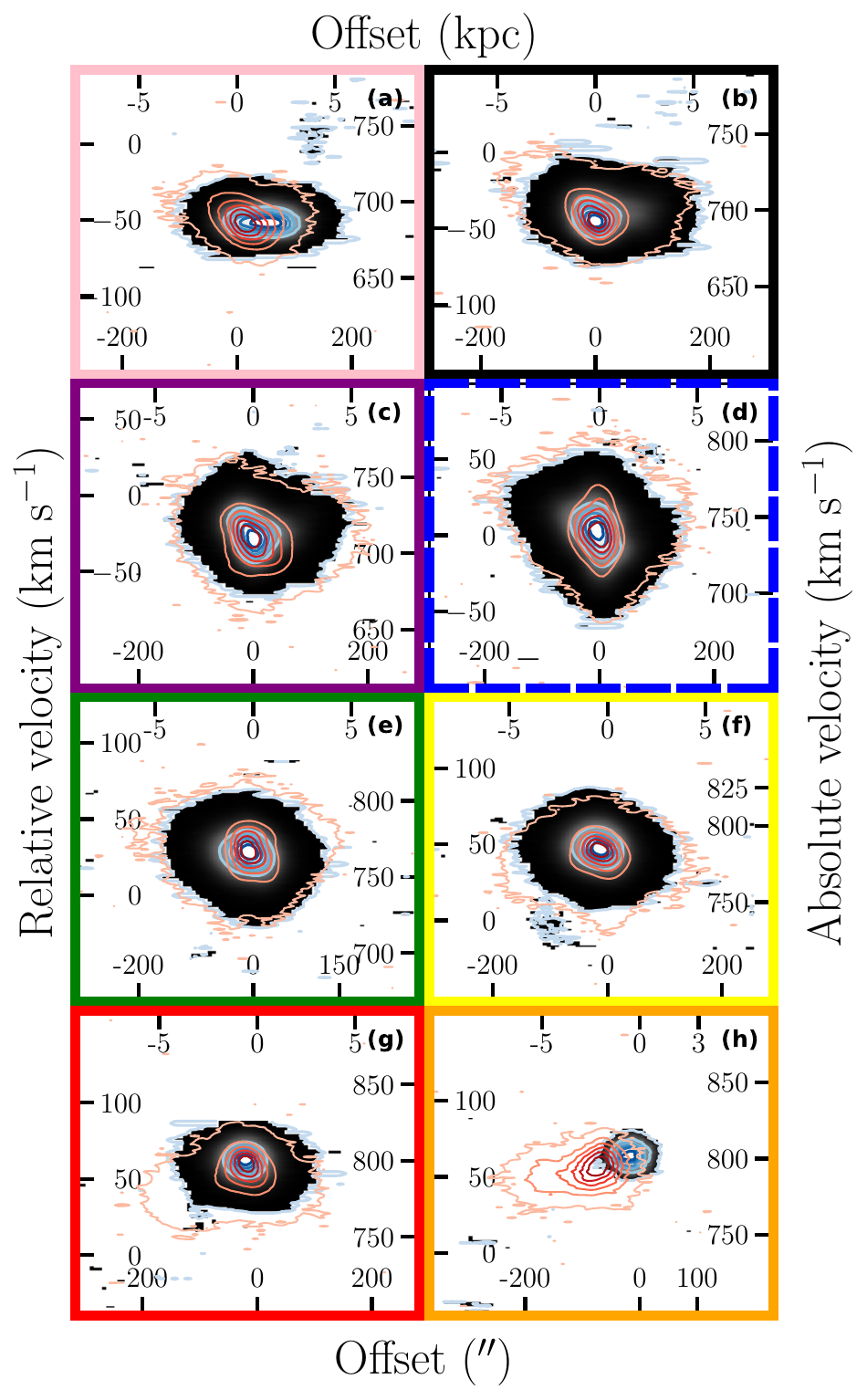}
	\caption{\refrep{PVD slices where the inclination warp has been removed from the model.}}
	\label{subfig:pvd_no_inc}
	\end{subfigure} 
	
	\begin{subfigure}{0.8\textwidth}
    \centering
	\includegraphics[width=\textwidth]{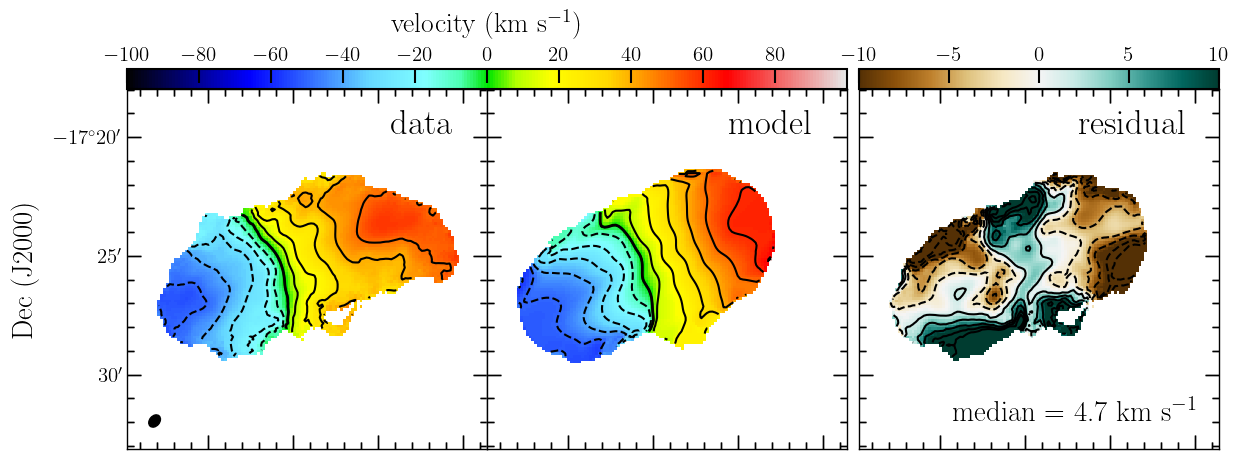}
	\caption{\refrep{Velocity residuals for the final model, where the change in position angle has been removed.}}
	\label{subfig:mom1_no_pa}
	\end{subfigure}
	\begin{subfigure}{0.8\textwidth}
    \centering
	\includegraphics[width=\textwidth]{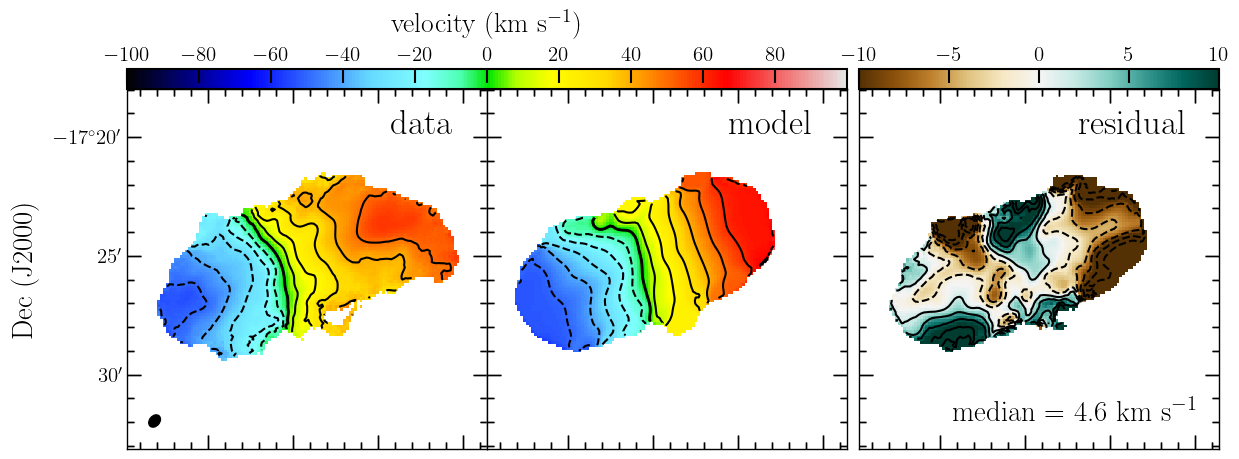}
	\caption{\refrep{Velocity residuals for the final model, where the inclination warp has been removed.}}
	\label{subfig:mom1_no_inc}
	\end{subfigure}	
\end{figure*}
	
\begin{figure*}\ContinuedFloat 

    \begin{subfigure}{0.4\textwidth}
    \centering
	\includegraphics[width=\textwidth]{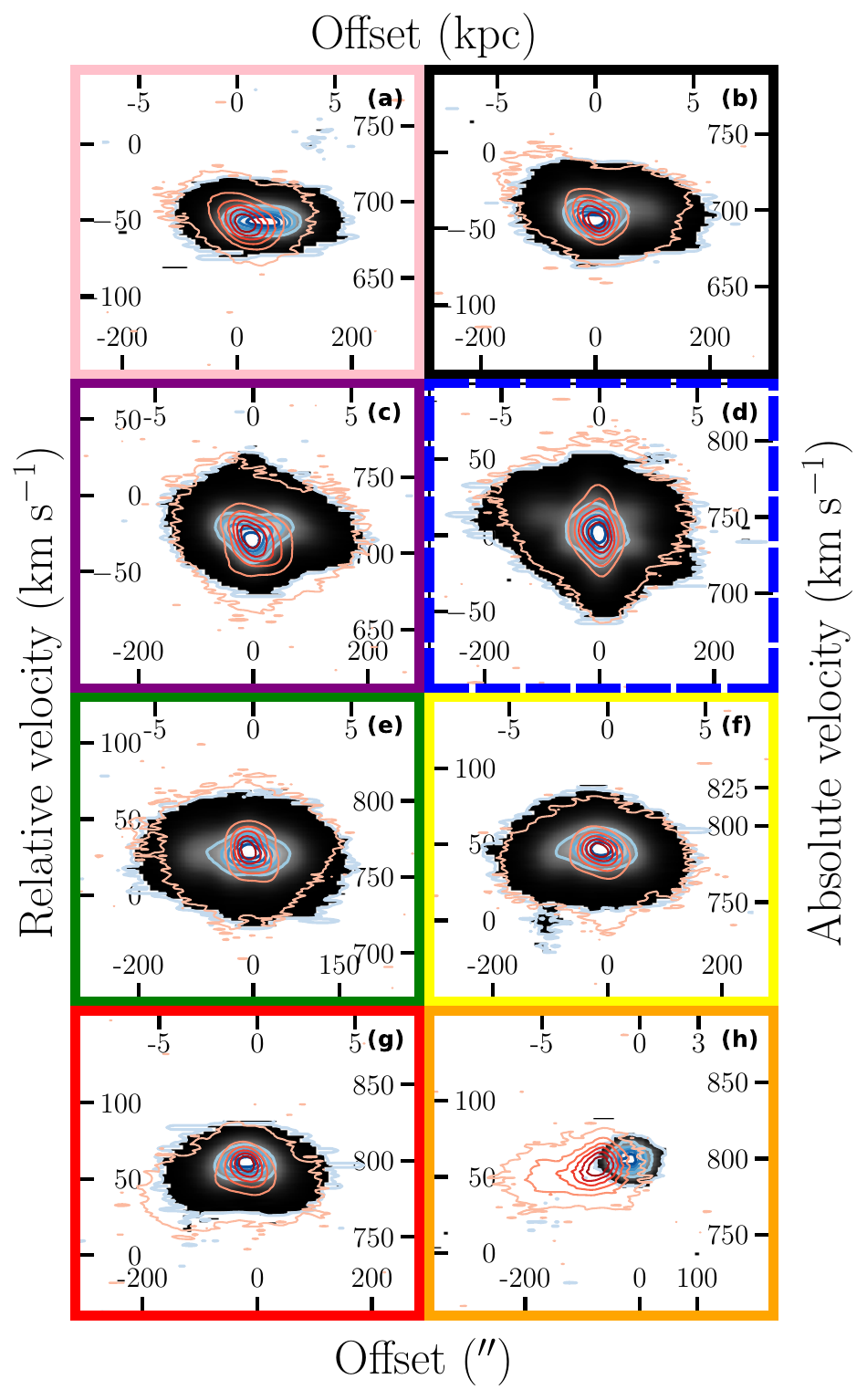}
	\caption{\refrep{PVD slices for the final model with the radial motions removed.}}
	\label{subfig:pvd_no_vrad}
	\end{subfigure}	
	 \begin{subfigure}{0.4\textwidth}
    \centering
	\includegraphics[width=\textwidth]{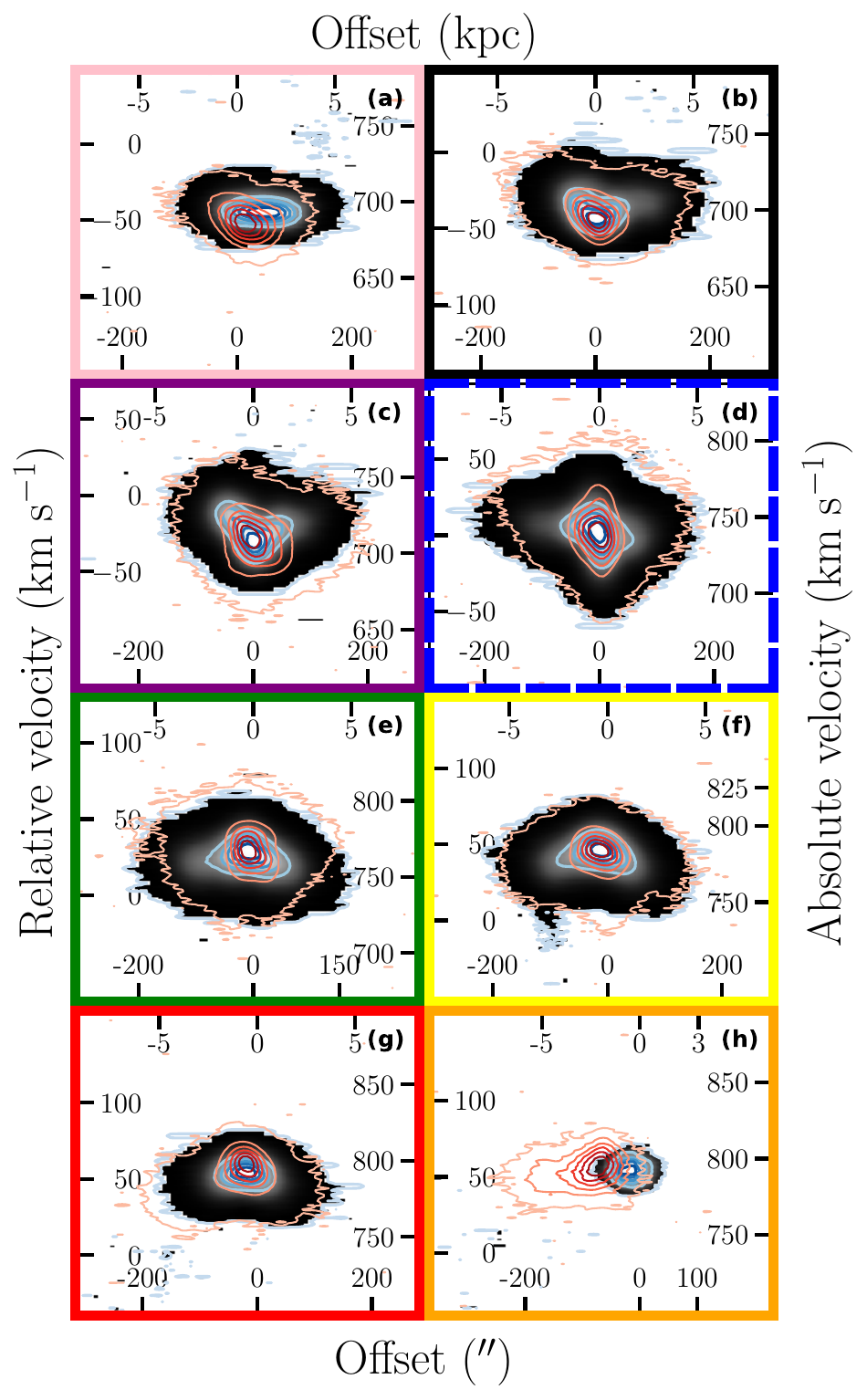}
	\caption{\refrep{PVD slices for the final model with the change in systemic velocity removed.}}
	\label{subfig:pvd_no_vsys}
	\end{subfigure} 
	
	\begin{subfigure}{0.8\textwidth}
    \centering
	\includegraphics[width=\textwidth]{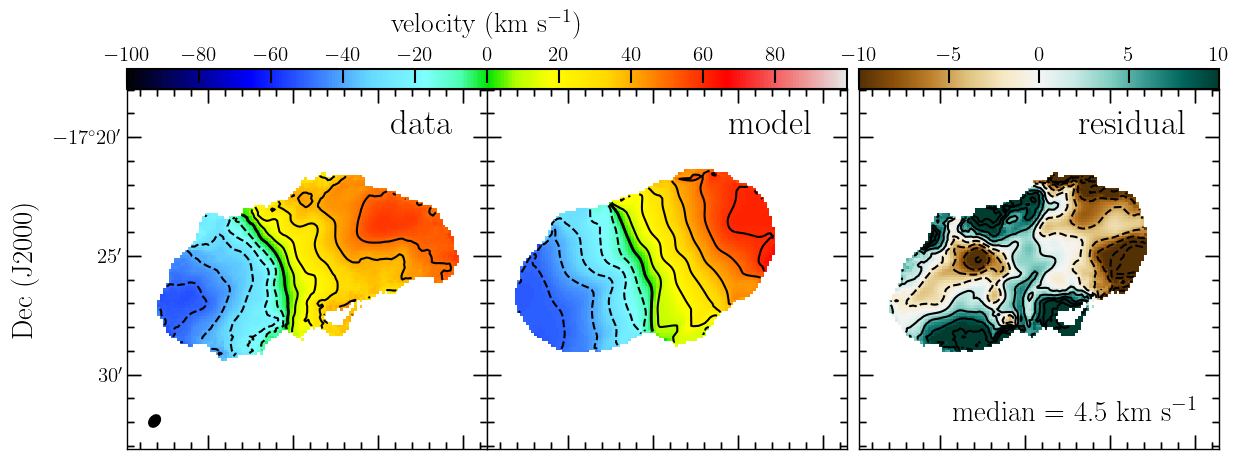}
	\caption{\refrep{Velocity residuals for the final model, where the radial motions have been removed.}}
	\label{subfig:mom1_no_vrad}
	\end{subfigure}	
	\begin{subfigure}{0.8\textwidth}
    \centering
	\includegraphics[width=\textwidth]{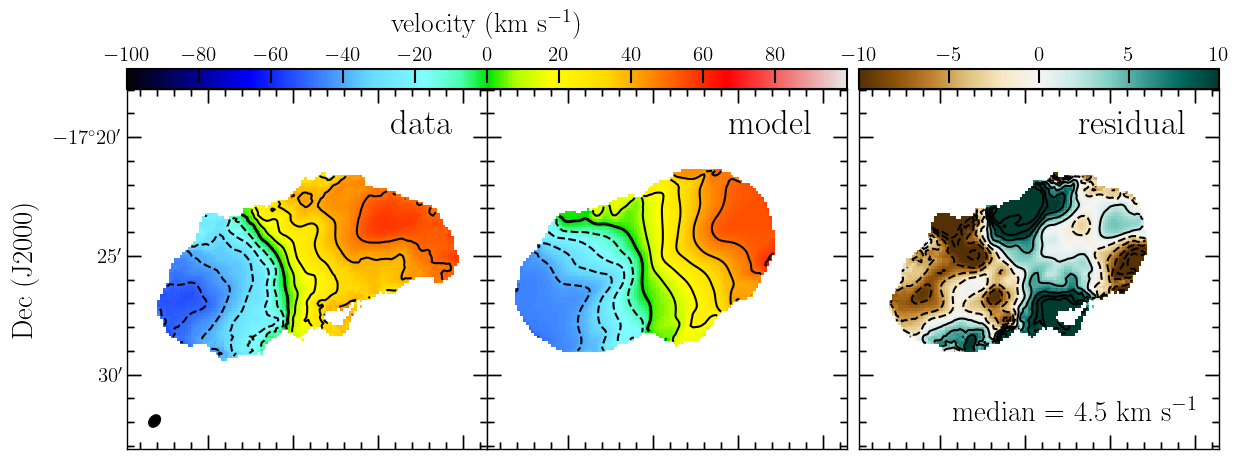}
	\caption{\refrep{Velocity residuals for the final model, where the change in systemic velocity has been removed.}}
	\label{subfig:mom1_no_vsys}
	\end{subfigure}
\end{figure*}

\begin{figure*}\ContinuedFloat 

    \begin{subfigure}{0.4\textwidth}
    \centering
	\includegraphics[width=\textwidth]{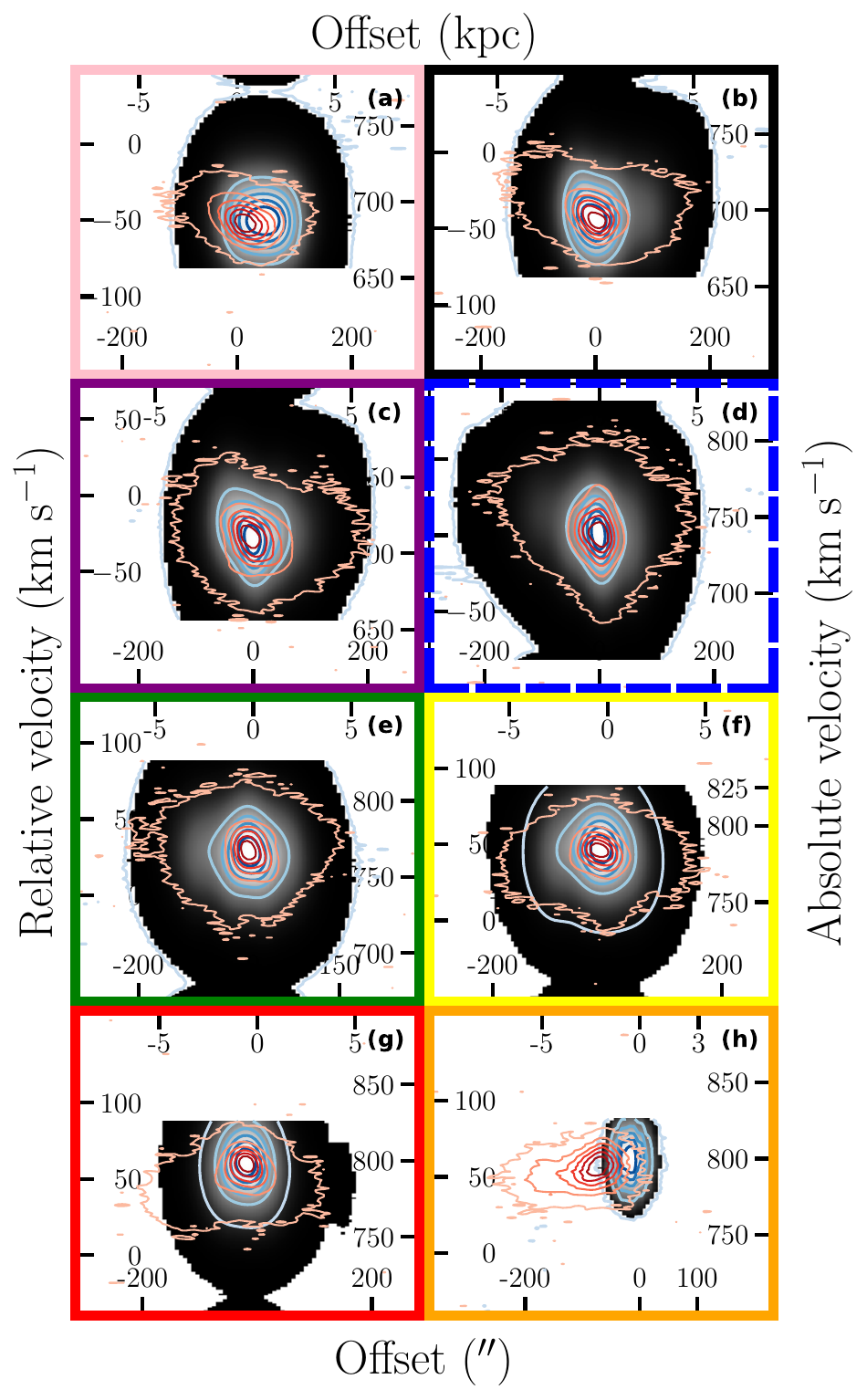}
	\caption{\refrep{PVD slices for the final model, but with a constant velocity dispersion of 15 km~s$^{-1}$, better representing the observed moment two values in the right-hand panel of Figure \ref{fig:moment_maps}}.}
	\label{subfig:pvd_mom2}
	\end{subfigure}	
	
	\begin{subfigure}{0.8\textwidth}
    \centering
	\includegraphics[width=\textwidth]{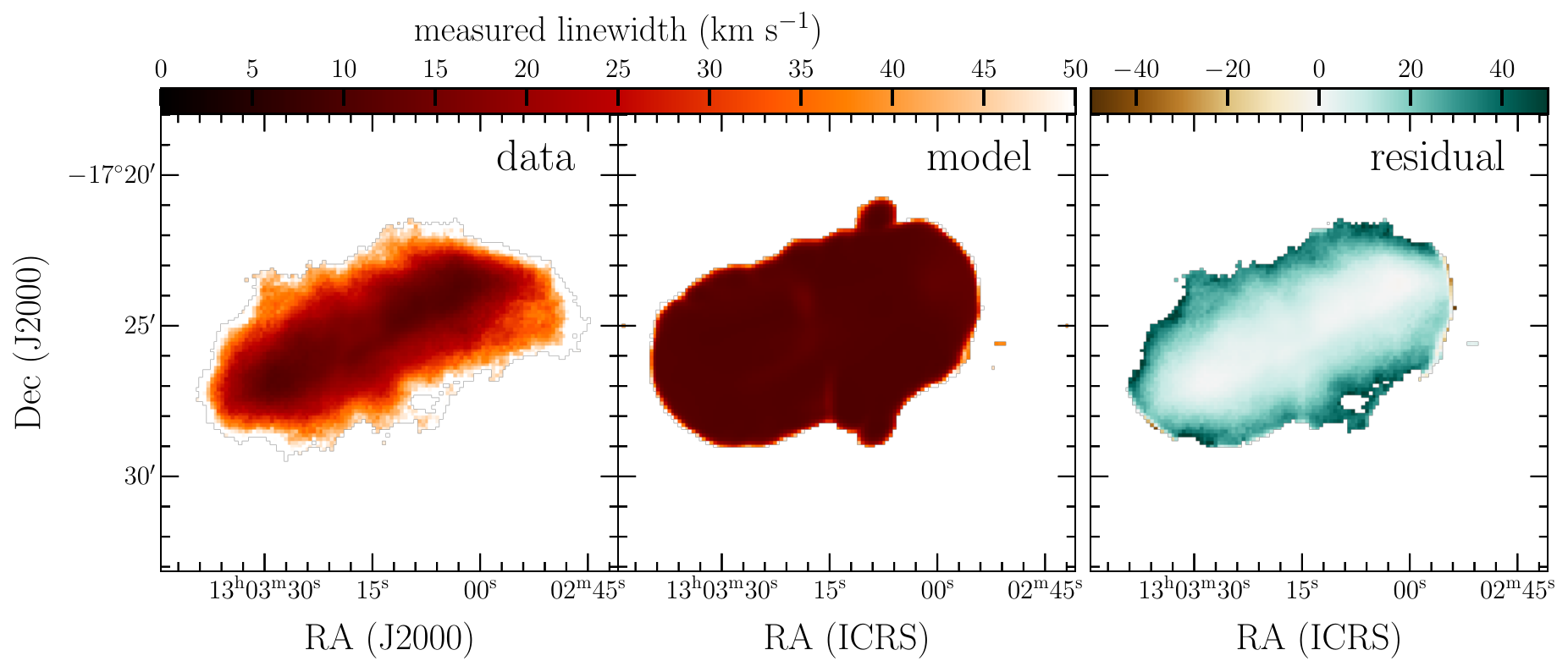}
	\caption{\refrep{Velocity residuals for the final model, but with a constant velocity dispersion of 15 km~s$^{-1}$.}}
	\label{subfig:mom2_mom2}
	\end{subfigure}	
	
	\caption{\refrep{Position-velocity diagrams parallel to the minor axis, and velocity maps, for alternative \texttt{TiRiFiC} models. The data for UGCA~320 is overlaid as red contours. For similar figures showing the final model, the data, and the paths along which the PVDs were extracted, we refer the reader to Figure \ref{fig:pvd_slices_final_model}}.}
	\label{fig:alternative_models}
	\end{figure*}



\bsp	
\label{lastpage}
\end{document}